\documentclass[prb,twocolumn,longbibliography,superscriptaddress,nofootinbib,floatfix]{revtex4-2}
\usepackage{graphicx}
\usepackage{textgreek}

\usepackage{amsmath}
\usepackage{amssymb}
\usepackage{hyperref}
\usepackage[caption=false]{subfig}
\usepackage[capitalize]{cleveref}
\usepackage{braket}
\usepackage{mathtools}
\usepackage{interval}
\intervalconfig{soft open fences}

\AtBeginDocument{

}

\DeclareMathOperator{\tr}{tr}

\DeclareMathOperator{\sgn}{sgn}
\DeclareMathOperator{\diag}{diag}
\DeclareMathOperator*{\argmin}{argmin}
\DeclarePairedDelimiter\abs{\lvert}{\rvert}

\begin{document}
\title{Unconventional discontinuous transitions in a 2D system with spin and valley degrees of freedom}
\author{Zachary M. Raines}
\affiliation{School of Physics and Astronomy and William I.
    Fine Theoretical Physics Institute, University of Minnesota, Minneapolis, MN 55455, USA}
\author{Leonid I. Glazman}
\affiliation{Department of Physics, Yale University, New Haven, CT 06520, USA}
\author{Andrey V. Chubukov}
\affiliation{School of Physics and Astronomy and William I. Fine Theoretical Physics Institute, University of Minnesota, Minneapolis, MN 55455, USA}
\begin{abstract}
We analyze the transition into the most favorable ordered state for a system of 2D fermions with spin and valley degrees of freedom.
We show that for
a range of rotationally invariant dispersions, the
ordering transition is
highly unconventional:  the associated susceptibility diverges (or almost diverges) at the transition, yet immediately below it the system jumps discontinuously into a fully polarized state.
We analyze
the dispersion of the longitudinal and transverse
collective modes
in different parameter regimes above and below the transition.
Additionally, we consider ordering in a system
with full $\mathrm{SU(4)}$ symmetry and show that
there is a cascade of discontinuous transitions into a set of states, which includes a quarter-metal, a half-metal and a three-quarter metal.
We compare our results with the data for biased bilayer and tri-layer graphene.
\end{abstract}

\maketitle

\section{Introduction}

Many interesting quasi-two-dimensional materials are described by a low energy theory with effectively conserved spin and valley isospin degrees of freedom.
The prototypical example is mono-layer graphene, whose low-energy theory is phrased in terms of two-fold spin-degenerate Dirac cones in the vicinity of the $K$ and $K'$ points of the Brillouin zone.
The $\mathfrak{su}(2)$ algebra of the valley isospin, in analogy to the usual spin algebra, has motivated many works considering ``valleytronics'' applications~\cite{Schaibley2016} which generalize earlier proposals for ``spin-tronics'' in semiconductors~\cite{Zutic2004}.
More generally, the existence of a larger low-energy spin/valley manifold with an approximate $\mathrm{SU(4)}$ symmetry presents opportunities for new phenomena allowed by the more complicated structure of the symmetry group.

Interest in two-dimensional multi-valley materials grew further with the discovery of correlated insulating states and superconductivity in twisted bilayer graphene (TBG)~\cite{Bistritzer2010,Cao2018,Cao2018a,Andrei2020}.
Here, among
other phenomena of interest,
studies of inverse compressibility have found~\cite{Xie2019,Wong2020,Zondiner2020}
a cascade of symmetry-breaking transitions between different isospin orders occurring as a function of electron density.
This discovery instigated a number of works on isospin symmetry breaking~\cite{Szabo2022,Chichinadze2020a,*Chichinadze2021,Dong2023,Dong2023a,*Dong2023aa}.

More recently work on Bernal-stacked bilayer graphene (BBG) and rhombohedral tri-layer graphene (RTG)  in displacement field has shown density tuned transitions between states with different breaking of the spin-valley isospin symmetry~\cite{Seiler2022,Zhou2022a,Zhou2021,DeLaBarrera2022,Seiler2023,Arp2023,Holleis2023,Zhang2023a,Blinov2023,*Blinov2023a}.
These ordered states remain metallic, but exhibit different numbers and sizes of Fermi surfaces,
reminiscent of itinerant ferromagnetism.
The competition between these states has been investigated theoretically within a Stoner-like framework to determine which type of order
develops first as a dimensionless coupling increases~\cite{ghazaryan2021unconventional,chatterjee2022inter,Chichinadze2022,Chichinadze2022a,You2022,Szabo2022,Xie2023,Dong2023,Dong2023b}.

The data on BBG and RTG revealed two features, which are seemingly incompatible with each other~(Refs.~\onlinecite{Zhou2021,Zhou2022a,DeLaBarrera2022,Arp2023,Holleis2023,Zhang2023a}).
On one hand, there are clear indications of soft boson excitations on the ``paramagnetic'' side of the transition;  on the other, the transitions are strongly first order.
This likely holds also for TBG, although the evidence is less certain.
First-order transitions into valley polarized/spin unpolarized and spin and valley polarized states, accompanied by
the growth of the corresponding susceptibilities
have been reported in experiments on a 2D electron gas
AlAs in certain ranges of the density-controlled
gas parameter
$r_s$~\cite{Hossain2021,*Hossain2022,*Hossain2020}.
First-order transitions into the same ordered states have been detected in variational Monte-Carlo calculations~\cite{Valenti2023}.

In this communication we argue that these two features are actually generic to Stoner-like
spin/isospin ordering of two-dimensional (2D) fermions with
a parabolic dispersion and, more generally, with power-law dispersion $\epsilon \sim k^{2\alpha}$, over a range of $\alpha$.
We consider a two-band model of spin-full fermions in 2D with circular Fermi surfaces and valley index associated with each band, and with short range intra-band and inter-band interactions.
We investigate how the behavior near a symmetry breaking transition depends on the global curvature of the electronic density of states~\cite{Shimizu1964} and analyze isospin collective modes in the vicinity of the transition, both in the disordered and in the ordered states.
We apply our results to twisted and un-twisted graphene-based materials.

\subsection{Summary of results}

\begin{figure}
    \centering
    \includegraphics[width=\linewidth]{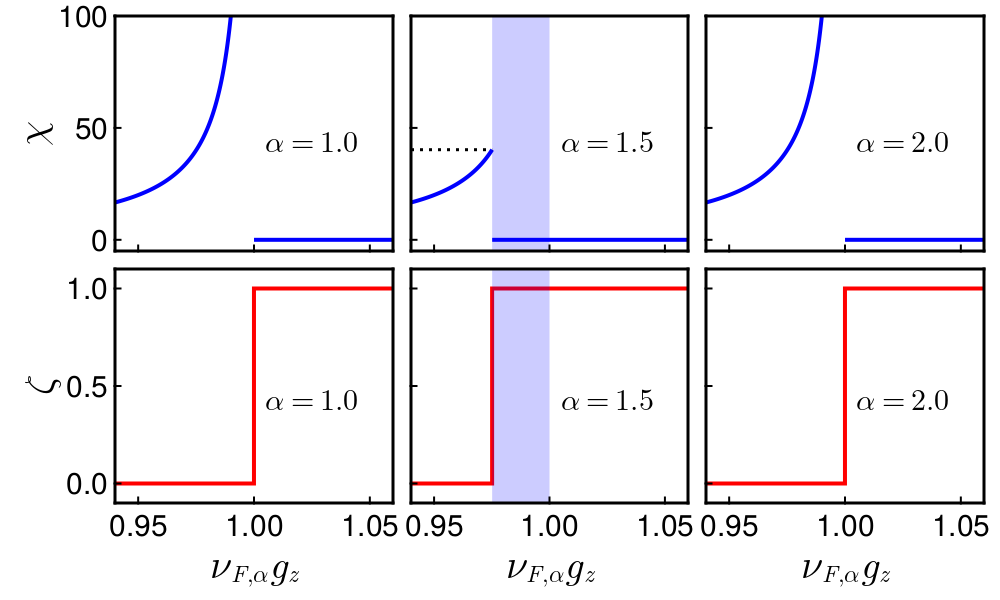}
    \caption{
    (Color online) The static valley isospin susceptibility $\chi$ (top panel) and polarization $\zeta$ (bottom panel) as a function of the dimensionless coupling $\nu_{F, \alpha}g_{z}$ for a selection of $\alpha$s between 1 and 2.
    The range of couplings where the
    system is bistable
    is highlighted in blue.
    For both $\alpha=1,2$ the susceptibility diverges at the critical point and there is only ever one local minimum.
    For generic $1\leq\alpha\leq2$ the behavior is qualitatively similar to $\alpha=1.5$.
    The susceptibility can increase by an order of magnitude or more before the divergence is cut off by the transition.
    \label{fig:chi_and_zeta}
    }
\end{figure}

The model we consider allows two types of charge instabilities: valley polarization (VP) and inter-valley coherence (IVC), and three types of spin instabilities: valley-symmetric ferromagnetism (FM), valley-staggered ferromagnetism (SFM), and
inter-valley spin-coherence order (sIVC).
When interactions favor one of these orders to develop prior to others, one would expect the system to undergo a second-order Stoner transition, below which the corresponding order parameter (polarization) gradually increases with the deviation from the instability, leading to a gradual evolution of the fillings of isospin bands.
This is what happens in three dimensions (3D).
In two-dimensions (2D) we found that for power law dispersions $k^{2}$ and $k^4$, there is instead a discontinuous transition from an unpolarized state with four (spin and valley) equally filled bands to a half-metal state with two bands fully depleted
(Figs\.~\ref{fig:schematic-fillings}b and~\ref{fig:su4_anisotropic}).
The transition is first order, yet it occurs at
exactly the critical coupling for the Stoner instability.
This leads to a situation where the isospin susceptibility
diverges at criticality, as would normally be expected for a second-order transition, yet immediately
beyond
the transition the polarization jumps to the maximum possible value
(\cref{fig:chi_and_zeta}a).
Upon further increase of the coupling, the system likely develops another transition of the same nature into a quarter-metal state
(\cref{fig:su4_anisotropic}).

For power-law dispersion $\epsilon (k) \propto k^{2\alpha}$, we found that the isospin order emerges continuously when $\alpha <1$ and $\alpha >2$ and
discontinuously when $1 \leq \alpha \leq 2$
(\cref{fig:su2_phase}).
Despite this distinction, for all $\alpha$ the isospin order parameter susceptibility strongly increases upon approaching the transition from the
paramagnetic
side and
on the ordered side the order parameter rapidly evolves near the critical point and very quickly reaches the largest possible value, much like
for $\alpha =1$ and $2$.

\begin{figure}
    \centering
    \includegraphics[width=\linewidth]{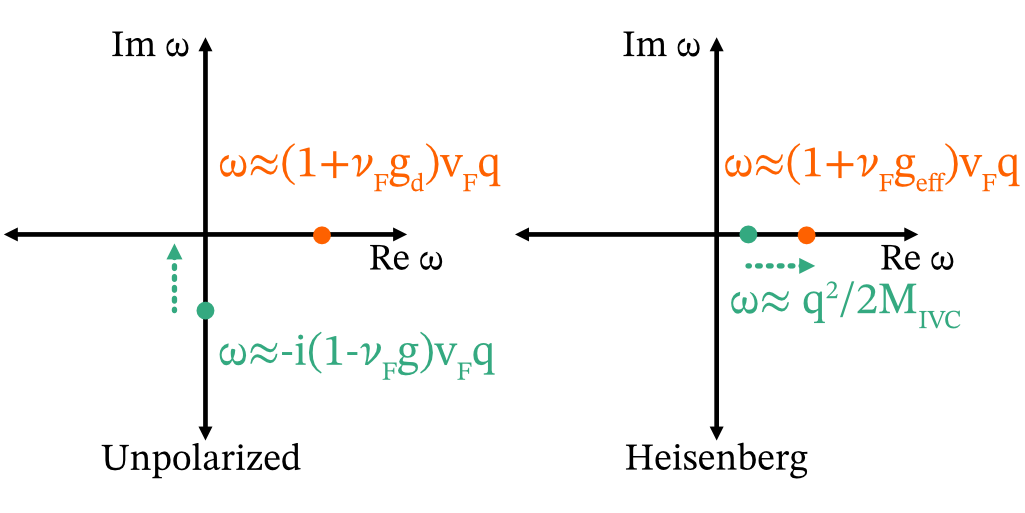}
    \caption{(Color online) Schematic behavior of the charge collective mode poles on both sides of the isospin transition for the
        isotropic case $g_{VP} = g_{IVC} =g$.
        The frequencies of the modes are shown in the complex plane for a fixed value of $q\ll k_{F}$.
        In the unpolarized state (left panel) there is a three-times degenerate,
        overdamped (exponentially decaying)
        isospin mode
        (green)
        whose `velocity' moves toward the real
        axis
        as the dimensionless coupling approaches $1$.
        In the ordered state (right panel),
        there is a propagating magnon mode, quadratic in $q$ (two poles in the susceptibility at $\pm \omega$),
        but the  longitudinal mode disappears.
        The picture is qualitatively similar for all three types of anisotropy: the ordered state hosts
        propagating transverse modes,
        but no
        longitudinal mode.  In both normal and ordered states, there is a propagating plasmon mode, whose frequency is located	just below the real axis (red dot).  The dispersion of the plasmon mode changes  in the ordered state, but the mode remains propagating.
        \label{fig:heisen-poles}}
\end{figure}

The unusual character of the Stoner transition in 2D can also be detected by analyzing the dispersion of collective modes in the particle-hole channel.
In the disordered (paramagnetic state'' the collective mode in the channel where order develops is an overdamped ``zero-sound'' mode.
It is located in the lower half-plane of complex frequency, indicating that a paramagnetic state is stable.
The velocity of the mode decreases as the system approaches a Stoner instability and approaches zero at the instability.
In 3D, where the Stoner transition is continuous, the mode (the longitudinal one for a multi-component order parameter) bounces back into the lower-half plane and its velocity gradually increases with the polarization.
In 2D, however, the mode continues into the upper half-plane for any partial polarization, indicating that a partially polarized state is unstable.
The exception is when
the polarization jumps to the maximum possible value.
In this case,
the longitudinal mode of the order parameter disappears, while all other collective modes remain in the lower half-plane of frequency
(\cref{fig:heisen-poles}).
The structure of the transverse modes depends on the type of order.

In the case where the
interaction in all spin-valley channels is equally strong,
the model becomes fully $\mathrm{SU(4)}$ symmetric.
We derived the full Landau functional without assuming that the $\mathrm{SU(4)}$ order parameter is weak.
The functional is rather involved and contains terms with even and odd powers of the order parameter.
We analyzed this functional and found that for a $k^2$ dispersion, susceptibilities for all order parameters diverge as the coupling reaches its critical value, yet the transition is discontinuous into a quarter-metal state
(\cref{fig:su4-phase}).
For dispersions between $k^{2}$ and $k^{4}$, there is a set of first-order transitions
with increasing fermionic density into a quarter-metal state, a half-metal state, and a three-quarter-metal state
(\cref{fig:density-cascade}a).
Each transition is accompanied by a near-divergence of the corresponding susceptibility.
For dispersions greater than $k^{4}$, and smaller than $k^2$, there exist intermediate partially polarized phases, but only in a very narrow range of densities
(\cref{fig:su4-phase} and~\ref{fig:density-cascade}b).
In all cases the behavior of the system in the normal state near the transition is determined by fermions near the Fermi surface,
while in the ordered state it is determined by high-energy fermions, even in the vicinity of the transition.
In each ordered state there are transverse Goldstone modes (four in a half-metal and three in a quarter-metal and in three-quarter metal), overdamped longitudinal modes (six in a three-quarter metal and three in a half-metal), and the plasmon mode describing fluctuations of the total density.
All modes are located in the lower half-plane of complex frequency.

These considerations are different from non-analytic aspects of a Hertz-Millis-Moriya-like description of itinerant ferromagnetism, which may lead to a weak first order transition already in 3D~\cite{Hertz1976,*Millis1993,*Moriya2012,Vojta1999,Chubukov2004,Efremov2008,Chubukov2009}.

\subsection{Outline}
The outline of the paper is as follows.
In \cref{sec:model} we outline the family of models we will consider
and discuss the set of potential spin/valley orders.
In \cref{sec:mf}, we consider VP and IVC orders.
In \cref{sec:collective}, we obtain and analyze the isospin collective modes in the VP and IVC ordered states.
Then, in \cref{sec:su4}, we analyze the ordering transition in a fully $\mathrm{SU(4)}$ invariant model with the same family of dispersions $\epsilon (k) \propto k^{2\alpha}$.
We derive the full Landau functional without assuming that the $\mathrm{SU(4)}$ order parameter is weak and analyze the ordered states.
In \cref{sec:conclusion} we summarize our results and apply them to interpret the data on the cascade of phase transitions in TBG, BBG and RTG.\@

\section{Model}\label{sec:model}

\begin{figure}
    \centering
    \includegraphics[width=0.6\linewidth]{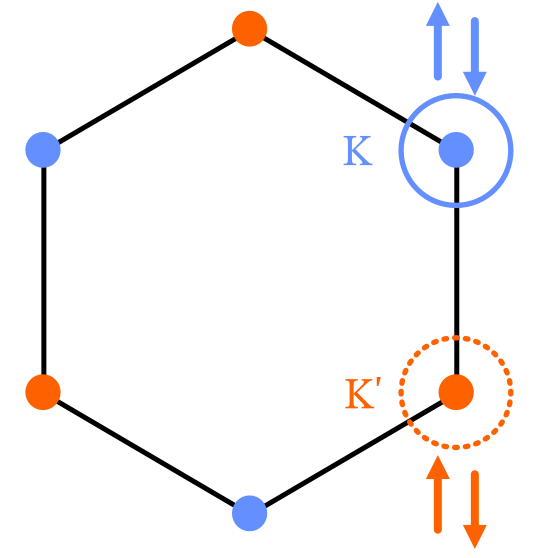}
    \caption{(Color online) Schematic illustration of the first Brillouin zone of mono-layer graphene, an archetypal example of a
        whose low-energy model involves spin and valley degrees of freedom.
        Inequivalent the low energy degrees of freedom are centered about the $K$ (blue) and $K'$ (orange) points.
        If we take the Fermi level to be in the conduction band, there is a Fermi surface at both $K$ and $K'$ points, and low energy degrees of freedom are labeled by both a valley index $\tau=K,K'$ as well as spin.\label{fig:graphene}
    }
\end{figure}

We neglect lattice warping effects and consider a low-energy model of spin-full 2D fermions with two equivalent bands, one per valley, with a rotationally symmetric power-law dispersion centered at $K$ and $K'$, and with short-ranged interactions.
The effective action is
\begin{multline}
    S =  -\sum_{k\tau}
    \bar{\psi}_{k\tau\sigma}(i\epsilon_{n} - \epsilon(\abs{\mathbf{k}}) + \mu)\psi_{k\tau\sigma}\\
    + \frac{1}{2}T^{2}
    \smashoperator{\sum_{\substack{kk'q\sigma\sigma'\\
                \alpha\beta\delta\gamma
            }}} g^{\alpha\beta\delta\gamma}\bar{\psi}_{k+\frac{q}{2}\alpha\sigma}\psi_{k-\frac{q}{2}\beta\sigma}
    \bar{\psi}_{k'-\frac{q}{2}\delta\sigma'}\psi_{k'+\frac{q}{2}\gamma\sigma'}
    \label{eq:S0}
\end{multline}
where $k = (\epsilon_{n},\mathbf{k})$, $q=(\omega_{m}, \mathbf{q})$, and fermions have spin
and valley degrees of freedom.
Summation over repeated spin and valley indices is implied.
We write $T \sum_{\omega_m}$ for notational convenience, although in this work we consider $T=0$, when $T \sum_{\omega_m} = \int d\omega_m/(2\pi)$.
The momentum $\mathbf{k}$ is measured with respect to the center of the band, e.g., for graphene $\mathbf{k}$ is measured with respect to the $K$ ($K'$) points for the $\tau=+$ ($-$) fermions (cf.~\cref{fig:graphene}).
\begin{figure}
    \centering
    \includegraphics[width=0.8\linewidth]{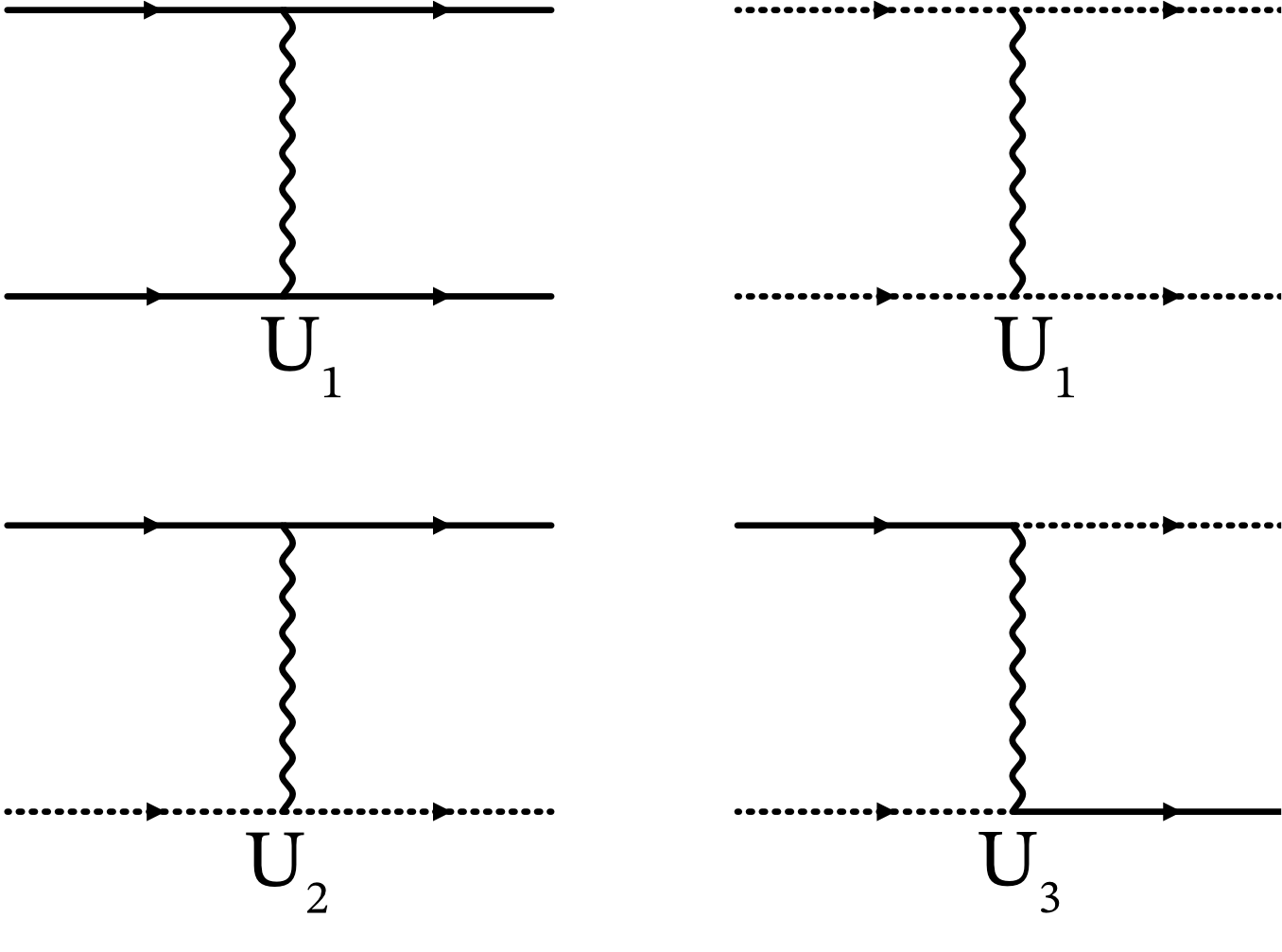}
    \caption{Feynman diagrams for interactions between the fermions in the two valleys included in the model \cref{eq:S0}.
        Solid lines indicate fermions in the $\tau=+$ valley while dashed lines indicate fermions in the $\tau=-$ valley.
        The top two panels are density-density interactions for particles in the same valley, $U_{1}$, while the bottom left describes the interaction between density in one valley and density in the other, valley $U_{2}$.
        These correspond to interactions with momentum transfer much smaller than the valley separation.
        The final diagram describes scattering of particles between valleys, $U_{3}$, corresponding to momentum transfers of order the valley separation scale.
        \label{fig:density-interactions}}
\end{figure}
The components of $g^{\alpha\beta\delta\gamma}$ are intra- and inter-valley density-density interactions $U_1$ and $U_{2}$ and inter-valley
iso-spin
exchange interaction $U_3$, depicted in~\cref{fig:density-interactions}
\begin{multline}
    g^{\alpha\beta\delta\gamma} = U_{1}\delta^{\alpha\beta}\delta^{\delta\gamma}
    + U_{2} \left(
    \tau^{\alpha\gamma}_{+}\tau_{-}^{\delta\beta}
    + \tau^{\alpha\gamma}_{-}\tau_{+}^{\delta\beta}
    \right)\\
    + U_{3} \left(
    \tau^{\alpha\beta}_{+}\tau_{-}^{\delta\gamma}
    + \tau^{\alpha\beta}_{-}\tau_{+}^{\delta\gamma}
    \right).
    \label{eq:gdef}
\end{multline}
All $U_i$
are properly screened Coulomb interactions.
Density-density interactions $U_1$ and $U_2$ have a small momentum transfer, within a single valley, while the
inter-valley isospin exchange interaction $U_3$ has a much larger momentum transfer close to $K-K'$.
For a generic $r_s = O(1)$, the Thomas-Fermi screening momentum $\kappa$ is of order $k_F$,
which is much smaller than $K-K'$.
In this situation,
$U_3$ is far smaller than $U_1$ and $U_2$.
The latter two are the same if only particle-hole bubble diagrams are kept for screening, but generally differ somewhat once one adds additional contributions, e.g., the renormalization of the interaction in the particle-particle channel.

Below we approximate $U_1$ and $U_2$ as static, i.e., include only the static screening of the Coulomb interaction.
Our reasoning is based on the observation (see a footnote below) that the density-density interactions, which appear in the order parameter susceptibilities and in the dispersions of the corresponding collective modes, are at characteristic momenta of order $k_F$ (and hence of order $\kappa$), even if the order is a homogeneous one, with momentum $q=0$.
For such interactions, using static screening is justified, at least qualitatively.
The Coulomb interaction at $q \to 0$, for which the dynamical screening is crucial, cancels out in the susceptibilities and the dispersion of the collective modes.
The only collective excitation for which this does not happen and one has to keep the full dynamical screening of the Coulomb interaction, is the one describing fluctuations of the total density.
If we included only the static screening, we would find that this fluctuation is a propagating zero sound mode with a linear dispersion.
In reality, fluctuations of a total density of a system of charged fermions are plasmons.
Total density fluctuations will not play any substantial role in our analysis, and for simplicity of presentation, below we will keep interactions as static in the analysis of collective excitations in all channels, including the total density channel.
We will keep reminding, however, that collective modes in the total density channel are actually plasmons.

\Cref{eq:S0} can also be written in a fully anti-symmetrized form to explicitly treat all channels on the same footing
\begin{multline}
    S =  -\sum_{k\tau}\bar{\psi}_{k}(i\epsilon_{n} - \epsilon(\abs{\mathbf{k}}) + \mu)\hat{\Gamma}_0\psi_{k}\\
    + \frac{1}{8}T^{2} \sum_{kk'q} g_{d}\bar{\psi}_{k+\frac{q}{2}}\hat{\Gamma}_0\psi_{k-\frac{q}{2}}\bar{\psi}_{k'-\frac{q}{2}}\hat{\Gamma}_0\psi_{k'+\frac{q}{2}}\\
    - \frac{1}{8}T^{2}\sum_{kk'q} \sum^{15}_{\gamma=1} g_{\gamma}\bar{\psi}_{k+\frac{q}{2}}\hat{\Gamma}_{\gamma}\psi_{k-\frac{q}{2}}\bar{\psi}_{k'-\frac{q}{2}}\hat{\Gamma}_{\gamma}\psi_{k'+\frac{q}{2}}.
    \label{eq:Smodel}
\end{multline}
Here $g_d = U_1+2U_2 - U_3$ is the interaction in the total density channel,
$\hat{\Gamma}_0=\sigma_0\tau_0$ is the identity matrix,
$\hat{\Gamma}_{\gamma}$ are the generators of rotations in the 4-component spin-valley space and $g_{\gamma}$ are the corresponding couplings, all expressed as linear combinations of $U_{1}$, $U_{2}$ and $U_{3}$.
The fifteen generators of SU(4) correspond to the five types of spin/valley order, which can develop in the model~\cite{Chichinadze2022a,Dong2023b}:
one-component VP ($\tau_z \sigma_0$, $g_{\text{VP}} = 2U_2-U_1-U_3$),
three-component FM ($\sigma_{x,y,z}\tau_0$, $g_{FM} = U_1 +U_3$),
three-component SFM ($\sigma_{x,y,z}\tau_{z}$, $g_{SFM} = U_1 -U_3$),
two-component IVC
($\tau_{x,y} \sigma_0$,  $g_{\text{IVC}} = U_2-2U_3$),
and six-component sIVC ($\sigma_{x,y,z}\tau_{x,y}$, $g_{sIVC} = U_2$).
\footnote{In a more accurate treatment, the combination $2U_2 - U_1$ in $g_{\text{VP}}$,
    which is associated with a  $q=0$  order,
    should be replaced by ${\bar U}_1  + 2(U_2 (0) -U_1(0))$,
    where ${\bar U}_i$ is the Coulomb interaction averaged over momenta of order $k_F$, and $U_i (0)$ are the interactions at zero momentum transfer. The interactions $U_1 (0)$ and $U_2 (0)$ remain identical as long as they are screened by
    the dynamical $\Pi (q \to 0)$ and cancel out in $g_{\text{VP}}$.
    These interactions do become different once one includes other renormalizations,t e.g., the insertions of particle-particle polarizations.
    For the latter, however, typical momenta are of order $k_F$, and static screening is sufficient.}

For our case, where $\tau$ describes a valley degree of freedom, the IVC and SFM states are time reversal invariant while the VP, FM and sIVC states spontaneously break time reversal symmetry.
As defined,
$g_d >0$, hence there
is no instability in the total density channel (the one that leads to phase separation).

\cref{eq:Smodel} is particularly useful for the purely $\mathrm{SU(4)}$ symmetric case,
where all $g_{\gamma}$ are equal.
This holds when the two interactions with small momentum transfer, $U_1$ and $U_2$ are equal and the interaction $U_3$ with a finite momentum transfer is set to zero.
We consider this case in \cref{sec:su4}, but before that we
consider a
simpler case when $U_1$ and $U_2$ are not equal, and $U_3$ is finite~\cite{Aleiner2007,*Kharitonov2012,*Raines2021,Chichinadze2022a,Dong2023b}.
In this situation, the couplings in different channels are generally non-equal, and the corresponding instabilities can be treated independent of each other.

\section{isospin transition for non-equal couplings}\label{sec:mf}

\begin{figure}
    \centering
    \includegraphics[width=0.7 \linewidth]{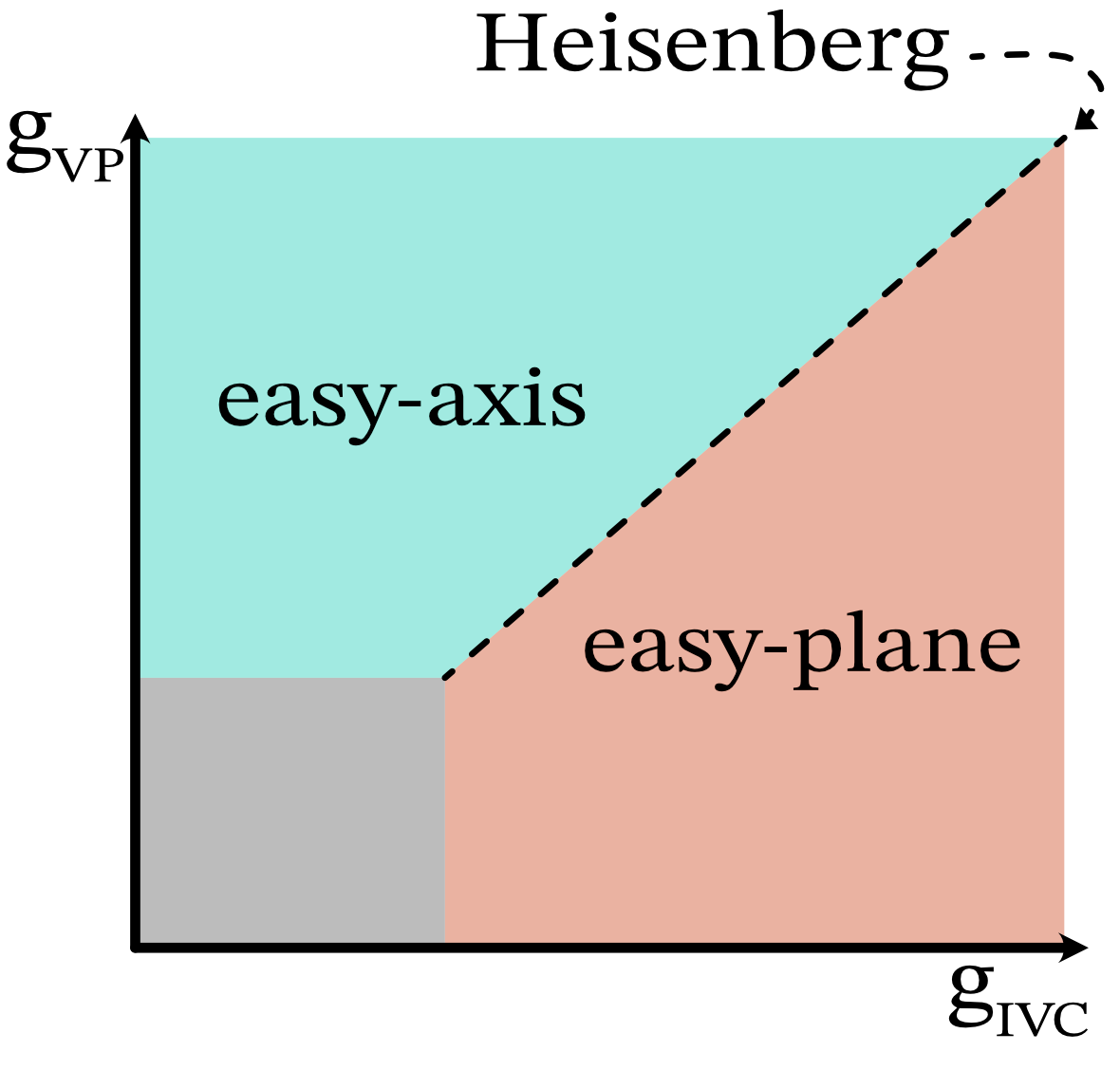}
    \caption{
        (Color online)
        The three types of valley isospin states we consider in terms of two independent
        couplings $g_\text{VP}$ and $g_\text{IVC}$.
        The easy-plane and
        easy axis states occur when $g_{\text{IVC}}$ or $g_{\text{VP}}$ is the larger, respectively.
        The Heisenberg state is on the boundary between the easy-axis and easy-plane states when $g_{\text{VP}} = g_{\text{VC}}$.
        We orient our isospin coordinates such that the strongest instability is along $\hat{\tau}_{z}$.
        \label{fig:gphase}}
\end{figure}
To be concrete, we will focus on instabilities in the charge IVC and VP channels;
we imagine a situation where the interactions in these channels are largest, and interactions in
other (spin)
channels can be treated as small perturbations~\footnote{
    Note that we have chosen the valley isospin (IVC and VP) channels as a simple illustrative example, but the results hold equivalently when the strongest interactions are in any $\mathrm{SU(2)}$ subspace of the four-dimensional valley and spin manifold.
}.
We treat the IVC and VP instabilities and corresponding orders in a unified language by choosing a basis of valley isospin $\tau$  such that the leading instability is always in the $\hat{\tau}_{z}$ channel.
We denote the creation operators of this basis $c^{\dagger}_{\mathbf{k}\lambda\sigma}$ where $\lambda=\pm$.~\footnote{
    In the VP state, $\lambda$ labels the bands associated with a given valley, while in the IVC case
    it labels linear superpositions
    of these bands.}
For the VP case this
coincides with our original basis, $\psi_{\pm\sigma}=c_{\pm\sigma}$,
for the IVC case
$\psi_{\sigma}=e^{-i\hat{\tau}_{y}\pi/2}c_{\sigma}$.
In terms of the $c$ operators, the
VP and IVC orders can be treated on equal footings, with the effective action
\begin{multline}
    S =  -\sum_{k \lambda}\bar{c}_{k}(i\epsilon_{n} - \epsilon(\abs{\mathbf{k}}) + \mu)\hat{\tau}_{0}\hat{\sigma}_{0}c_{k}\\
    + \frac{1}{8}T^{2} \sum_{kk'q } g_{d}\bar{c}_{k+\frac{q}{2}\sigma}\hat{\tau}_{0}\hat{\sigma}_{0}c_{k-\frac{q}{2}\sigma}\bar{c}_{k'-\frac{q}{2}\sigma'}\hat{\tau}_{0}\hat{\sigma}_{0}c_{k'+\frac{q}{2}\sigma'}\\
    - \frac{1}{8}T^{2}\sum_{kk'q} \sum_{i\in\{x,y,z\}} g_{i}\bar{c}_{k+\frac{q}{2}}\hat{\tau}_{i}\hat{\sigma}_{0}c_{k-\frac{q}{2}}\bar{c}_{k'-\frac{q}{2}}\hat{\tau}_{i}\hat{\sigma}_{0}
    c_{k'+\frac{q}{2}}.
    \label{eq:Ssimp}
\end{multline}
Here one of the couplings $g_{x,y,z}$
is $g_{\text{VP}}$ and two others are $g_{\text{IVC}}$, the choice depends on which order develops first.
We call the regime where $g_{\text{VP}} > g_{\text{IVC}}$ \emph{easy-axis} (Ising), the one where $g_{\text{IVC}} > g_{\text{VP}}$ \emph{easy-plane}, and the one where $g_{\text{VP}} = g_{\text{IVC}}$ \emph{Heisenberg}.
These names reflect that dimensionality of the order parameter manifold in each case.
These regions are depicted schematically in~\cref{fig:gphase}.
The relations between $g_{\text{VP}}, g_{\text{IVC}}$ and $g_x$, $g_y$ and $g_z$ are given in \cref{tab:isospin-interaction}.
Note that by our convention, $g_{z}$ is always the strongest isospin interaction.

\begin{table}
    \centering
    \begin{tabular}{l|c|c|c|c}
                            & Interactions                     & $g_{x}$          & $g_{y}$          & $g_{z}$          \\
        \hline
        \textbf{Easy-axis}  & $g_{\text{VP}} > g_{\text{IVC}}$ & $g_{\text{IVC}}$ & $g_{\text{IVC}}$ & $g_{\text{VP}}$  \\
        \textbf{Easy-plane} & $g_{\text{VP}} < g_{\text{IVC}}$ & $g_{\text{VP}}$  & $g_{\text{IVC}}$ & $g_{\text{IVC}}$ \\
        \textbf{Heisenberg} & $g_{\text{VP}} = g_{\text{IVC}}$ & $g$              & $g$              & $g$              \\
    \end{tabular}
    \caption{Interaction constants $g_{x,y,z}$ in the rotated basis for each type of isospin ordered state in terms of inter-valley coherence and valley polarization constant $g_{\text{IVC}}$ and $g_{\text{VP}}$.\label{tab:isospin-interaction}}
\end{table}

\subsection{Isospin transition}\label{sec:mf_1}

We begin by considering the nature of the $T=0$ isospin transition.
By our convention,
the order parameter is given by the expectation value
\begin{equation}
    \Delta \equiv
    \frac{1}{4}
    g_{z} T\sum_{k}\Braket{\bar{c}_{k\sigma}\hat{\tau}_{z}c_{k\sigma}}.
    \label{eq:self-consistency}
\end{equation}
As such,
the self-consistency equation
takes the same form for the easy-axis, easy-plane, and Heisenberg cases.
The order parameter $\Delta$ acts as a valley-dependent modification of the Fermi level, $E_{F\lambda}= \bar{\mu} \pm \Delta$ for the two $\lambda$ bands.
Here we write $\bar{\mu}$ to highlight the fact that the chemical potential in the ordered state must change with $\Delta$ in order for the total density to be conserved (i.e., in general $\bar{\mu}\neq\mu$).

Let $n_{\lambda}$  denote the density per spin in each band.
The total fermionic density is $n \equiv 4n_{0} = \sum_{\sigma} \sum_{\lambda=\pm} n_{\lambda}$.
In the unpolarized state all bands are degenerate and thus have equal densities $n_{\lambda} = n_{0}$,
but in the ordered phase an imbalance develops between the bands.
We parameterize the band densities $n_\lambda \equiv n_0 (1 + \lambda \zeta)$ with the polarization $-1\leq \zeta \leq 1$.

The mean-field internal energy is simply the sum of the kinetic and potential terms
\begin{equation}
    U_{MF} = 2\sum_{\lambda} u_{K}(n_{\lambda})
    -
    \frac{1}{2}
    g_{z}(
    \sum_{\lambda} \lambda n_{\lambda})^{2}
    \label{eq:Umf}
\end{equation}
The expectation value of the order parameter $\Delta$ is related to the band densities as
$\abs{\Delta} = (1/2) g_{z}\sum_{\lambda}\lambda n_{\lambda}$ (see \cref{eq:self-consistency}).
The kinetic energy of each band can be obtained by integrating the corresponding density dependent Fermi-energy
\begin{equation}
    u_{K}(n_{\lambda})  = \int_{0}^{n_{\lambda}}dn' E_{F}(n') = \int_{0}^{n_{\lambda}}dn' \epsilon(\sqrt{4\pi n'}).
    \label{eq:kinetic-2d-rot}
\end{equation}

\begin{figure}
    \centering
    \subfloat[\label{fig:su2disordered}]{\includegraphics[width=0.3\linewidth]{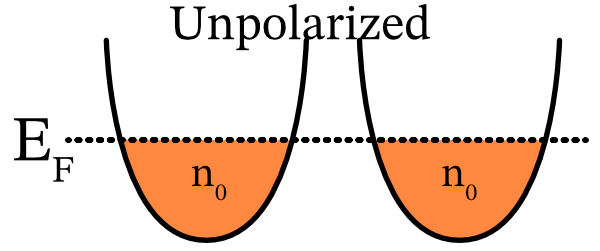}}
    \subfloat[\label{fig:su2full}]{\includegraphics[width=0.3\linewidth]{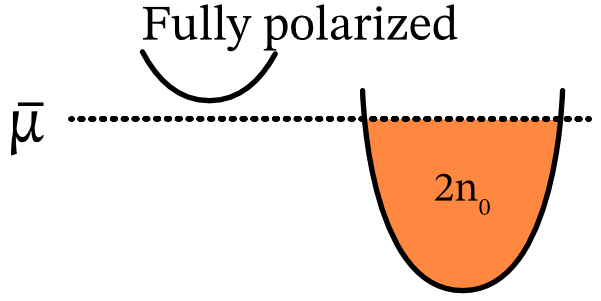}}
    \subfloat[\label{fig:su2partial}]{\includegraphics[width=0.3\linewidth]{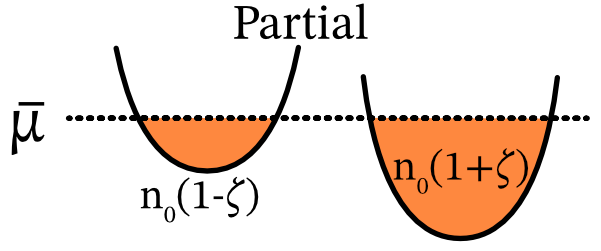}}
    \caption{(Color online) Schematic filling of the spin degenerate valley isospin bands for the (a) normal, (b) fully polarized, and (c) partial polarized states.
        \label{fig:schematic-fillings}}
\end{figure}
Already, we notice something special about the case of a parabolic dispersion.
Letting $\epsilon(\abs{\mathbf{k}}) = k^{2}/2m$ we may straightforwardly evaluate
\begin{equation}
    u_{K} (n_{\lambda}) = \int_{0}^{n_{\lambda}}dn' \frac{4\pi n'}{2m} = \frac{1}{2\nu_{F,1}} n_{\lambda}^{2},
\end{equation}
where $\nu_{F,1} = m/2\pi$ is the density of states per spin and valley (the sub-index $1$ in $\nu_{F,1}$ stands for
the exponent $\alpha =1$ in the dispersion $k^{2\alpha}$, see below).
In the unpolarized state, the Fermi momentum in each band is $k^2_F = 4\pi n_0$, and the common chemical potential is $\mu = n_0/\nu_{F,1}$.
Since the potential term in \cref{eq:Umf} is also quadratic, the energy is thus \emph{exactly} quadratic in the polarization $\zeta$.
If we evaluate the energy difference between the normal and ordered states we find the dimensionless Landau internal energy
\begin{equation}
    \delta u \equiv \nu_{F, 1}\frac{U_{MF}  - U_{N}}{4n_{0}^{2}} = \frac{1}{2}\left(1
    - \nu_{F,1} g_{z}\right)\zeta^{2}.
    \label{eq:deltau-parabolic}
\end{equation}
A critical point, where the unpolarized state becomes unstable,
is set by the usual Stoner criterion $\nu_{F,1} g_{z,cr}= 1$.
At the same time, immediately upon crossing this point, the system's energy is minimized by setting
$\zeta$
to the largest possible value, since the energy profile is simply an inverted parabola.
We thus have a situation where the susceptibility in the VP or IVC channel diverges as the critical point is approached, and
just beyond it there is a discontinuous jump of the order parameter to a full polarization state with $\zeta = \pm 1$.
For such $\zeta$, $n_+ =0$ or $n_{-} =0$, i.e.,  one of the bands becomes fully depleted immediately beyond the critical point.
This can be also seen by comparing the magnitude of the order parameter
at the critical point, $g_{z}=g_{z,cr}$,
$\Delta =  n_0 g_{z, cr} = n_0/\nu_{F,1}$ with the chemical potential at this point $\bar{\mu} = k_{F}^2/(2m) = n_0/\nu_{F,1}$.
We see that the two are equal and therefore the Fermi level of one of the bands $E_{F} = \bar{\mu} - \Delta$ becomes 0.
For larger $g_z$, the Fermi level of the occupied band remains constant at twice the unpolarized state value, $E_{F} = \bar{\mu} + \Delta = 2n_{0}/\nu_{F,1}$, due to conservation of charge, while the Fermi level of the other band, $E_F = \bar{\mu} -\Delta = -2n_{0}(g_{z}-1/\nu_{F, 1})$, continues to decrease below the band bottom.
We show this in \cref{fig:su2full}.

\begin{figure}
    \centering
    \includegraphics[width=\linewidth]{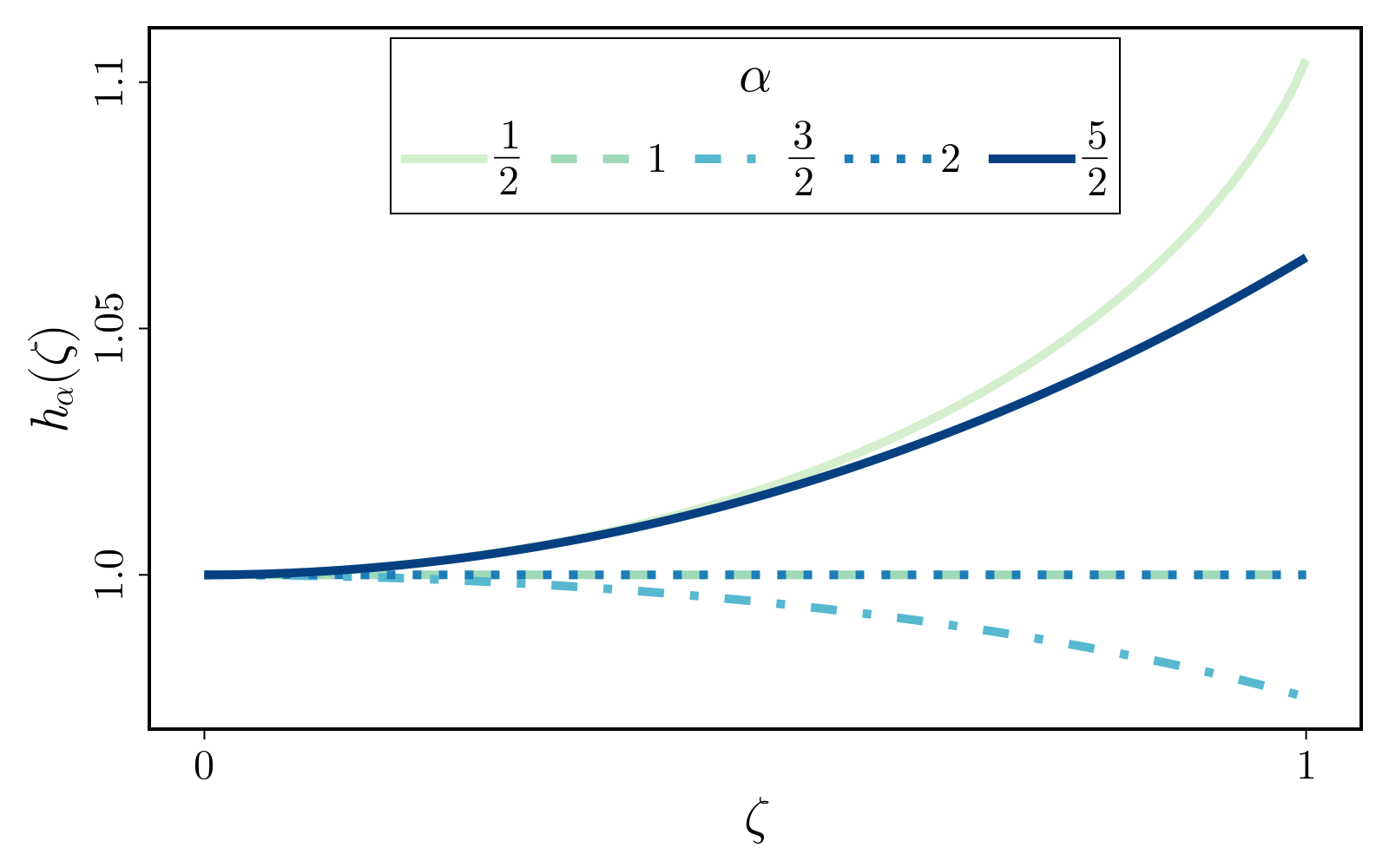}
    \caption{The function $h_{\alpha}(\zeta)$, \cref{eq:h}, for a variety of power law dispersions $\epsilon \sim k^{2\alpha}$.
        The leading valley isospin instability occurs at the value of $\zeta$ which minimizes $h(\zeta)$.
        For $1<\alpha<2$ $h$ has minima at $\zeta = \pm 1$, while for $\alpha < 1$ or $\alpha > 2$ it has a minimum at $\zeta=0$.
        \label{fig:hzeta}
    }
\end{figure}

The fact that the point where the susceptibility diverges and the first-order transition point exactly coincide is a special property of the quadratic dispersion and the $T=0$ limit~(see \cref{sec:hs-deriv} for more details), but we now show that a
similar behavior holds for a range of dispersions.
For this we consider the family of power-law dispersions
\begin{equation}
    \epsilon(\abs{\mathbf{k}}) = c\left(\frac{k^{2}}{4\pi}\right)^{\alpha}, \quad \alpha > 0.
    \label{eq:eps-power}
\end{equation}
For such $\epsilon(\abs{\mathbf{k}})$ the kinetic energy \cref{eq:kinetic-2d-rot} is
\begin{equation}
    u_{K}(n_{\lambda}) = \frac{c}{\alpha + 1}n_{\lambda}^{\alpha+1} = \frac{cn_{0}^{\alpha+1}}{\alpha + 1}(1+\lambda\zeta)^{\alpha+1}.
\end{equation}
Proceeding as in the parabolic case \cref{eq:deltau-parabolic} we write the dimensionless Landau energy in the form
\begin{equation}
    \delta u
    =
    \frac{1}{2}
    \left(
    h(\zeta)
    -
    \nu_{F,\alpha} g_{z}
    \right)\zeta^{2}.
    \label{eq:delta-u}
\end{equation}
where
\begin{equation}
    \nu_{F,\alpha} = \frac{1}{c \alpha n^{\alpha-1}_{0}}
    \label{eq:normal-dos}
\end{equation}
and
\begin{equation}
    h(\zeta)  \equiv
    \frac{1}{\alpha( \alpha+1)}\frac{
        (1+\zeta)^{\alpha+1}
        + (1-\zeta)^{\alpha+1}
        -2
    }{\zeta^{2}}.
    \label{eq:h}
\end{equation}
Writing the energy in this form allows us to extract the behavior of the instability in terms of the properties of $h(\zeta)$;
the leading instability is to a state which minimizes $h(\zeta)$.
We plot $h(\zeta)$ for different values of $\alpha$ in \cref{fig:hzeta}.
For any $\alpha >0$, $h(0) =1$, hence the usual Stoner criterion, i.e., the point where the susceptibility diverges, is $g_z \nu_{F,\alpha} =1$.

We first notice that for $\alpha=2$, i.e., for a $k^4$ dispersion, \cref{eq:h} yields $h (\zeta) =1$ identically, the same as for $k^2$ dispersion.
Accordingly, the ordering transition happens exactly in the same way as for a parabolic dispersion: the isospin susceptibility diverges at $g_{z,cr} = 1/\nu_{F,2}$, and immediately
beyond
the transition the polarization jumps to the largest possible value, and one of the two bands becomes depleted.

We next consider a generic $\alpha$.
Let us start from a small value of $g_{z}$ such that $\delta u$ is positive for all $\zeta \in \interval{0}{1}$.
As we make $g_{z}$ larger, the system will experience a transition when the minimum of the term in parentheses in~\cref{eq:delta-u} crosses zero.
Infinitesimally on the other side of the transition the energy will be minimized by a state with polarization $\bar{\zeta} = \argmin_{\zeta \in \interval{-1}{1}}h(\zeta)$.
For $1\leq\alpha\leq2$, $h(\zeta)$ is a monotonically decreasing function
of $|\zeta|$ in the interval $\linterval{0}{1}$.
Therefore, its minimum is always at $\abs{\bar{\zeta}}=1$.
This implies that the system
undergoes a first-order transition at a threshold value of the dimensionless coupling
\begin{equation}
    \nu_{F,\alpha} g_{z,cr} =
    h(1)=
    \frac{2}{\alpha(\alpha+1)}\left(2^{\alpha}-1\right).
    \label{eq:gt}
\end{equation}
For $\alpha =1$ and $\alpha =2$, $h(1) = h(0) =1$, and \cref{eq:gt} coincides with the Stoner criterion $g_{z,cr} = 1/\nu_{F,\alpha}$.
For intermediate $1< \alpha <2$, $h(1) <h(0)$ as $h(\zeta)$ is a monotonically decreasing function, and so the instability occurs at smaller values of the interaction~\footnote{
    This behavior can already be seen at the level of the Landau theory by noting the negativity of the quartic term for this range of dispersions~\cite{Tremblay2021}.
}
Recalling that we defined the band occupations as $n_{\pm} = n_{0} (1\pm\zeta)$, we see that
the first-order transition is into a fully polarized state, where one band is completely filled while the other is completely depleted, like in the case of a parabolic dispersion.
One can again verify that
beyond the transition the magnitude of the order parameter $\abs{\Delta}$
exceeds the chemical potential $\bar{\mu}$ so that Fermi level of one of the bands $E_{F\pm}= \bar{\mu} \pm\Delta$ falls at or below the band bottom, like in \cref{fig:su2full}.

\begin{figure}
    \centering

    \includegraphics[width=\linewidth]{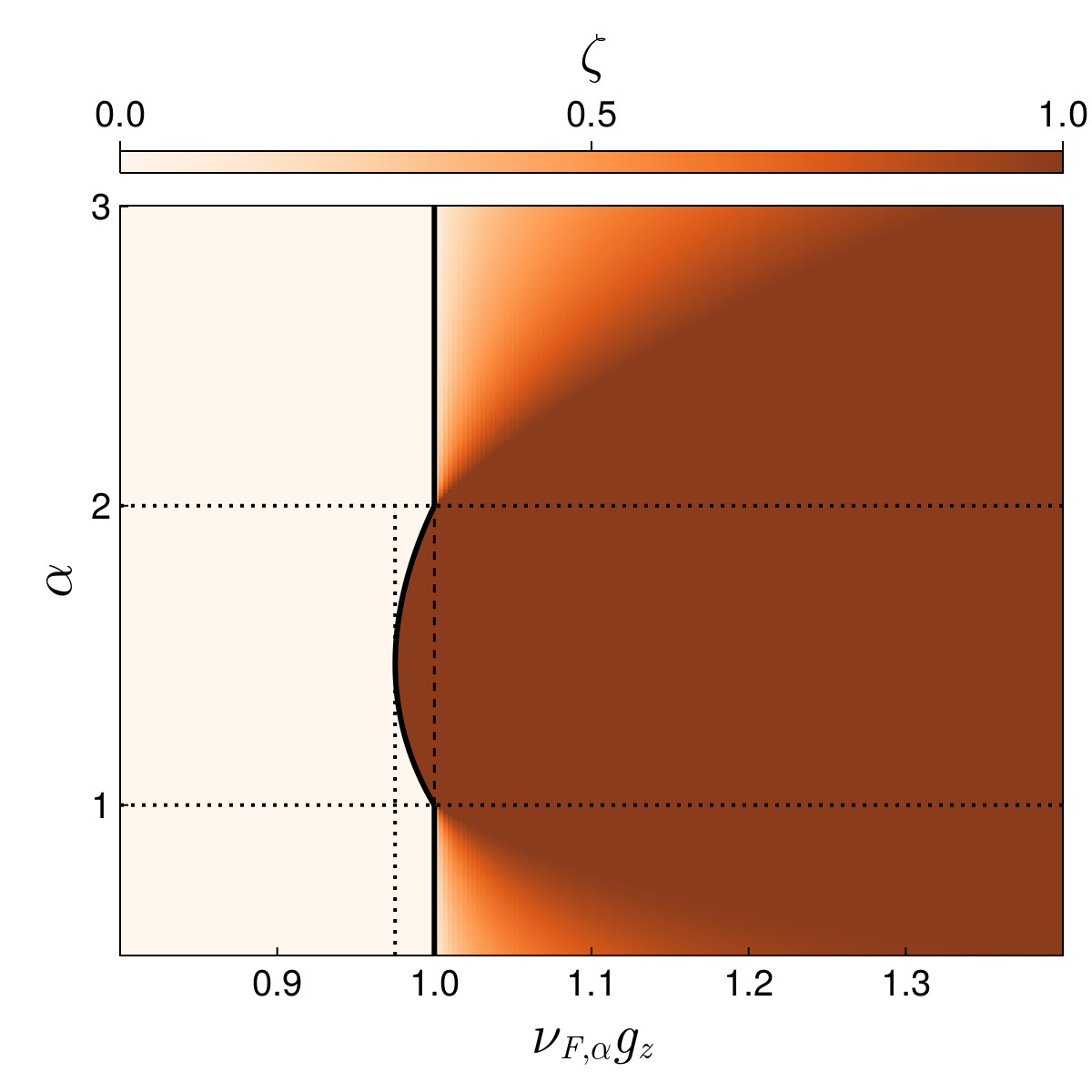}
    \caption{(Color online)
    Phase diagram as a function $\alpha$, the power of the electron dispersion $\epsilon\sim k^{2\alpha}$, and the dimensionless coupling strength $\nu_{F,\alpha}g$.
    The color scale shows the polarization $\zeta$.
    For $1\leq\alpha\leq2$ this is a first order transition from the unpolarized state \cref{fig:su2disordered}, $\zeta=0$, to a state of maximum polarization \cref{fig:su2full}, $\abs{\zeta}=1$, while for $\alpha < 1$ or $\alpha >2$ there is a second order transition as given by the Stoner criterion $\nu_{F,\alpha}g_{z}=1$ to a partially polarized state \cref{fig:su2partial}.
    Even in the case of a second order transition, the width of the partially polarized region is confined to a narrow window in the vicinity of the Stoner coupling $g_{z}=1/\nu_{F,\alpha}$.
    \label{fig:su2_phase}
    }
\end{figure}

We found (see \cref{fig:su2_phase}) that for all $\alpha$ between $1$ and $2$,
$\nu_{F,\alpha} g_{z,cr}$ remains very close to one: the deviation $1 - \nu_{F,\alpha} g_{z,cr}$ does not exceed $0.025$.
For all practical purposes it looks like for all $1\leq \alpha \leq 2$ the
isospin susceptibility diverges at the onset of a first-order transition into a fully polarized state.

For values of $\alpha$ outside this range, i.e., $\alpha < 1$ or $\alpha>2$,
$h(\zeta)$ is a monotonically increasing function of $\zeta$.
In this situation,
the system undergoes a second-order phase transition at $\nu_{F ,\alpha} g_{z,cr}= 1$ as given by the Stoner criterion; this case includes Dirac-like linear dispersions~\footnote{
    The analysis performed here is valid away from the Dirac point, but near charge neutrality the contribution from the valence band must also be included.
} ($\alpha=1/2$) as well higher powers like the $k^{6}$ dispersion ($\alpha=3$) which was argued to hold in RTG in a displacement field~\cite{Koshino2009,Zhou2021}.
Yet, the largest value $|\zeta|=1$ is reached at $g_z$ only slightly larger than $1/\nu_{F,\alpha}$ (saturation is reached at $g_z =1.4/\nu_{F,\alpha}$ for $\alpha =0.5$ and $\alpha =3$).

It is worth noting that the Fermi surface in the fully polarized
isospin state $|\zeta|=1$ is spin degenerate.
Interactions in the spin channel are attractive, we just assumed that they are weaker than in VP and IVC channels.
If we keep increasing $\nu_{F, \alpha} g_z$, we very likely obtain the condition for a secondary transition, in which spin order develops (and the system possesses both isospin and spin order).
The effective action for the spin degrees of freedom
is qualitatively similar to \cref{eq:Ssimp},
hence the behavior is the same as the one we just described.
Namely, if $\alpha$ is between $1$ and $2$, the spin susceptibility strongly increases as the instability is approached, but immediately beyond it,
the spin order parameter jumps to its maximal value and one of the bands becomes fully depleted.
This gives rise to a quarter metal phase (see \cref{fig:su4_anisotropic} below).
We will return to this issue in \cref{sec:anisotropy}.

\subsection{Collective excitations}\label{sec:collective}

Having characterized the nature of the mean field transition we now examine
the isospin collective modes in the
unpolarized state and in
each of the three ordered states described in~\cref{sec:model}.
For simplicity we set the dispersion to be
$k^2$ (and re-label $\nu_{F,1}$ as simply $\nu_F$).
The results for other dispersions are
quite similar, except in the immediate vicinity of the transition.

\begin{figure}
    \centering
    \includegraphics[width=0.9\linewidth]{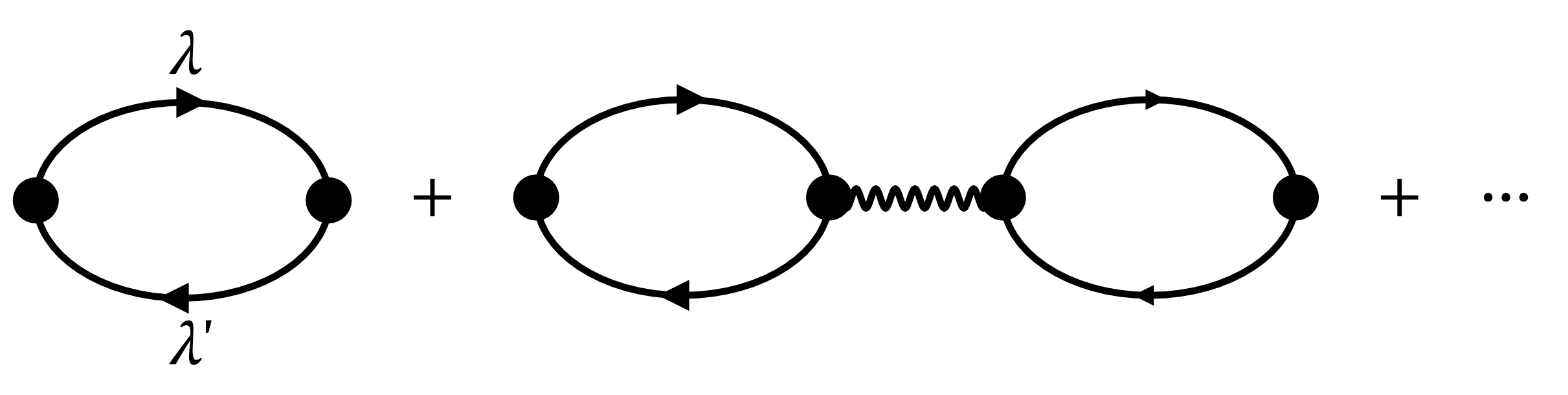}
    \caption{Diagrammatic representation of the RPA susceptibility in the valley isospin channel.
        Solid lines are the mean-field electron propagators in the band $\lambda$ and wavy lines are the isospin interaction~\cref{eq:Ssimp}.\label{fig:rpa}}
\end{figure}

The dispersions of charge collective modes in the VP and IVC channels
are obtained by solving $ \det D^{-1}(\omega,\mathbf{q}) = 0$, where
\begin{equation}
    D^{-1}_{\mu\nu}(\omega, \mathbf{q}) \equiv \frac{\delta_{\mu\nu}}{g_{\mu}}  - \chi^{R}_{\mu\nu}(\omega, \mathbf{q})
    \label{eq:rpa-pole}
\end{equation}
and $\chi^{R}_{\mu\nu}(\omega, \mathbf{q})$ are the retarded charge
susceptibilities in the valley isospin channels (see \cref{fig:rpa}).
The indices $\mu,\nu$ take the values $0,1,2,3$.
The latter three values label
the isospin components
$\hat{\tau}_i$ in \cref{eq:Ssimp} (the corresponding couplings $g_\mu$
are $g_{x}, g_y$, and $g_z$), and the $\mu=0$ component labels
the total density $\hat{\tau}_0$ (the corresponding coupling is $-g_d$).
We have to include the total density channel $\hat{\tau}_0$ as in the ordered state fluctuations in this channel couple to fluctuations in the valley staggered density channel $\hat{\tau}_{z}$.

We calculate $\chi$ on the Matsubara axis and then convert to real frequencies.
We have
\begin{equation}
    \chi_{\mu\nu} (i\omega_{m},\mathbf{q}) = -
    \frac{T}{2} \sum_{k\lambda\lambda'} G_{\lambda}(k+q)G_{\lambda'}(k)\tr[\hat{P}_\lambda\hat{\tau}_{\mu}\hat{P}_{\lambda'}\hat{\tau}_{\nu}]
\end{equation}
where
\begin{equation}
    G_{\lambda} (k) =\frac{1}{i\epsilon_{n} - \epsilon_{k} + E_{F\lambda}}
\end{equation}
is the Green's function of band $\lambda$
and $\hat{P}_{\pm}=(1/2)(\hat{\tau}_{0}\pm\hat{\tau}_{z})$ are projectors for the $\pm$-bands.
Performing the Matsubara sum and analytically continuing to real frequency we find
\begin{multline}
    \chi^{R}_{\mu\nu}(\omega, \mathbf{q})
    =
    \frac{1}{2} \sum_{\lambda\lambda'}\int \frac{d\mathbf{k}}{(2\pi)^{2}}\tr[\hat{P}_\lambda\hat{\tau}_{\mu}\hat{P}_{\lambda'}\hat{\tau}_{\nu}]
    \\
    \times
    \frac{n_{F}(\epsilon_{\mathbf{k}+\mathbf{q}/2} - E_{F\lambda}) - n_{F}(\epsilon_{\mathbf{k}-\mathbf{q}/2} - E_{F\lambda'})}{\omega + i 0- (\epsilon_{\mathbf{k}+\mathbf{q}/2} - \epsilon_{\mathbf{k}-\mathbf{q}/2}) + E_{F\lambda} -E_{F\lambda'}}.
    \label{eq:chi-normal}
\end{multline}

We first consider the unpolarized state, where $E_{F\lambda} = E_{F}$.
In the usual fashion, we expand $\chi^{R}_{\mu\nu} (\omega, \mathbf{q})$ to lowest order in $q/k_{F}$:
\begin{equation}
    \chi^{R}_{\mu\nu}(\omega, \mathbf{q})
    =-
    \nu_{F}\delta_{\mu\nu}
    \oint_{FS}     \frac{\mathbf{v}_{F}\cdot \mathbf{q}}{\omega + i 0- \mathbf{v}_{F}\cdot\mathbf{q}}.
    \label{eq:retarded-bubble}
\end{equation}
Defining $s=\omega/v_{F}q$ as the independent variable, after angular integration, we can write the mode condition~\cref{eq:rpa-pole}
as
\begin{equation}
    \frac{1}{
        \nu_{F}g_\mu}
    = 1
    -
    \frac{i s}{\sqrt{1 - (s+i0)^{2}}}.
    \label{eq:normal-state-modes}
\end{equation}
This is the typical expression one obtains when solving for zero-sound modes in two-dimensions~\cite{Klein2020}.
For the collective mode in the critical channel with coupling $\nu_F g_z \leq 1$,
we can define $\delta = \nu_{F}g_{z}-1$ and trivially solve \cref{eq:normal-state-modes} for $s \approx i\delta$, i.e., $\omega \approx i \delta v_{F}q$,
This is an overdamped
mode, similar to an overdamped spin mode near a ferromagnetic Stoner instability in a one-valley system~\cite{Klein2020}.
As long as $\delta < 0$, the mode is located in the lower half-plane of frequency, which implies that the zero-sound mode decays in time.
The static susceptibility $\chi_{zz}(\omega=0,\mathbf{q}\to0)$ diverges as $1/\delta$ as is expected at the conventional Stoner transition.
The critical mode is non-degenerate when $g_z = g_{\text{VP}}$,
doubly degenerate when $g_z = g_{\text{IVC}}$ as in this case $g_y=g_z$, and
triply degenerate when $g_{\text{VP}} = g_{\text{IVC}}$.
In the non-critical channel (VP for $0<g_x < g_z=g_y$ and IVC for
$0<g_x=g_y < g_z$,
see \cref{tab:isospin-interaction}), the collective mode is either overdamped with a positive velocity, or is a more subtle ``mirage'' mode~\cite{Klein2020}.
In the total density channel
the mode is a conventional propagating zero-sound mode, if we keep $g_d$ as a constant, but we remind that in reality this mode is a plasmon.

When $\delta$ becomes positive, the pole at $s \approx i\delta$ crosses into the
upper half-plane, indicating that the unpolarized state becomes unstable.
Conventional wisdom, supported by calculations in 3D
(see \cref{sec:analytic-collective})
is that introducing a small but finite value of the order parameter $\Delta$ moves the zero sound pole back into the lower half-plane of frequency.
However, in 2D we find that
the zero-sound pole remains in the upper-half plane for any value of $|\Delta| < \bar{\mu}$, that is as long as both bands have non-zero occupation the system remains unstable to perturbations (see \cref{sec:analytic-collective} for more details).
The system becomes stable only when $|\Delta| \geq \bar{\mu}$, i.e., when
one of the bands becomes fully depleted (two if we include fermion spin).
Below we will consider the collective modes in the fully polarized half-metal state, obtained in \cref{sec:mf}, where one band per spin is occupied and one is empty.

Prior to our work, the collective mode in the upper half-plane at $\delta >0$ has been discussed in Refs.~\onlinecite{Ma2024};
there the authors
conjectured that this implies that the order develops with a finite momentum, despite that the instability itself is at $q=0$.
We argue instead that the order is homogeneous, but the magnitude of the order parameter instantly jumps to a finite value.

In the fully polarized ordered state at $T=0$, the Fermi functions in \cref{eq:chi-normal} are non-zero only for the occupied band, which label as ($+$).
We have in this situation
\begin{multline}
    \chi^{R}_{\mu\nu}(\omega, \mathbf{q})
    =
    \frac{1}{2}
    \sum_{\lambda\lambda'}\int \frac{d\mathbf{k}}{(2\pi)^{2}}\tr[\hat{P}_\lambda\hat{\tau}_{\mu}\hat{P}_{\lambda'}\hat{\tau}_{\nu}]\\
    \times\frac{\Theta(E_{F+}-\epsilon_{\mathbf{k}+\mathbf{q}/2})\delta_{\lambda,+} - \Theta(E_{F+}-\epsilon_{\mathbf{k}-\mathbf{q}/2})\delta_{\lambda'+}}{\omega + i 0- \mathbf{v} \cdot \mathbf{q}  + E_{F\lambda} -E_{F\lambda'}}.
    \label{eq:chiR}
\end{multline}
Explicitly evaluating the matrix trace $\tr[\hat{P}_\lambda\hat{\tau}_{\mu}\hat{P}_{\lambda'}\hat{\tau}_{\nu}]$, we find
\begin{equation}
    \chi^{R}_{\mu\nu}(\omega,\mathbf{q})
    =
    \left(
    \begin{smallmatrix}
        \chi^R_{L} & 0                                   & 0                                   & \chi^R_{L}\\
        0                              & \chi^{R}_{+-}+ \chi^{R}_{-+}& i (\chi^{R}_{+-}  - \chi^{R}_{-+})& 0                              \\
        0                              & -i (\chi^{R}_{+-}  - \chi^{R}_{-+})&\chi^{R}_{+-}+ \chi^{R}_{-+} & 0                              \\
        \chi^R_{L} & 0                                   & 0                                   & \chi^R_{L}
    \end{smallmatrix}\right)_{\mu\nu}
    \label{eq:long-trans}
\end{equation}
where we have defined the longitudinal and transverse susceptibilities as
\begin{multline}
    \chi^{R} _{L}(\omega, \mathbf{q}) \equiv \chi^{R}_{++}(\omega, \mathbf{q})\\
    =
    \frac{1}{2}\int \frac{d\mathbf{k}}{(2\pi)^{2}}
    \frac{\Theta(E_{F+}-\epsilon_{\mathbf{k}+\mathbf{q}/2}) - \Theta(E_{F+}-\epsilon_{\mathbf{k}-\mathbf{q}/2})}{\omega + i 0- \mathbf{v} \cdot \mathbf{q}}.
    \label{eq:long-susc}
\end{multline}
and
\begin{multline}
    \chi^{R}_{-+}(\omega,\mathbf{q}) = \chi^{A}_{+-}(-\omega,-\mathbf{q})\equiv\\
    =
    -\frac{1}{2}
    \int \frac{d\mathbf{k}}{(2\pi)^{2}}
    \frac{\Theta(E_{F+}-\epsilon_{\mathbf{k}-\mathbf{q}/2})}{\omega + i 0- \mathbf{v} \cdot \mathbf{q}  - 2\Delta}.
\end{multline}
Plugging \cref{eq:long-trans} into \cref{eq:rpa-pole}
we find
that $D^{-1}$ decouples into a longitudinal $0-3$ sector
\begin{equation}
    \hat{D}^{-1}_{L}
    = \begin{pmatrix}
        -\frac{1}{g_{d}} - \chi^{R}_{L} & -\chi^{R}_{L}                \\
        -\chi^{R}_{L}                   & \frac{1}{g_{z}}-\chi^{R}_{L}
    \end{pmatrix},
    \label{eq:rpa-long}
\end{equation}
and a transverse $1-2$ sector
\begin{equation}
    \hat{D}^{-1}_{T} =
    \begin{pmatrix}
        \frac{1}{g_{x}} - \chi^{R}_{+-} - \chi^{R}_{-+} & -i(\chi^{R}_{+-}-\chi^{R}_{-+})                 \\
        i(\chi^{R}_{+-} - \chi^{R}_{-+})                & \frac{1}{g_{y}} - \chi^{R}_{+-} - \chi^{R}_{-+}
    \end{pmatrix}.
    \label{eq:rpa-trans}
\end{equation}

We independently
solved
for the poles of each sector.
The calculations are somewhat involved and we present them in \cref{app:coll_m}.
Here we list the results.
In the longitudinal sector, there is a
single propagating
mode --- a modified mode from the total density channel in the unpolarized state.
The equation for the dispersion of this mode is
\begin{equation}
    1
    -
    \frac{i s_{+}}{\sqrt{1 - (s_{+}+i0)^{2}}} =
    -\frac{1}{\nu_{F+} g_\text{eff}}
    \label{eq:long-mode}
\end{equation}
where $s_{+}\equiv \omega/v_{F-}q$ is defined in terms of the Fermi velocity of the occupied band, and
$g_\text{eff} = (g_d-g_z)/2$.
The effective interaction $g_{\text{eff}}$ is positive
($g_\text{eff} = U_1$ for VP and $g_\text{eff} =(U_1 + U_2 + U_3)/2$ for IVC).
We emphasize that this is the only collective charge mode in the longitudinal sector
for all three cases;
because one band is fully depleted, valley density and total density cannot fluctuate independently~\footnote{
    In reality this single mode is the plasmon mode, whose spectrum will be insensitive to the nature of the isospin transition.
}.
We therefore `lose' the
longitudinal valley-density fluctuation
mode when going from the unpolarized state to the full polarization state
(see \cref{sec:analytic-collective} for
a more detailed description of this).

The structure of the transverse modes depends on the type of order.
For the easy-axis polarization (VP order with $g_{x} = g_{y} = g_{\text{IVC}}$, $g_{z}=g_{\text{VP}}$ and
$g_{\text{VP}} > g_{\text{IVC}} > 0$),
there are two
poles in the susceptibility at $\omega = \pm \omega_\text{IVC}(q) - i 0$, describing
massive excitation with dispersion $\omega_\text{IVC}(q) =
    \left[\frac{n}{2} (g_{\text{VP}}- g_{\text{IVC}}) + \frac{q^{2}}{2M_\text{IVC}}\right]$,
where the expression for the kinetic mass $M_{IVC} >0$ is presented in \cref{app:coll_m}.
This massive mode is qualitatively similar to the Silin mode in a Fermi-liquid with an applied Zeeman field~\cite{Silin1958}.
The Silin mode is an excitation of a Fermi liquid in the presence of a Zeeman field and can be thought of as precession of the spin about the external field (for a generalization of Silin modes
to materials with spin and valley degrees of freedom see Ref.~[\onlinecite{Raines2022}]).
In our case, the transverse modes can also be viewed as precession modes as they are the circularly polarized eigenvectors of the valley isospin.
In distinction from Silin's case,
there is no external field, but rather the isospin precesses around its spontaneously generated expectation value.
For this reason, the precession frequency is interaction-dependent.
Indeed this mode softens as the isotropic point $g_{\text{VP}}=g_{\text{IVC}}$ is approached.

For the easy plane polarization (IVC order with  $g_{x}=g_{z}=g_{\text{IVC}}$, $g_{y}=g_{\text{VP}}$,
and $g_{\text{IVC}} > g_{\text{VP}}$),
there is a Goldstone mode describing rotation of the mean field state in the easy-plane.
This mode is linearly dispersing with $\omega_{GS} = v_{GS}q$, where $v_{GS} \propto (g_{\text{IVC}} - g_{\text{VP}})^{1/2}$.

In the Heisenberg case all couplings are equal, $g_{x}=g_{y}=g_{z} =g$.
This limit
may be reached from the easy-axis state solution taking $g_{\text{IVC}} \to g_{\text{VP}}$, or from the easy-plane solution taking $g_{\text{VP}}\to g_{\text{IVC}}$.
Either way, we obtain the expected quadratic magnon dispersion $\omega_{M}(q) =
    \frac{q^{2}}{2M_\text{IVC}}$.

\section{\texorpdfstring{$\mathrm{SU(4)}$-invariant case}{SU(4) invariant case}}
\label{sec:su4}

We now
consider the fully $\mathrm{SU(4)}$ symmetric case, in which
$g_{\gamma}$ \emph{for all $\mathrm{SU(4)}$ generators are equal
    to $g$}.
Now we have to consider all four bands on equal footing.
We
again rotate the electron operators $\psi=\hat{R}c$ into a basis, which is diagonal in both the normal and ordered states, and describe the ordered state in terms of the densities of each of four bands $n_{\lambda\rho}$, where we set $\lambda$ and $\rho$ to have plus or minus values ($\lambda=\pm,\rho=\pm$).~\footnote{
    The most natural choice for $\lambda$ and $\rho$ is spin and valley.
    However, due to the $\mathrm{SU(4)}$ invariance each index could in general be some linear superposition of spin and valley.}
We parameterize $n_{\lambda \rho}$ in terms of three variables
$\zeta_1, \zeta_2$ and $\zeta_3$, each subject to $-1 \leq \zeta_i \leq 1$~\footnote{This can be thought of as expressing the density imbalances in terms of
    traceless diagonal generators of
    $\mathrm{SU(4)}$, similar to~\cite{Chichinadze2022}.}
\begin{equation}
    n_{\lambda\rho} = n_{0}(1 + \lambda \zeta_1 + \rho \zeta_2 + \lambda \rho \zeta_3)
\end{equation}
where $n_{0}$ is the normal state density per spin and valley.
The total density $n = 4 n_0$.

In these notations, the ground state energy can be written as
\begin{equation}
    U_{MF}  = \sum_{\lambda\rho}u_{K}(n_{\lambda\rho}) - \frac{1}{8}g n^{2}\left(\zeta_1^{2} +\zeta_2^{2} + \zeta_3^{2}\right).
    \label{eq:su4-umf}
\end{equation}
For a parabolic dispersion, $u_{K}$ is quadratic in $\zeta_i$, hence
the
ground state energy is again
a quadratic function of $\zeta_1,\zeta_2,\zeta_3$.
In this case we expect to see the same behavior as in \cref{sec:mf}, with the susceptibility diverging at a coupling set by the Stoner criterion, followed by a transition in to a state of maximal polarization
satisfying $n=\sum_{\lambda\rho}n_{\lambda\rho}$.
We identify this state below.

For the power law model
with a generic $\alpha$,
the kinetic energy of each band is
\begin{equation}
    u_{K}(n_{\lambda\rho}) = \frac{cn_{0}^{\alpha+1}}{\alpha + 1}(1+\lambda \zeta_1 + \rho \zeta_2 + \lambda \rho \zeta_3)^{\alpha+1}.
\end{equation}
At  $\alpha \neq 1$,
this $ u_{K}(n_{\lambda\rho})$ contains terms with higher powers of $\zeta_i$, including odd-power terms (a cubic one, etc).
Such terms are allowed by the group structure of $\mathrm{SU(4)}$~\cite{Chichinadze2022}.

The difference in energy from the unpolarized state
is, for a generic $\alpha$,
\begin{multline}
    \delta u \equiv \nu_{F,\alpha}\frac{\delta U_{MF} }{4n^{2}_{0}}
    =
    \frac{1}{2}\\
    \times
    \Bigg(
    \frac{1}{2\alpha(\alpha+1)}\left[\sum_{
            \lambda\rho = \pm 1}
        (1 +\lambda \zeta_1 + \rho \zeta_2 + \lambda \rho \zeta_3)^{\alpha+1} - 4\right]
    \\
    -
    \nu_{F,\alpha} g
    (\zeta_1^{2} +\zeta_{2}^{2} + \zeta_3^{2})
    \Bigg)
    \label{eq:delta-u-su4}
\end{multline}
where,
we remind, $\nu_{F,\alpha} = 1/(c \alpha n^{\alpha -1}_0)$ is the normal state density of states per spin and valley (cf.~\cref{eq:normal-dos}).

We emphasize that \cref{eq:su4-umf,eq:delta-u-su4} are valid for arbitrary $\zeta_i$ subject to $|\zeta_i| <1$.
In this respect our approach goes beyond Refs.~\onlinecite{Chichinadze2022,Chichinadze2022a}, where $\delta u $ was obtained by expanding in
$\zeta_i$ to order $\zeta^4_i$.
These authors argued for a first order transition, but their arguments were based either on the presence of a cubic term~\cite{Chichinadze2022} or on a potentially negative coefficient for the quartic term~\cite{Chichinadze2022a}.
In our analysis below we analyze the full expression for $\delta u$ without assuming $\zeta_i$ to be small and using the fact that \cref{eq:delta-u-su4} sets exact relations between the coefficients for cubic, quartic, and all higher order terms in powers of $\zeta_i$.

\begin{figure}
    \centering
    \includegraphics[width=0.6\linewidth]{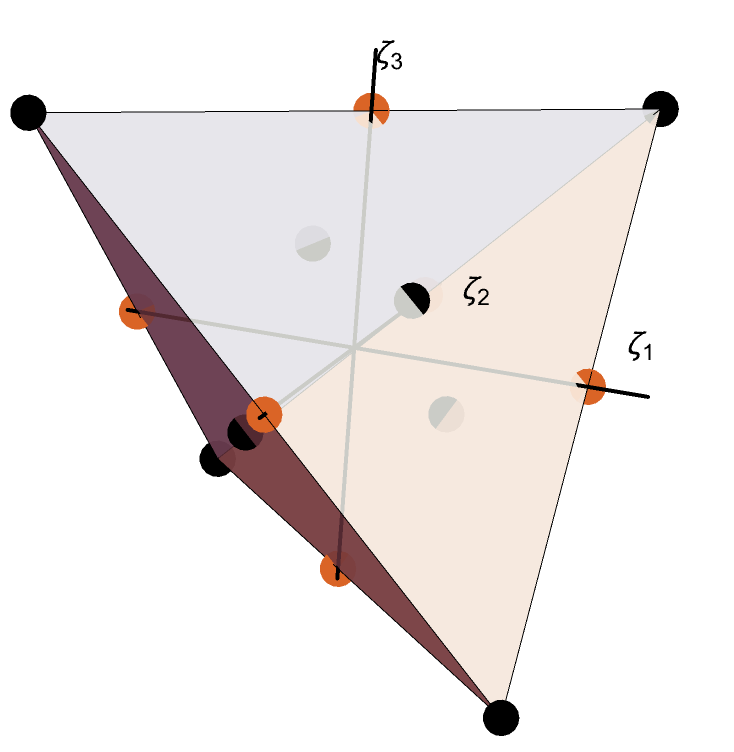}
    \caption{(Color online) The band imbalances $\zeta_1,\zeta_{2},\zeta_{3}$ are constrained to a tetrahedral region determined by the inequalities
    $n_{\lambda\rho} \geq 0$ and the conservation of density $\sum_{\lambda\rho}n_{\lambda\rho}=n$.
    The centers of each edge, marked with orange dots, correspond to states with $2$ occupied bands and $2$ empty bands.
    The centers of the faces and vertices of the tetrahedron, marked by black dots, correspond to states with either $1$ occupied and $3$ unoccupied bands or vice versa.
    \label{fig:su4region}}
\end{figure}

We now analyze $\delta u$.
In analogy with how we defined \cref{eq:h} for the
case of non-equivalent interactions,
we define the function
\begin{equation}
    h(\zeta_{1},\zeta_{2}, \zeta_{3})\equiv
    \frac{\sum_{\lambda\rho}(1 +\lambda \zeta_1 + \rho \zeta_2 + \lambda \rho \zeta_3)^{\alpha+1} - 4}{2\alpha(\alpha+1)(\zeta^{2}_{1}+\zeta^{2}_{2}+\zeta^{2}_{3})}
    \label{eq:h-def-su4}
\end{equation}
so that the energy can be written
\begin{equation}
    \delta u
    =
    \frac{1}{2}(\zeta_1^{2} +\zeta_{2}^{2} + \zeta_3^{2})
    \left(
    h(\zeta_{1},\zeta_{2},\zeta_{3})
    - \nu_{F,\alpha} g
    \right).
    \label{eq:du-su4}
\end{equation}
Upon reaching a threshold coupling, the first transition will be to the state that
minimizes $h$.
Here the region of variation of $h(\zeta_{1},\zeta_{2},\zeta_{3})$ is the tetrahedron, \cref{fig:su4region}, defined by
\begin{equation}
    0 \leq n_{\lambda\rho} = 1 + \lambda \zeta_{1} + \rho \zeta_{2} +\lambda \rho \zeta_{3} \leq 4.
\end{equation}
We can determine the nature of the ordered state by first Taylor expanding the
first term of \cref{eq:delta-u-su4}
as
\begin{multline}
    \frac{1}{2\alpha(\alpha+1)}\left[\sum_{\lambda\rho}(1 +\lambda \zeta_1 + \rho \zeta_2 + \lambda \rho \zeta_3)^{\alpha+1} - 4\right]\\
    = \sum_{l=2}\frac{c_{l}}{4}\sum_{\lambda\rho}(\lambda \zeta_1 + \rho \zeta_2 + \lambda\rho \zeta_3)^{l}
    \label{eq:kinetic-taylor-expansion}
\end{multline}
where we have used that the linear term vanishes after summing over $\lambda$ and $\rho$.
The coefficients $c_l$ are given by the recursion relations
\begin{equation}
    c_{2}  = 1, \quad c_{l+1} =\frac{\alpha +1 - l}{l+1}c_{l}.
\end{equation}

A straightforward analysis shows that there are three different regimes depending on the value of $\alpha$.
For $\alpha < 1$ all odd coefficients are negative $c_{2l+1} < 0$ while all even coefficients are positive $c_{2l}>0$.
For $1< \alpha <2$, all odd coefficients are positive and all even coefficients except for $c_{2}>0$ are negative.
Finally for $\alpha > 2$ all coefficients are positive.
To quartic order, the \cref{eq:delta-u-su4} takes a form previously considered in~\cite{Chichinadze2022}, but in our case with explicit values of the coefficients
\begin{multline}
    \delta u =
    \frac{1}{2}
    (1-\nu_{F,\alpha}g)(\zeta^{2}_{1}+\zeta^{2}_{2}+\zeta^{2}_{3})
    + (\alpha - 1) \zeta_{1}\zeta_{2}\zeta_{3}\\
    + \frac{(\alpha - 1)(\alpha-2)}{8}
    \left[(\zeta_{1}^{2}+\zeta^{2}_{2}+\zeta^{2}_{3})^{2}
        -
        \frac{2}{3}
        \left(
        \zeta_{1} ^{4}+ \zeta_{2}^{4}+ \zeta_{3}^{4}
        \right)
        \right]\\
    + \cdots.
    \label{eq:energy}
\end{multline}
We compare this expression with the ground state energy in Ref.~\onlinecite{Chichinadze2022} in \cref{app:comparison}.

\subsection{\texorpdfstring{Ordered states in terms of $\zeta_i$}{Ordered states in terms of \textzeta{}i}}

Analyzing
\cref{eq:energy},
we find that the even terms (quadratic and quartic) are extremized when $\zeta_1^{2},\zeta_2^{2},\zeta_3^{2}$ are each extremized, while the odd cubic term is extremized when $\zeta_{1}\zeta_{2}\zeta_{3}$ is extremized.
This tendency continues to hold for higher order terms in $\zeta_{i}$ as can be
seen by expanding the polynomial $(\lambda \zeta_1 + \rho \zeta_2 + \lambda\rho \zeta_3)^{l}$ in \cref{eq:kinetic-taylor-expansion}.
For odd powers of $l$ only terms of form $\zeta_{1}^{k_{1}}\zeta_{2}^{k_{2}}\zeta_{3}^{k_{3}}$ with $k_{i}$ odd survive as for any other combination the sum over band indices is zero.
Combining this with the fact that all $c_{l\in\text{odd}}$ have the same sign, we see that all the odd terms in \cref{eq:kinetic-taylor-expansion} favor states where $\sgn(\zeta_1\zeta_2\zeta_3)=\sgn(\alpha - 1)$.
For a similar reason, among even in $l$ terms, only the ones containing even powers of each of $\zeta_i$ survive.
Since all even terms, beyond the
quadratic one,
have the same sign they favor states which extremize each of $\zeta_1^{2},\zeta_2^{2},\zeta_3^{2}$.
For $1\leq\alpha\leq2$ this favors states which maximize $\zeta^{2}_{i}$ since all even terms are negative, and
for $\alpha < 1$ or $\alpha > 2$ this favors states which minimize $\zeta^{2}_{i}$.

The ground state of the system, for any particular value of $\alpha$ and $g$ is determined by the competition between the quadratic term, and higher order even and odd terms.
As we will see, in general,
the net result is a strong first order transition into a maximally ordered state
with a threshold value of the coupling $g$ close to $1/\nu_{F,\alpha}$.
We will see that this holds for all values of $\alpha$, in contrast to the case of
non-equal interactions, where a first-order transition holds only for $1 \leq \alpha \leq 2$.

We now consider separately each of the parameter regimes $\alpha < 1$, $1 \leq \alpha \leq 2$, and $\alpha > 2$.
In each range, we analyze the evolution of the system upon an increase of the interaction $g$.
In the next Section we also discuss the evolution upon the change of the electron density.

\subsubsection{\texorpdfstring{$1\leq \alpha \leq 2$}{1 <= \textalpha{} <= 2}}

\begin{figure}
    \centering
    \subfloat[]{\includegraphics[width=0.5\linewidth]{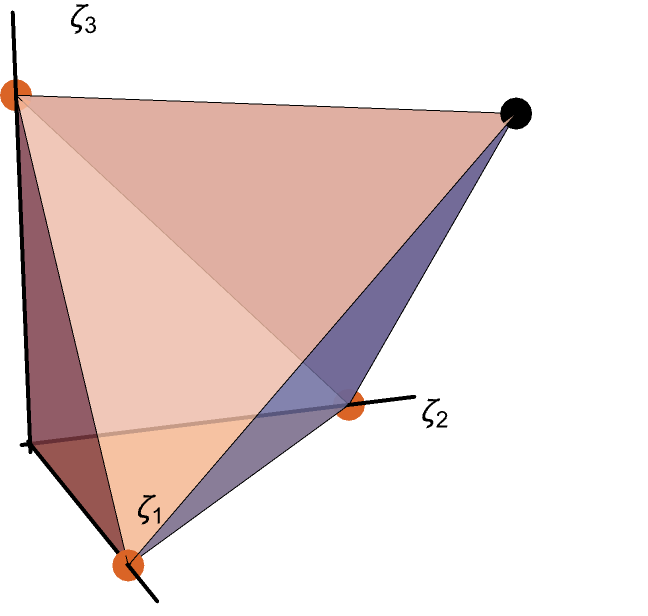}}
    \subfloat[\label{fig:region-zoom-b}]{\includegraphics[width=0.5\linewidth]{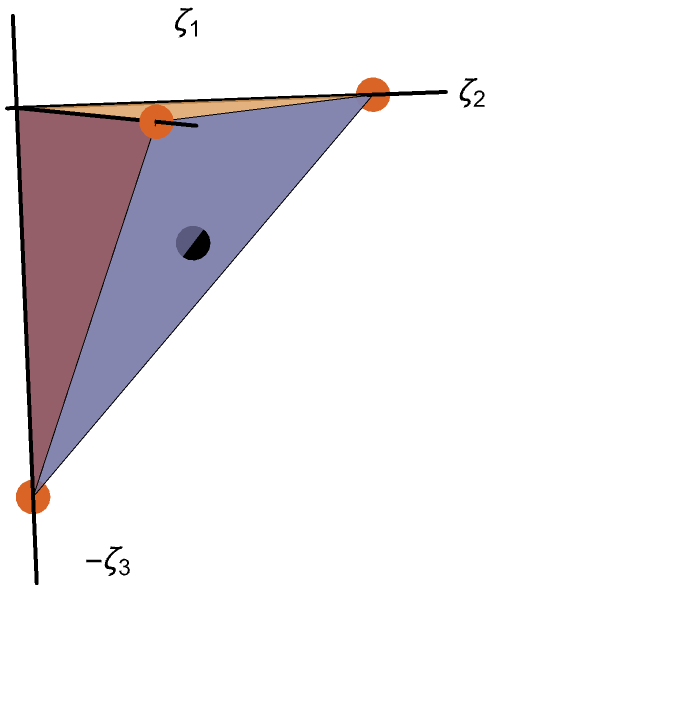}}
    \caption{(Color online) Primitive regions for minimization based upon the sign of $\zeta_{1}\zeta_{2}\zeta_{3}$ for
        (a) $\zeta_{1}\zeta_{2}\zeta_{3}\geq0$ and (b) $\zeta_{1}\zeta_{2}\zeta_{3}\leq0$.
        Other regions of the tetrahedron \cref{fig:su4region} are related to these by permutation of the bands.
        For $1\leq\alpha\leq2$, region (b) is favored, while for $\alpha>1$ or $\alpha<2$ region (a) is favored.
        \label{fig:region-zoom}
    }
\end{figure}

As we just discussed, for $1\leq\alpha\leq2$ the odd terms favor a state with $\zeta_{1}\zeta_{2}\zeta_{3}<0$, while the even terms are agnostic to the signs of $\zeta_{i}$.
Let us then take, without loss of generality, $\zeta_3 < 0 < \zeta_1 \leq \zeta_2$.
Conservation of the total density constrains potential solutions to lie within the region $0 < \zeta_1 < 1 + \zeta_3 - \zeta_2, 0 < \zeta_2 < 1 + \zeta_3$,
(see~\cref{fig:region-zoom-b}).
Minimizing $h (\zeta_1, \zeta_2, \zeta_3)$ from \cref{eq:h-def-su4}, we obtain $\zeta_1=\zeta_2=-\zeta_3=1/3$.
Such an order fully depletes one of the bands.
Concretely, the occupations of the bands are $\{\frac{4}{3}n_{0}, \frac{4}{3}n_{0}, \frac{4}{3}n_{0}, 0\}$.
This state is a three-quarter-metal
as it has $3$ occupied bands and $1$ empty band.
We designate this state as $(3,1)$.
The threshold coupling for the transition into the $(3,1)$ state is
\begin{equation}
    g_{(3,1)} =  \frac{6}{\nu_{F,\alpha} \alpha(\alpha+1)}\left[(4/3)^{\alpha} - 1\right].
    \label{eq:g31}
\end{equation}

As $g$ increases there is
another first-order
transition into a state with two occupied and two unoccupied bands;
in this state the band densities are $\{2n_{0}, 2n_{0}, 0, 0\}$.
This is a half-metal state,
which we designate as $(2, 2)$.
The emergence of the $(2,2)$ state
can be understood by noting that the energy \cref{eq:du-su4} is a product of two functions $\zeta^{2}_{1} + \zeta^{2}_{2}+\zeta^{2}_{3}$, and $h(\zeta_{1},\zeta_{2},\zeta_{3}) - \nu_{F,\alpha}g$.
The $(3,1)$ state is the minimum of the latter function but the $(2,2)$ state leads to a larger value of the former.
This means that once the coupling is strong enough
such that $h(1,0,0) - \nu_{F,\alpha} g$
becomes negative,
the $(2, 2)$ state becomes a local minimum of the energy.
As the magnitude of the coupling further increases, the energy of this local minimum decreases faster than that of the $(3,1)$ state, and at
\begin{equation}
    g_{(2,2)} = 3\frac{2^{\alpha}\left[1-(2/3)^{\alpha}\right]}{ \nu_{F,\alpha} \alpha(\alpha+1)}
    \label{eq:g22}
\end{equation}
the minima cross, leading to a first-order transition from the $(3,1)$ state to the $(2,2)$ state.

\begin{figure}
    \centering
    \includegraphics[width=\linewidth]{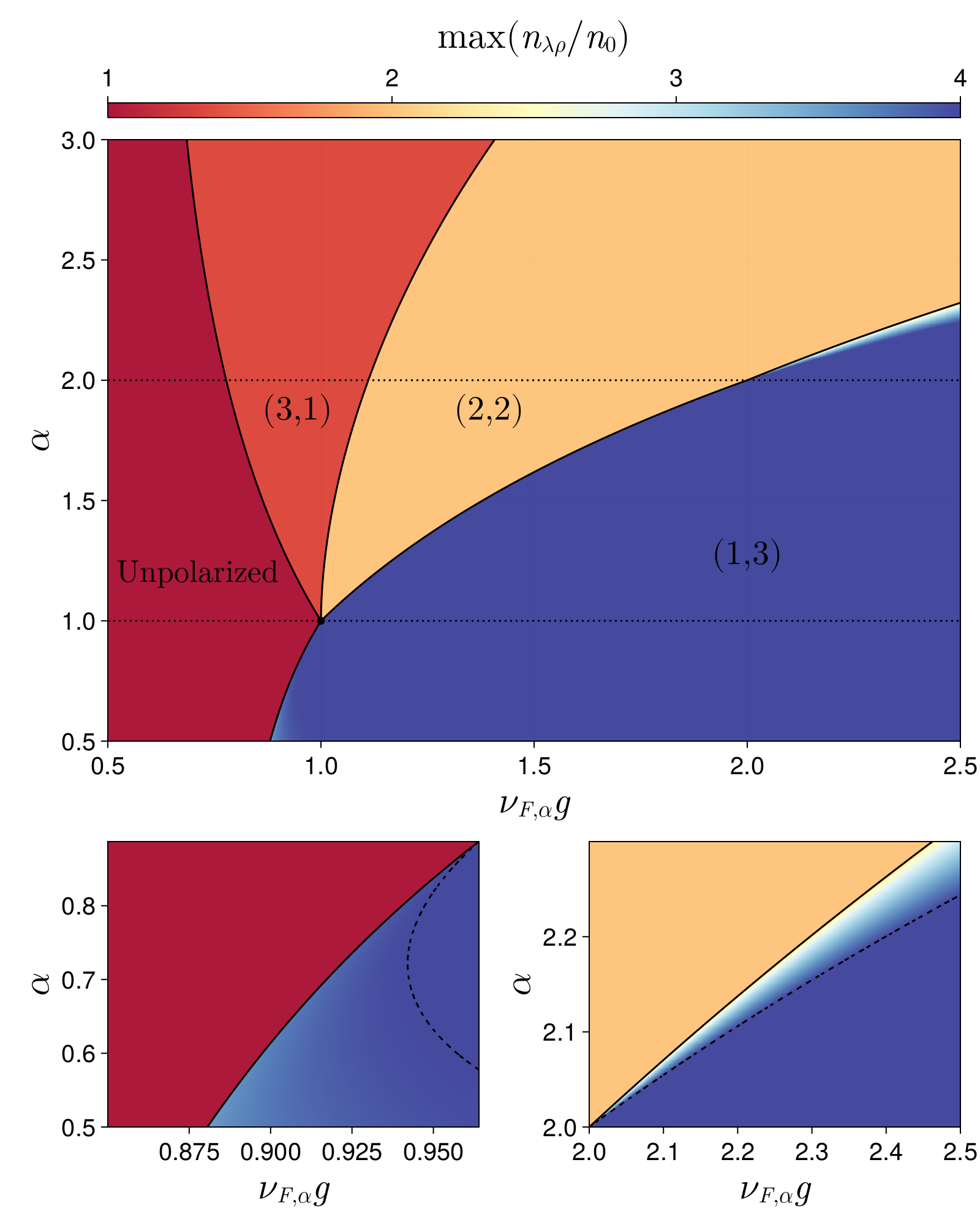}
    \caption{(Color online) Upper panel: Phase diagram of the
        $\mathrm{SU(4)}$ invariant model, showing the polarization states as a function of
        the power $\alpha$ appearing in the fermionic dispersion $k^{2\alpha}$  and the dimensionless coupling $\nu_{F,\alpha}g$.
        The coloring indicates the filling of the most occupied band $\max n_{\lambda\rho}$, compared to the unpolarized state filling $n_{0}$.
        Note that while the color scale is continuous, most of the plot is formed from solid colors.
        This is due to the full polarization nature of the ordered states.
        For $1<\alpha\leq2$ there are four phases separated by first order transitions, labeled here $(n_\text{occ}, n_\text{empty})$ in terms of
        the number of occupied and empty bands.
        For $\alpha \leq 1$ there is
        a direct first order transition between the unpolarized and $(3,1)$ phases.
        Below a threshold value $\alpha_{1}\approx0.89$, the transition is no longer to a fully polarized state but
        remains strongly
        first order, and the polarization quickly reaches the maximum value.
        For $\alpha > 2$, the transitions between the normal and $(3, 1)$ phase and $(3, 1)$ and $(2, 2)$ phases remain first order, but the transition between the $(2, 2)$ and $(1, 3)$ phases becomes second order.
        Lower panel: Zoomed view of the phase diagram showing the partially polarized phases --- the narrow
        regions
        of color gradient --- for $\alpha<\alpha_{1}$ (left) and $\alpha > 2$ (right).
        The coupling at which full polarization is reached is indicated by
        dashed lines.
        \label{fig:su4-phase}
    }
\end{figure}

Upon increasing $g$ further, there is
yet another first-order
transition to a
quarter-metal state with $1$ occupied band and $3$ unoccupied bands which we label as
the
$(1,3)$ state.
In this state the band densities are $\{4n_{0}, 0, 0, 0\}$.
As with the transition into the $(2,2)$ state, at some value of $g$,
$h(1,1,1) -  \nu_{F,\alpha}  g$ becomes negative, and the $(1,3)$ becomes a local minimum.
As $g$ further increases, the energies of $(2,2)$ and $(1,3)$ states cross at
\begin{equation}
    g_{(1,3)} = 2^{\alpha}\frac{2^{\alpha}-1}{\alpha(\alpha+1)\nu_{F,\alpha}},
    \label{eq:g13}
\end{equation}
leading to a first-order transition from the $(2,2)$ state to the $(1,3)$ state.
Note that the density of states for the occupied bands in the $(2,2)$ state is, according to \cref{eq:normal-dos},
\begin{equation}
    \nu^{(2,2)}_{F,\alpha}  =  \frac{1}{c \alpha (2n_{0})^{\alpha-1}} =  2^{1-\alpha}\nu_{F,\alpha},
\end{equation}
so we can rewrite \cref{eq:g13}
\begin{equation}
    g_{(1,3)} = 2\frac{2^{\alpha}-1}{\alpha(\alpha+1)\nu^{(2,2)}_{F,\alpha}},
\end{equation}
which is exactly the threshold value predicted for the full polarization transition in \cref{sec:mf}.
We can then intuitively understand the transition into the $(1,3)$ state as arising from a
Stoner instability of the occupied bands in the $(2,2)$ state.

\begin{figure}
    \centering
    \includegraphics[width=\linewidth]{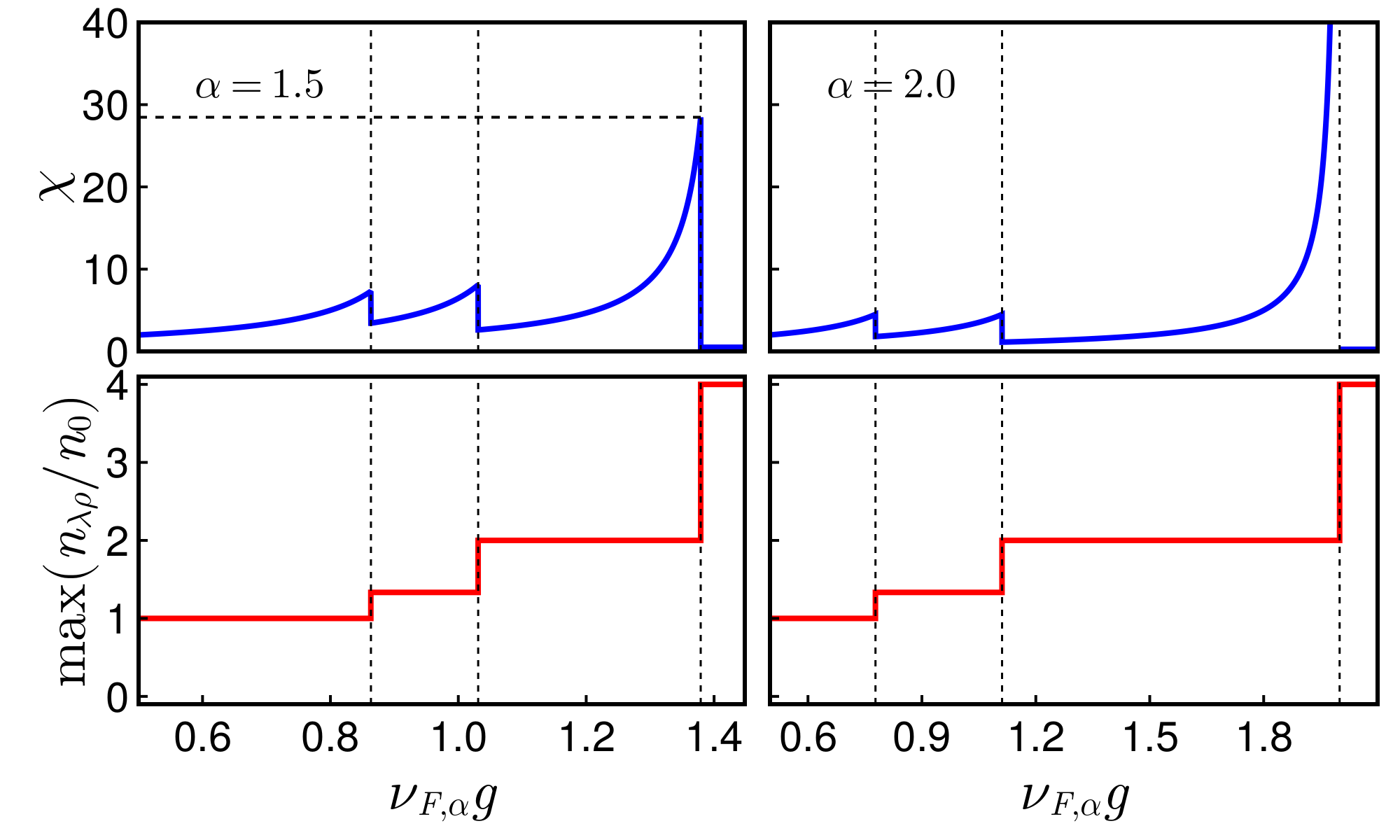}
    \caption{(Color online) The behavior of the order parameter susceptibility (top panel) and filling of the most occupied band (bottom panel) near each of the transitions for the SU(4) invariant case for representative $\alpha =1.5$ and $2$.
      Near the transitions into $3/4$ and $1/2$ metal, the susceptibility is enhanced, but remains finite for all $\alpha >1$.
      Near the transition into $1/4$ metal, the susceptibility is enhanced more strongly and diverges for $\alpha =2$.
      This is consistent with the analysis in \cref{sec:mf}.\label{fig:chi_n}
    }
\end{figure}

We show the phase diagram as a function of $\alpha$ and the coupling strength in \cref{fig:su4-phase}.
Note that the difference between $g_{(3,1)}$, $g_{(2,2)}$, and $g_{(1,3)}$, which determines the
width of the $(3,1)$ and $(2, 2)$ regions, vanishes at $\alpha \to 1$ (all three of  $g_{(3,1)}$, $g_{(2,2)}$, and $g_{(1,3)}$ become equal to $1/\nu_{F,1}$).
In this limit, the system undergoes a single strong first-order transition into the $(1,3)$ state.
This behavior can be clearly seen in \cref{fig:su4-phase}.
The width of the $(3,1)$ and $(2,2)$ regions grow with increasing $\alpha$,
at $\alpha\to2$ reaching
\begin{equation}
    g_{(3,1)} = \frac{7}{9\nu_{F,2}},\quad g_{(2,2)} = \frac{10}{9\nu_{F,2}},\quad g_{(1,3)} = \frac{2}{\nu_{F,2}}.
\end{equation}

In \cref{fig:chi_n} we show the behavior of the order parameter susceptibility vs $\nu_{F, \alpha} g$ for $\alpha =1.5$ and $\alpha =2$.   The
susceptibility gets strongly enhanced near each  transition. The enhancement is the strongest near the transition into a $1/4$ metal, where the susceptibility diverges at $\alpha =2$, consistent with out analysis in the previous section.

\subsubsection{\texorpdfstring{$\alpha <1$ and $\alpha >2$}{\textalpha < 1 and \textalpha >2}}

\begin{figure}
    \centering
    \includegraphics[width=\linewidth]{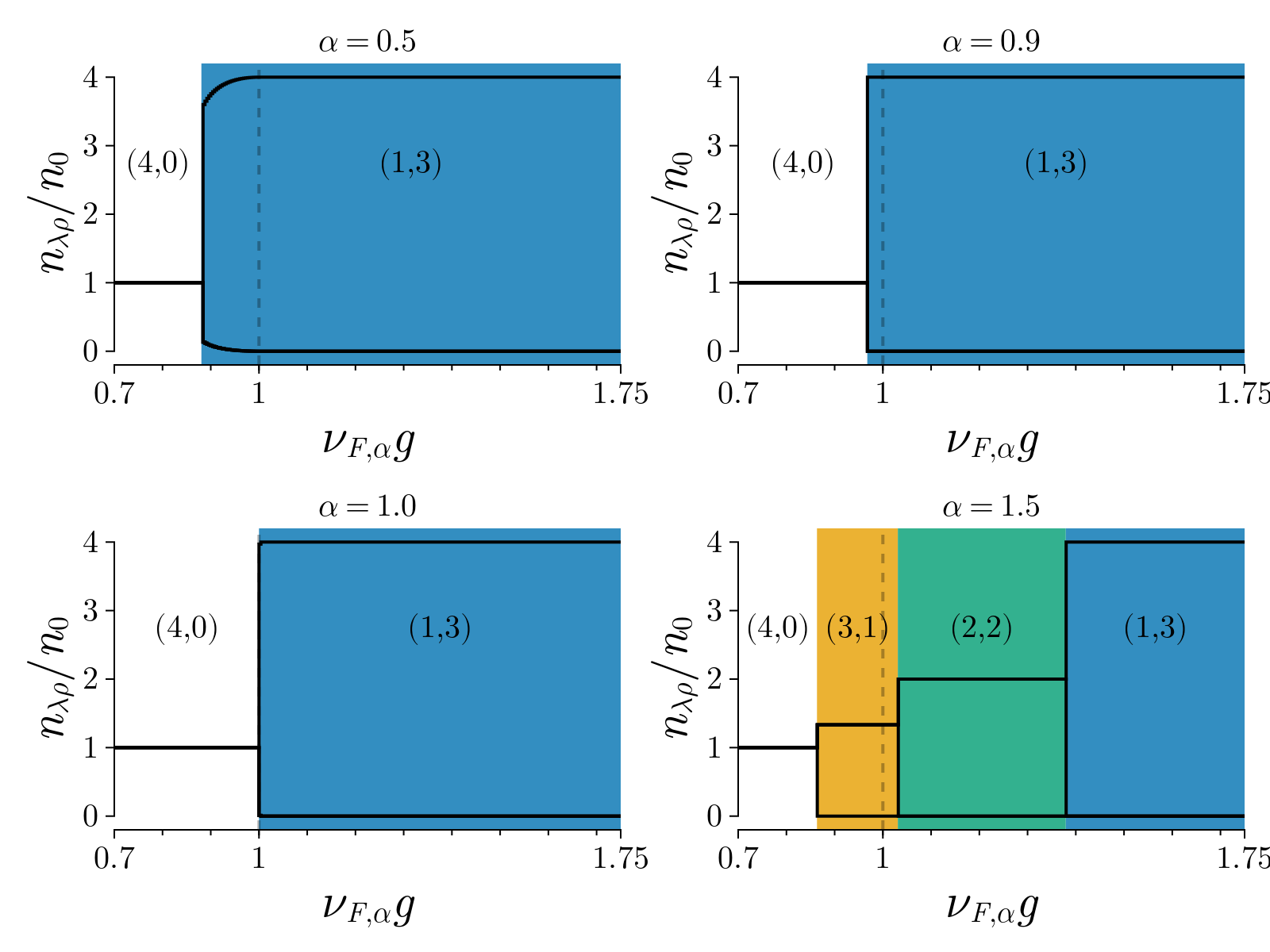}
    \caption{(Color online) Fillings of the fermion bands as a function of the dimensionless coupling $\nu_{F,\alpha} g$ for four different values of $\alpha$.
        For $\alpha \leq 1$ there is a single first order transition into a quarter-metal state $(1, 3)$
        (fully polarized for $\alpha > 0.89$ and ``almost'' fully polarized for $\alpha <0.89$),
        while for $1 < \alpha \leq 2$ there
        is a cascade of transitions into a three-quarter metal $(3, 1)$, half-metal $(2, 2)$, and quarter-metal $(1, 3)$.
        \label{fig:fillings}
    }
\end{figure}

We next
present
the results for two other ranges of $\alpha$.
For $\alpha < 1$, there is a single first-order transition (see \cref{fig:fillings}).
For values of $\alpha$ near $1$ the transition is
into the fully polarized $(1, 3)$ state.
However there is a threshold value below which the transition is instead to a large but finite polarization, such that there is one band containing most of the density, and three nearly empty bands.
This occurs when
$\left.(d/dz)h(z,z,z)\right|_{z=1} = 0$
which is satisfied at
\begin{equation}
    \alpha_1 = \frac{5}{3} + \frac{1}{2\ln 2} W_{-1}\left(-\frac{2^{2/3}\ln 2}{3}\right) \approx 0.89
\end{equation}
where $W_{-1}$ is the $-1$ branch of the Lambert W function.
For $\alpha < \alpha_{1}$ a local minimum of $h$ forms at the state $(\zeta, \zeta, \zeta)$, $\zeta < 1$.
This can be seen as a small sliver of decreased density at the bottom of the $g_{(1,3)}$ line in \cref{fig:su4-phase}.
We emphasize that such a partially polarized state can only be obtained by analyzing the full $\delta u$, without expanding it in powers of $\zeta_i$.
A previous study~\cite{Chichinadze2022}, which used this expansion, did not detect partially polarized states.

For $\alpha > 2$, the first order transitions into the $(3,1)$ and $(2,2)$ states persist, occurring at the same threshold couplings as a function of $\alpha$ given \cref{eq:g31,eq:g22}.
In contrast the transition from the $(2,2)$ state to the $(1,3)$ instead becomes a second order transition into a state with fillings
$\{2n_{0}+\delta n, 2n_{0}-\delta n, 0, 0\}$.
This occurs at
\begin{equation}
    g^{*}_{(1,3)}=2^{\alpha-1}/\nu_{F,\alpha}.
    \label{eq:g13star}
\end{equation}
As we did for $1\leq\alpha\leq2$ case,
comparing \cref{eq:g13star} with the density of states in the $(2,2)$, we can view it as the Stoner criterion $(2,2)$ state $g^{*}_{(1,3)}\nu^{(2,2)}_{F}=1$.
This aligns with our statement in \cref{sec:mf} that the Stoner transition is second order for $\alpha > 2$.

\subsection{Collective excitations}

The calculation of the collective modes for the $(3,1)$, $(2,2)$, and $(1,3)$ states closely parallels that in \cref{sec:collective}.
In particular the mode equation has the same structure as \cref{eq:rpa-pole}
\begin{equation}
    \begin{gathered}
        D^{-1}_{\gamma\gamma'}(\omega, \mathbf{q}) \equiv \frac{\delta_{\gamma\gamma'}(1-\delta_{\gamma0})}{g}  - \chi^{R}_{\gamma\gamma'}(\omega, \mathbf{q}),\\
        \det D^{-1}(\omega,\mathbf{q}) = 0.
    \end{gathered}
\end{equation}
where $\gamma,\gamma'\in0\ldots15$ index the identity matrix ($\gamma=0$) and all $15$ generators of $\mathrm{SU(4)}$,
and
\begin{multline}
    \chi^{R}_{\gamma\gamma'}(\omega, \mathbf{q})
    =\frac{1}{2}\sum_{oo'}\int \frac{d\mathbf{k}}{(2\pi)^{2}}\tr[\hat{P}_o\hat{\Gamma}_{\gamma}\hat{P}_{o'}\hat{\tau}_{\gamma'}]\\
    \times\frac{\Theta(E_{F+}-\epsilon_{\mathbf{k}+\mathbf{q}/2})\delta_{o,+} - \Theta(E_{F+}-\epsilon_{\mathbf{k}-\mathbf{q}/2})\delta_{o'+}}{\omega + i 0- \mathbf{v} \cdot \mathbf{q}  + E_{Fo} -E_{Fo'}}
\end{multline}
where the sub-index $o$ in $\hat{P}_o$ takes the values $o=+$ for the occupied subspace and $o=-$ for the unoccupied subspace.
Concretely for the $(3,1)$, $(2,2)$, and $(1, 3)$ states
\begin{equation}
    \begin{gathered}
        \hat{P}^{(3,1)}_{+} = \diag(1,1,1,0),\quad
        \hat{P}^{(3,1)}_{-} = \diag(0,0,0,1),\\
        \hat{P}^{(2,2)}_{+} = \diag(1,1,0,0),\quad
        \hat{P}^{(2,2)}_{-} = \diag(0,0,1,1)\\
        \hat{P}^{(1,3)}_{+} = \diag(1,0,0,0),\quad
        \hat{P}^{(1,3)}_{-} = \diag(0,1,1,1).
    \end{gathered}
\end{equation}

The determinant condition is then expressed for the $(3,1)$ state as
\begin{equation}
    \chi_{L}(1-2g\chi_{L})^{8}(1-2g\chi_{+-})^{3}(1-2g\chi_{-+})^{3} = 0
    \label{eq:collective-3-1}
\end{equation}
for the $(2,2)$ state as
\begin{equation}
    \chi_{L}(1-2g\chi_{L})^{3}(1-2g\chi_{+-})^{4}(1-2g\chi_{-+})^{4} = 0
    \label{eq:collective-2-2}
\end{equation}
and for the $(1,3)$ state
\begin{equation}
    \chi_{L}(1-2g\chi_{+-})^{3}(1-2g\chi_{-+})^{3} = 0
    \label{eq:collective-1-3}
\end{equation}
where the susceptibilities $\chi_{L},\chi_{\pm\mp}$ are of the same form as~\cref{eq:long-susc,eq:chi-pm}.
In \cref{eq:collective-3-1} the factors with exponents $1$, $8$, $3$ and $3$ correspond to, respectively, particle-hole excitations in the total density channel
(a plasmon for the actual Coulomb interaction),
$8$ overdamped longitudinal modes of relative density fluctuations of the occupied Fermi surfaces, and $3$ propagating Goldstone modes with a quadratic spectrum.
Likewise the factors with exponents $1$, $3$, $4$, $4$ in \cref{eq:collective-2-2} correspond to particle hole excitations in the total density channel, $3$ over-damped longitudinal modes, and
$4$  quadratic
Goldstone modes.
And in \cref{eq:collective-1-3} the exponents $1$, $3$, and $3$ correspond to particle hole excitations of the occupied Fermi surface, and
$3$ quadratic Goldstone modes.
In all states the fields for the $n$ Goldstone modes are complex and there are $n$ poles in the susceptibility at positive $\omega_i$ and $n$ at $-\omega_i$.
This is
consistent with the expectation from the coset construction for the corresponding orders~\cite{Watanabe2011}.
The total number of the poles for the collective modes
is 15 for the $(3, 1)$ state, 12 for the $(2, 2)$ state, and 7 for the $(1, 3)$ state.
This is smaller than $16$ poles in the paramagnetic state (15 collective coordinates for $\mathrm{SU(4)}$ and the $l=0$ mode of fluctuations of the total density).
The reduction comes about because one purely imaginary pole for the $(3, 1)$ state, four such poles for the $(2, 2)$ state, and 9 such poles for the $(1, 3)$ state disappear when one (or two or three) bands get fully depleted.

The overdamped longitudinal modes
($8$ or $3$)
can be understood as
overdamped
``zero-sound'' modes for a system with three Fermi surfaces  (the $(3, 1)$ state), two Fermi surfaces (the $(2, 2)$ state), or a single Fermi surface (the $(1, 3)$ state).

\subsection{Effect of anisotropy}
\label{sec:anisotropy}

\begin{figure}
    \centering
    \includegraphics[width=\linewidth]{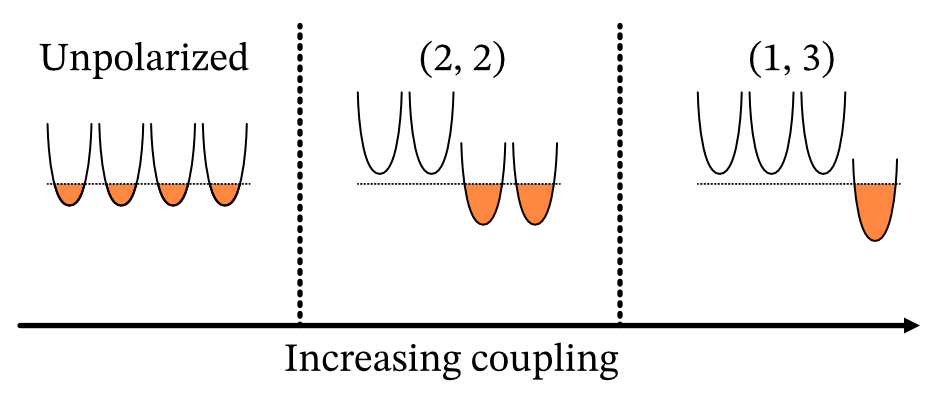}
    \caption{(Color online)
        Schematic depiction of the cascade of observed phases for sufficiently strong anisotropy.
        Breaking of the $\mathrm{SU(4)}$ symmetry shrinks the $(3,1)$ region and enlarges the $(2, 2)$ region, generically leading to the sequency $(4,0) \to (2,2)\to(1,3)$ as the strength of interactions is increased.
        \label{fig:su4_anisotropic}
    }
\end{figure}

When $\mathrm{SU(4)}$ symmetry is broken, the system behavior near the instability of the unpolarized state gradually becomes the same as in \cref{sec:mf}.
Namely, for $\alpha >1$, the range of the $(3,1)$ phase shrinks and eventually disappears, and for $\alpha <1$, there appears a range of the $(2,2)$ phase next to the unpolarized state.
In distinction to the discussion in \cref{sec:mf}, where we neglected the spin degree of freedom, in the anisotropic valley/spin $\mathrm{SU(4)}$ model, interactions in both valley and spin channels are attractive, hence once the interaction increases beyond the leading instability, there must be a second transition in the ``other'' isospin component (i.e., spin component, if the leading instability is in the valley sector, and valley component, if the leading instability is in the spin sector).
This necessary leads to a second transition into the $(1,3)$ state.
For $1 < \alpha <2$, both transitions are first-order into fully polarized spin/valley or spin and valley states, and for both the corresponding susceptibility nearly diverges at the onset of 1st order transition (truly diverges for $\alpha =1$ and $\alpha =2$).
We show the sequence of the ordered states in \cref{fig:su4_anisotropic}.

\section{Discussion and conclusion}
\label{sec:conclusion}

In this communication we have derived the isospin transitions and associated collective mode spectra for a model of two-dimensional electrons with spin and valley degrees of freedom.

We first considered instabilities in the valley isospin subspace --- valley polarization and inter-valley coherence --- assuming a situation where these occur before any spin instability.
We showed that a system with short ranged interactions and power-law fermionic dispersions $k^{2\alpha}$ with $1\leq \alpha \leq 2$ exhibits a discontinuous transition to a fully isospin polarized half-metal state, which fully depletes two out of four bands (one per spin).
Although the transition is discontinuous, the transition point is in close proximity to the critical coupling determined by the Stoner criterion, leading to an ``almost-divergent'' susceptibility immediately above the transition point --- such a behavior has also been
reported for
AlAs~\cite{Hossain2020}.
Beyond the transition, the full polarization of the ordered state leaves a strong imprint on the collective modes of the system.
Since one of the bands (per spin) is fully depleted, all that remains in the longitudinal sector is an overdamped ``sound''-like mode of the other band.
In the transverse sector we found that the structure of the modes depends on the nature of the instability.
For two-component inter-valley coherence order (the easy-plane case in our notations) the transverse sector contains a Goldstone mode associated with rotation in the easy-plane, while for valley polarization order (easy-axis case) the transverse modes are gapped precessions of isospin about its quantization axis.
In the fully $SU(2)$ symmetric case one recovers gapless, quadratic, magnons for the transverse excitations.

In the fully $\mathrm{SU(4)}$-invariant case, including both spin and valley channels, there is
a sequence of first-order transitions,
as a function of the dimensionless interaction,
for $1\leq \alpha \leq 2$ between a metallic state and three-quarter metal, half-metal, and quarter-metal states.
We also found partly polarized states with strongly depleted but not empty bands.

Each fractional metal state hosts a number of longitudinal modes analogous to the usual zero-sound modes of a multi-component Fermi-liquid, as well as the transverse
propagating
Goldstone modes of the symmetry broken state.
For the $(3,1)$ state these are the
propagating
total density mode (plasmon), $6$
overdamped
longitudinal modes associated with the quantum numbers of the occupied Fermi surfaces, and
$3$ transverse propagating Goldstone modes~\footnote{
    Each Goldstone mode gives rise to two poles in the susceptibility, at $\omega_i$ and $-\omega_i$.
    $3$ Goldstone modes in the coset construction counting scheme~\cite{Watanabe2011} correspond to $6$ poles}.
In the $(2, 2)$ state these are the propagating density mode (plasmon), $3$ overdamped
longitudinal modes --- the analog of the
spin `zero-sound' modes in a conventional Fermi liquid~\cite{*[{see e.g., }] [{}] Lifsic2006, *Nozieres1999} --- and
$4$ transverse propagating Goldstone modes.
In the $(1, 3)$ state there is the propagating density mode (plasmon) and
$3$ transverse propagating Goldstone modes.

\subsection{Comparison with BBG and RTG}

\begin{figure}
    \centering
    \subfloat[\label{fig:12cut}]{\includegraphics[width=\linewidth]{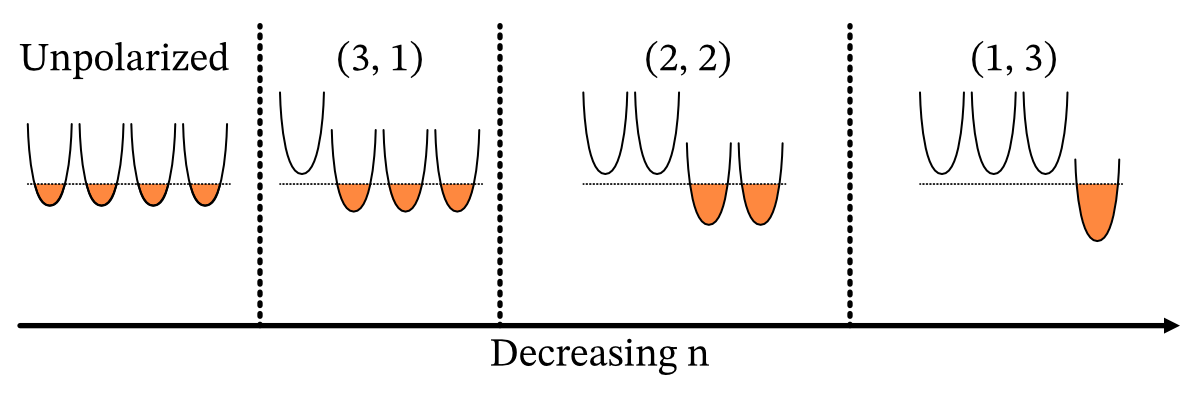}}\\
    \subfloat[\label{fig:34cut}]{\includegraphics[width=\linewidth]{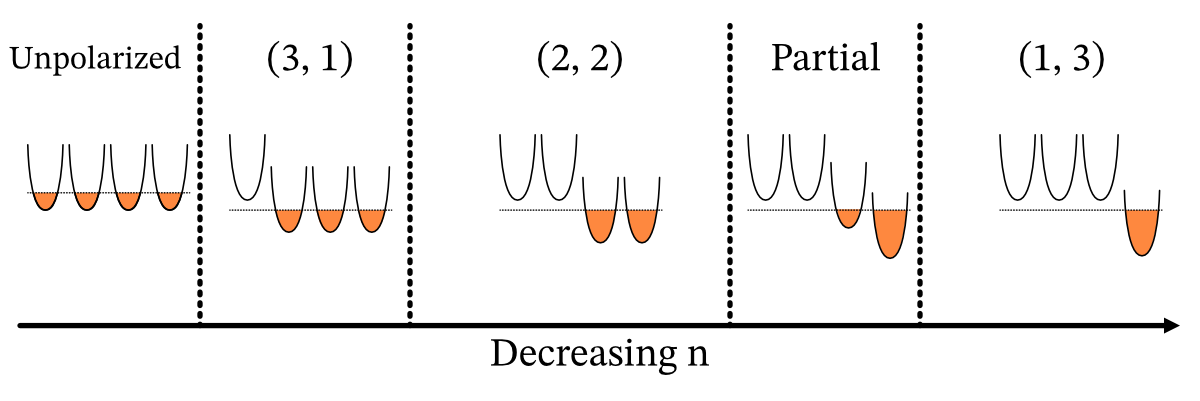}}
    \caption{(Color online)
        Schematic depiction of the cascade of observed phases as density is decreased for (a) $1 \leq \alpha \le 2$ and (b) $\alpha > 2$.
        The former is qualitatively applicable to the case of BBG in displacement field for densities above the van Hove filling, while the latter similarly applies to RTG in displacement field.
        Note that decreasing $n$ corresponds to increasing $\nu_{F,\alpha}g$ in \cref{fig:su4-phase} for this range of dispersions.
        In both cases the order of instabilities is consistent with what is experimentally observed.
        Additionally for $\alpha > 2$ there is a partially polarized phase which exists between the $(2, 2)$ and $(1, 3)$ phases.
        \label{fig:density-cascade}}
\end{figure}

For comparison with experiments in BBG and RTG we note that we have two parameters in our theory: the exponent $\alpha$ and the dimensionless coupling $g \nu_{F,\alpha}$.
The value of $\alpha$ is controlled, to a certain extent, by the displacement field $D$ (larger $D$ corresponds to larger $\alpha$).
Indeed the dispersion of BBG in displacement field, in absence of trigonal warping, can be approximated as $\epsilon \approx \sqrt{(k^{2}/2m)^{2} + D^{2}}$~\cite{McCann2013,Dong2023}, which interpolates between the $\alpha=2$ case at small $k$ and $\alpha=1$ at large $k$.
The dimensionless coupling is controlled by fermion density via $\nu_{F,\alpha}$.
For
$\alpha > 1$, which we are mostly interested in, $\nu_{F,\alpha}$ is a decreasing function of density, hence at a fixed $g$ the dimensionless $g \nu_{F,\alpha}$ increases as the system approaches charge neutrality point.
From this perspective, our results predict that as density decreases
at a fixed $D$,
the system undergoes a sequence of first order transitions from a spin and valley symmetric $(4,0)$ metallic state at larger densities to $(3,1)$, $(2,2)$ and $(1,3)$ states,
as shown in \cref{fig:density-cascade}.
This is consistent with what has been observed experimentally in BBG and RTG~\cite{Zhou2021,Seiler2022,Zhou2022a,DeLaBarrera2022,Seiler2023,Arp2023,Holleis2023,Zhang2023a},
modulo that the $(3,1)$ state has not been detected yet.
Probably, the density interval for this state is narrow.
Additionally, anisotropy of the interactions or spin-orbit coupling~\cite{Xie2023,Szabo2022}, which both break the $\mathrm{SU(4)}$ symmetry may further suppress the $(3, 1)$ phase.
As an extreme case, when the valley interactions are much stronger than the spin interactions, we may consider the case of \cref{sec:mf}, with a small residual spin interaction.
Then the interaction in the valley channel drives a transition to a fully valley isospin polarized $(2,2)$ state as density decreases, and at smaller densities the residual spin interaction drives a Stoner instability of the occupied valley into a $(1,3)$ spin \emph{and} valley polarized state.

The sequence of transitions between spin/valley ordered states has been obtained in earlier
theoretical works for TBG~\cite{Zondiner2020,Chichinadze2022}, where moir\'e physics gives rise to a flat band, and BBG/RTG~\cite{Zhou2021,Chichinadze2022,Chichinadze2022a,Rakhmanov2023,*Rakhmanov2023a,*Rakhmanov2023b}, where
the conduction (valence) bands are flat at the bottom (top), but the bandwidth is large, of order of an eV.
The novel aspect of our results with respect to BBG and RTG is the derivation and analysis of the full Landau functional for competing spin and valley order parameters, without assuming that order parameter magnitudes are small or selecting \emph{a priori} a particular spin/valley order.
As a consequence, we found a sequence of strong first order transitions into half-metal, quarter-metal, and three-quarter metal states, which fully deplete some bands.
Additionally, we found intermediate states with depleted but not empty
bands.
Such intermediate states have also been observed in experiments~\cite{Zhou2021,Zhou2022a,DeLaBarrera2022, Arp2023,Holleis2023,Zhang2023a}.
Overall, our results indicate that the relevant physics for the cascade of transitions is the global curvature properties of the density of states, which are qualitatively captured by the simple power law model.
One may also consider the effect of the displacement field as increasing the value of $\alpha$ at a fixed density.
In our theory, this broadens the density ranges for the ordered states upon increasing $D$.
This is also consistent with the data.

Further, in our analysis above we assumed that the dimensionless coupling $g \nu_{F,\alpha}$ monotonically increases
with decreasing density.
In reality, at smaller densities, trigonal warping becomes important, and there is a van Hove singularity, at which the free-fermion dispersion undergoes a topological transition from a single Fermi surface to three smaller pockets (12 in total if we count spin and valley).
Near the transition the dispersion cannot be approximated by a simple power-law, and $\nu_{F,\alpha}$ develops a van Hove singularity.
At smaller dopings, our analysis has to be modified to the new dispersion.
Experiments do show that near charge neutrality the quarter-metal state yields to a spin/valley symmetric $(4,0)$ state.
We believe this happens because the effective interaction gets weaker in the 12 pocket phase and near charge neutrality becomes below the threshold for spin/valley order.

\subsection{Comparison with TBG}

The analysis of spin/valley order in TBG is more involved as it requires knowledge of the full dispersion in the flat band.
Nevertheless, our results shed some light on the interplay between metallic states at fractional electronic fillings and Chern-insulating states near integer fillings.
Namely, metallic behavior at low $T$ at some generic non-integer filling likely can be described by confining to low-energy states with $\epsilon_k$ near the chemical potential $\mu$ and neglecting fermionic excitations on the other side of the Dirac point, which energetically are at a finite distance from $\mu$.
If spin/valley order developed continuously due to, e.g., singular density of states near van Hove points~\cite{Chichinadze2020a,*Chichinadze2021,Chichinadze2022}, it would split the states near the Fermi surface, but would not affect excitations near the Dirac points.
It would then be difficult to describe how such an order can lead to a Chern insulator.
Our results show that although the onset of an order can be detected within low-energy theory, from near-divergence of the corresponding susceptibility, the magnitude of the spin/valley order parameter jumps to its maximal possible value.
Such an order affects fermions at all energies, including states on the other side of the Dirac point.
This is the prerequisite for Chern-insulating behavior.

\subsection{Further considerations}

A number of additional aspects need to be considered in applying our results to biased BBG and RTG.\@
First, a finite bandwidth also plays a role at large densities: as the total density becomes large enough, the fractional-metal states are unable to accommodate all of the charges leading to emergence of additional small pockets in the other bands.
There also remain interesting possibilities of inter-pocket collective modes~\cite{Dong2023}.
A more detailed analysis of the cascade of instabilities can be performed in the same way as we did for the power law model once the chemical potential of the bands $\mu(n)$ has been expressed in terms of the density.

Second, the calculations in this work were performed at zero temperature.
At finite temperature there will be some characteristic density $E_{F}(n_{+})\sim T$ such that the low temperature approximation no longer applies to the minority band.
This may also cause the polarization to saturate at some value $|\zeta| < 1$, although for temperatures low compared to the normal state Fermi temperature we still have $1-|\zeta| \ll 1$.

Finally, it should be noted that the results obtained above do not explicitly depend on the dimensionality of the system: our results hold in general for an electronic dispersion $k^{\alpha d}$ in $d$ dimensions.
Here we have focused on the 2D case as there are several readily available materials spanning the range of dispersions we have discussed; in $d=3$ the $1\leq\alpha\leq2$ range is for dispersions between $k^{3}$ and $k^{6}$.
We expect a 3D system with e.g., dispersion $\epsilon \sim k^{4}$ to exhibit discontinuous full-polarization transitions as outlined above.

\begin{acknowledgments}
We thank E. Berg, Z. Dong, D. Chichinadze, L. Classen, I. Esterlis, F. Guinea, S-S Lee, L. Levitov, H. Ma, A.MacDonald, S. Nadj-Perge, Y. Oreg, A. Paramekanti, A-M Tremblay, Y. Wang, T. Weitz, J. Wilson, and A. Young for fruitful discussions and feedback.
The work of AVC was supported by U.S.
Department of Energy, Office of Science, Basic Energy Sciences, under Award No. DE-SC0014402;
LIG acknowledges the support by NSF Grant No. DMR-2410182 and by the Office of Naval Research (ONR) under Award No. N00014-22-1-2764.
Part of the work was performed while the authors visited the Kavli Institute for Theoretical Physics (KITP).
KITP is supported in part by grant NSF PHY-1748958.
\end{acknowledgments}

\bibliography{references.gen}

\begin{thebibliography}{60}%
\makeatletter
\providecommand \@ifxundefined [1]{%
 \@ifx{#1\undefined}
}%
\providecommand \@ifnum [1]{%
 \ifnum #1\expandafter \@firstoftwo
 \else \expandafter \@secondoftwo
 \fi
}%
\providecommand \@ifx [1]{%
 \ifx #1\expandafter \@firstoftwo
 \else \expandafter \@secondoftwo
 \fi
}%
\providecommand \natexlab [1]{#1}%
\providecommand \enquote  [1]{``#1''}%
\providecommand \bibnamefont  [1]{#1}%
\providecommand \bibfnamefont [1]{#1}%
\providecommand \citenamefont [1]{#1}%
\providecommand \href@noop [0]{\@secondoftwo}%
\providecommand \href [0]{\begingroup \@sanitize@url \@href}%
\providecommand \@href[1]{\@@startlink{#1}\@@href}%
\providecommand \@@href[1]{\endgroup#1\@@endlink}%
\providecommand \@sanitize@url [0]{\catcode `\\12\catcode `\$12\catcode
  `\&12\catcode `\#12\catcode `\^12\catcode `\_12\catcode `\%12\relax}%
\providecommand \@@startlink[1]{}%
\providecommand \@@endlink[0]{}%
\providecommand \url  [0]{\begingroup\@sanitize@url \@url }%
\providecommand \@url [1]{\endgroup\@href {#1}{\urlprefix }}%
\providecommand \urlprefix  [0]{URL }%
\providecommand \Eprint [0]{\href }%
\providecommand \doibase [0]{https://doi.org/}%
\providecommand \selectlanguage [0]{\@gobble}%
\providecommand \bibinfo  [0]{\@secondoftwo}%
\providecommand \bibfield  [0]{\@secondoftwo}%
\providecommand \translation [1]{[#1]}%
\providecommand \BibitemOpen [0]{}%
\providecommand \bibitemStop [0]{}%
\providecommand \bibitemNoStop [0]{.\EOS\space}%
\providecommand \EOS [0]{\spacefactor3000\relax}%
\providecommand \BibitemShut  [1]{\csname bibitem#1\endcsname}%
\let\auto@bib@innerbib\@empty
\bibitem [{\citenamefont {Schaibley}\ \emph {et~al.}(2016)\citenamefont
  {Schaibley}, \citenamefont {Yu}, \citenamefont {Clark}, \citenamefont
  {Rivera}, \citenamefont {Ross}, \citenamefont {Seyler}, \citenamefont {Yao},\
  and\ \citenamefont {Xu}}]{Schaibley2016}%
  \BibitemOpen
  \bibfield  {author} {\bibinfo {author} {\bibfnamefont {J.~R.}\ \bibnamefont
  {Schaibley}}, \bibinfo {author} {\bibfnamefont {H.}~\bibnamefont {Yu}},
  \bibinfo {author} {\bibfnamefont {G.}~\bibnamefont {Clark}}, \bibinfo
  {author} {\bibfnamefont {P.}~\bibnamefont {Rivera}}, \bibinfo {author}
  {\bibfnamefont {J.~S.}\ \bibnamefont {Ross}}, \bibinfo {author}
  {\bibfnamefont {K.~L.}\ \bibnamefont {Seyler}}, \bibinfo {author}
  {\bibfnamefont {W.}~\bibnamefont {Yao}},\ and\ \bibinfo {author}
  {\bibfnamefont {X.}~\bibnamefont {Xu}},\ }\bibfield  {title} {\bibinfo
  {title} {Valleytronics in {{2D}} materials},\ }\bibfield  {journal} {\bibinfo
   {journal} {Nat. Rev. Mater.}\ }\textbf {\bibinfo {volume} {1}},\ \href
  {https://doi.org/10.1038/natrevmats.2016.55} {10.1038/natrevmats.2016.55}
  (\bibinfo {year} {2016})\BibitemShut {NoStop}%
\bibitem [{\citenamefont {{\v Z}uti{\'c}}\ \emph {et~al.}(2004)\citenamefont
  {{\v Z}uti{\'c}}, \citenamefont {Fabian},\ and\ \citenamefont
  {Das~Sarma}}]{Zutic2004}%
  \BibitemOpen
  \bibfield  {author} {\bibinfo {author} {\bibfnamefont {I.}~\bibnamefont {{\v
  Z}uti{\'c}}}, \bibinfo {author} {\bibfnamefont {J.}~\bibnamefont {Fabian}},\
  and\ \bibinfo {author} {\bibfnamefont {S.}~\bibnamefont {Das~Sarma}},\
  }\bibfield  {title} {\bibinfo {title} {Spintronics: {{Fundamentals}} and
  applications},\ }\href@noop {} {\bibfield  {journal} {\bibinfo  {journal}
  {Rev. Mod. Phys.}\ }\textbf {\bibinfo {volume} {76}},\ \bibinfo {pages} {323}
  (\bibinfo {year} {2004})}\BibitemShut {NoStop}%
\bibitem [{\citenamefont {Bistritzer}\ and\ \citenamefont
  {MacDonald}(2010)}]{Bistritzer2010}%
  \BibitemOpen
  \bibfield  {author} {\bibinfo {author} {\bibfnamefont {R.}~\bibnamefont
  {Bistritzer}}\ and\ \bibinfo {author} {\bibfnamefont {A.~H.}\ \bibnamefont
  {MacDonald}},\ }\bibfield  {title} {\bibinfo {title} {Transport between
  twisted graphene layers},\ }\href
  {https://doi.org/10.1103/PhysRevB.81.245412} {\bibfield  {journal} {\bibinfo
  {journal} {Phys. Rev. B}\ }\textbf {\bibinfo {volume} {81}},\ \bibinfo
  {pages} {245412} (\bibinfo {year} {2010})}\BibitemShut {NoStop}%
\bibitem [{\citenamefont {Cao}\ \emph {et~al.}(2018{\natexlab{a}})\citenamefont
  {Cao}, \citenamefont {Fatemi}, \citenamefont {Demir}, \citenamefont {Fang},
  \citenamefont {Tomarken}, \citenamefont {Luo}, \citenamefont
  {{Sanchez-Yamagishi}}, \citenamefont {Watanabe}, \citenamefont {Taniguchi},
  \citenamefont {Kaxiras}, \citenamefont {Ashoori},\ and\ \citenamefont
  {{Jarillo-Herrero}}}]{Cao2018}%
  \BibitemOpen
  \bibfield  {author} {\bibinfo {author} {\bibfnamefont {Y.}~\bibnamefont
  {Cao}}, \bibinfo {author} {\bibfnamefont {V.}~\bibnamefont {Fatemi}},
  \bibinfo {author} {\bibfnamefont {A.}~\bibnamefont {Demir}}, \bibinfo
  {author} {\bibfnamefont {S.}~\bibnamefont {Fang}}, \bibinfo {author}
  {\bibfnamefont {S.~L.}\ \bibnamefont {Tomarken}}, \bibinfo {author}
  {\bibfnamefont {J.~Y.}\ \bibnamefont {Luo}}, \bibinfo {author} {\bibfnamefont
  {J.~D.}\ \bibnamefont {{Sanchez-Yamagishi}}}, \bibinfo {author}
  {\bibfnamefont {K.}~\bibnamefont {Watanabe}}, \bibinfo {author}
  {\bibfnamefont {T.}~\bibnamefont {Taniguchi}}, \bibinfo {author}
  {\bibfnamefont {E.}~\bibnamefont {Kaxiras}}, \bibinfo {author} {\bibfnamefont
  {R.~C.}\ \bibnamefont {Ashoori}},\ and\ \bibinfo {author} {\bibfnamefont
  {P.}~\bibnamefont {{Jarillo-Herrero}}},\ }\bibfield  {title} {\bibinfo
  {title} {Correlated insulator behaviour at half-filling in magic-angle
  graphene superlattices},\ }\href {https://doi.org/10.1038/nature26154}
  {\bibfield  {journal} {\bibinfo  {journal} {Nature}\ }\textbf {\bibinfo
  {volume} {556}},\ \bibinfo {pages} {80} (\bibinfo {year}
  {2018}{\natexlab{a}})}\BibitemShut {NoStop}%
\bibitem [{\citenamefont {Cao}\ \emph {et~al.}(2018{\natexlab{b}})\citenamefont
  {Cao}, \citenamefont {Fatemi}, \citenamefont {Fang}, \citenamefont
  {Watanabe}, \citenamefont {Taniguchi}, \citenamefont {Kaxiras},\ and\
  \citenamefont {{Jarillo-Herrero}}}]{Cao2018a}%
  \BibitemOpen
  \bibfield  {author} {\bibinfo {author} {\bibfnamefont {Y.}~\bibnamefont
  {Cao}}, \bibinfo {author} {\bibfnamefont {V.}~\bibnamefont {Fatemi}},
  \bibinfo {author} {\bibfnamefont {S.}~\bibnamefont {Fang}}, \bibinfo {author}
  {\bibfnamefont {K.}~\bibnamefont {Watanabe}}, \bibinfo {author}
  {\bibfnamefont {T.}~\bibnamefont {Taniguchi}}, \bibinfo {author}
  {\bibfnamefont {E.}~\bibnamefont {Kaxiras}},\ and\ \bibinfo {author}
  {\bibfnamefont {P.}~\bibnamefont {{Jarillo-Herrero}}},\ }\bibfield  {title}
  {\bibinfo {title} {Unconventional superconductivity in magic-angle graphene
  superlattices},\ }\href {https://doi.org/10.1038/nature26160} {\bibfield
  {journal} {\bibinfo  {journal} {Nature}\ }\textbf {\bibinfo {volume} {556}},\
  \bibinfo {pages} {43} (\bibinfo {year} {2018}{\natexlab{b}})}\BibitemShut
  {NoStop}%
\bibitem [{\citenamefont {Andrei}\ and\ \citenamefont
  {MacDonald}(2020)}]{Andrei2020}%
  \BibitemOpen
  \bibfield  {author} {\bibinfo {author} {\bibfnamefont {E.~Y.}\ \bibnamefont
  {Andrei}}\ and\ \bibinfo {author} {\bibfnamefont {A.~H.}\ \bibnamefont
  {MacDonald}},\ }\bibfield  {title} {\bibinfo {title} {Graphene bilayers with
  a twist},\ }\href {https://doi.org/10.1038/s41563-020-00840-0} {\bibfield
  {journal} {\bibinfo  {journal} {Nat. Mater.}\ }\textbf {\bibinfo {volume}
  {19}},\ \bibinfo {pages} {1265} (\bibinfo {year} {2020})}\BibitemShut
  {NoStop}%
\bibitem [{\citenamefont {Xie}\ \emph {et~al.}(2019)\citenamefont {Xie},
  \citenamefont {Lian}, \citenamefont {J{\"a}ck}, \citenamefont {Liu},
  \citenamefont {Chiu}, \citenamefont {Watanabe}, \citenamefont {Taniguchi},
  \citenamefont {Bernevig},\ and\ \citenamefont {Yazdani}}]{Xie2019}%
  \BibitemOpen
  \bibfield  {author} {\bibinfo {author} {\bibfnamefont {Y.}~\bibnamefont
  {Xie}}, \bibinfo {author} {\bibfnamefont {B.}~\bibnamefont {Lian}}, \bibinfo
  {author} {\bibfnamefont {B.}~\bibnamefont {J{\"a}ck}}, \bibinfo {author}
  {\bibfnamefont {X.}~\bibnamefont {Liu}}, \bibinfo {author} {\bibfnamefont
  {C.-L.}\ \bibnamefont {Chiu}}, \bibinfo {author} {\bibfnamefont
  {K.}~\bibnamefont {Watanabe}}, \bibinfo {author} {\bibfnamefont
  {T.}~\bibnamefont {Taniguchi}}, \bibinfo {author} {\bibfnamefont {B.~A.}\
  \bibnamefont {Bernevig}},\ and\ \bibinfo {author} {\bibfnamefont
  {A.}~\bibnamefont {Yazdani}},\ }\bibfield  {title} {\bibinfo {title}
  {Spectroscopic signatures of many-body correlations in magic-angle twisted
  bilayer graphene},\ }\href {https://doi.org/10.1038/s41586-019-1422-x}
  {\bibfield  {journal} {\bibinfo  {journal} {Nature}\ }\textbf {\bibinfo
  {volume} {572}},\ \bibinfo {pages} {101} (\bibinfo {year}
  {2019})}\BibitemShut {NoStop}%
\bibitem [{\citenamefont {Wong}\ \emph {et~al.}(2020)\citenamefont {Wong},
  \citenamefont {Nuckolls}, \citenamefont {Oh}, \citenamefont {Lian},
  \citenamefont {Xie}, \citenamefont {Jeon}, \citenamefont {Watanabe},
  \citenamefont {Taniguchi}, \citenamefont {Bernevig},\ and\ \citenamefont
  {Yazdani}}]{Wong2020}%
  \BibitemOpen
  \bibfield  {author} {\bibinfo {author} {\bibfnamefont {D.}~\bibnamefont
  {Wong}}, \bibinfo {author} {\bibfnamefont {K.~P.}\ \bibnamefont {Nuckolls}},
  \bibinfo {author} {\bibfnamefont {M.}~\bibnamefont {Oh}}, \bibinfo {author}
  {\bibfnamefont {B.}~\bibnamefont {Lian}}, \bibinfo {author} {\bibfnamefont
  {Y.}~\bibnamefont {Xie}}, \bibinfo {author} {\bibfnamefont {S.}~\bibnamefont
  {Jeon}}, \bibinfo {author} {\bibfnamefont {K.}~\bibnamefont {Watanabe}},
  \bibinfo {author} {\bibfnamefont {T.}~\bibnamefont {Taniguchi}}, \bibinfo
  {author} {\bibfnamefont {B.~A.}\ \bibnamefont {Bernevig}},\ and\ \bibinfo
  {author} {\bibfnamefont {A.}~\bibnamefont {Yazdani}},\ }\bibfield  {title}
  {\bibinfo {title} {Cascade of electronic transitions in magic-angle twisted
  bilayer graphene},\ }\href {https://doi.org/10.1038/s41586-020-2339-0}
  {\bibfield  {journal} {\bibinfo  {journal} {Nature}\ }\textbf {\bibinfo
  {volume} {582}},\ \bibinfo {pages} {198} (\bibinfo {year}
  {2020})}\BibitemShut {NoStop}%
\bibitem [{\citenamefont {Zondiner}\ \emph {et~al.}(2020)\citenamefont
  {Zondiner}, \citenamefont {Rozen}, \citenamefont {{Rodan-Legrain}},
  \citenamefont {Cao}, \citenamefont {Queiroz}, \citenamefont {Taniguchi},
  \citenamefont {Watanabe}, \citenamefont {Oreg}, \citenamefont {{von Oppen}},
  \citenamefont {Stern}, \citenamefont {Berg}, \citenamefont
  {{Jarillo-Herrero}},\ and\ \citenamefont {Ilani}}]{Zondiner2020}%
  \BibitemOpen
  \bibfield  {author} {\bibinfo {author} {\bibfnamefont {U.}~\bibnamefont
  {Zondiner}}, \bibinfo {author} {\bibfnamefont {A.}~\bibnamefont {Rozen}},
  \bibinfo {author} {\bibfnamefont {D.}~\bibnamefont {{Rodan-Legrain}}},
  \bibinfo {author} {\bibfnamefont {Y.}~\bibnamefont {Cao}}, \bibinfo {author}
  {\bibfnamefont {R.}~\bibnamefont {Queiroz}}, \bibinfo {author} {\bibfnamefont
  {T.}~\bibnamefont {Taniguchi}}, \bibinfo {author} {\bibfnamefont
  {K.}~\bibnamefont {Watanabe}}, \bibinfo {author} {\bibfnamefont
  {Y.}~\bibnamefont {Oreg}}, \bibinfo {author} {\bibfnamefont {F.}~\bibnamefont
  {{von Oppen}}}, \bibinfo {author} {\bibfnamefont {A.}~\bibnamefont {Stern}},
  \bibinfo {author} {\bibfnamefont {E.}~\bibnamefont {Berg}}, \bibinfo {author}
  {\bibfnamefont {P.}~\bibnamefont {{Jarillo-Herrero}}},\ and\ \bibinfo
  {author} {\bibfnamefont {S.}~\bibnamefont {Ilani}},\ }\bibfield  {title}
  {\bibinfo {title} {Cascade of phase transitions and {{Dirac}} revivals in
  magic-angle graphene.},\ }\href {https://doi.org/10.1038/s41586-020-2373-y}
  {\bibfield  {journal} {\bibinfo  {journal} {Nature}\ }\textbf {\bibinfo
  {volume} {582}},\ \bibinfo {pages} {203} (\bibinfo {year}
  {2020})}\BibitemShut {NoStop}%
\bibitem [{\citenamefont {Szab{\'o}}\ and\ \citenamefont
  {Roy}(2022)}]{Szabo2022}%
  \BibitemOpen
  \bibfield  {author} {\bibinfo {author} {\bibfnamefont {A.~L.}\ \bibnamefont
  {Szab{\'o}}}\ and\ \bibinfo {author} {\bibfnamefont {B.}~\bibnamefont
  {Roy}},\ }\bibfield  {title} {\bibinfo {title} {Competing orders and cascade
  of degeneracy lifting in doped {{Bernal}} bilayer graphene},\ }\href
  {https://doi.org/10.1103/PhysRevB.105.L201107} {\bibfield  {journal}
  {\bibinfo  {journal} {Phys. Rev. B}\ }\textbf {\bibinfo {volume} {105}},\
  \bibinfo {pages} {L201107} (\bibinfo {year} {2022})}\BibitemShut {NoStop}%
\bibitem [{\citenamefont {Chichinadze}\ \emph {et~al.}(2020)\citenamefont
  {Chichinadze}, \citenamefont {Classen},\ and\ \citenamefont
  {Chubukov}}]{Chichinadze2020a}%
  \BibitemOpen
  \bibfield  {author} {\bibinfo {author} {\bibfnamefont {D.~V.}\ \bibnamefont
  {Chichinadze}}, \bibinfo {author} {\bibfnamefont {L.}~\bibnamefont
  {Classen}},\ and\ \bibinfo {author} {\bibfnamefont {A.~V.}\ \bibnamefont
  {Chubukov}},\ }\bibfield  {title} {\bibinfo {title} {Valley magnetism,
  nematicity, and density wave orders in twisted bilayer graphene},\ }\href
  {https://doi.org/10.1103/physrevb.102.125120} {\bibfield  {journal} {\bibinfo
   {journal} {Phys. Rev. B}\ }\textbf {\bibinfo {volume} {102}},\ \bibinfo
  {pages} {125120} (\bibinfo {year} {2020})}\BibitemShut {NoStop}%
\bibitem [{\citenamefont {Chichinadze}\ \emph {et~al.}(2021)\citenamefont
  {Chichinadze}, \citenamefont {Classen},\ and\ \citenamefont
  {Chubukov}}]{Chichinadze2021}%
  \BibitemOpen
  \bibfield  {author} {\bibinfo {author} {\bibfnamefont {D.~V.}\ \bibnamefont
  {Chichinadze}}, \bibinfo {author} {\bibfnamefont {L.}~\bibnamefont
  {Classen}},\ and\ \bibinfo {author} {\bibfnamefont {A.~V.}\ \bibnamefont
  {Chubukov}},\ }\bibfield  {title} {\bibinfo {title} {Erratum: {{Valley}}
  magnetism, nematicity, and density wave orders in twisted bilayer graphene
  [{{Phys}}. {{Rev}}. {{B}} 102, 125120 (2020)]},\ }\href
  {https://doi.org/10.1103/PhysRevB.103.039901} {\bibfield  {journal} {\bibinfo
   {journal} {Phys. Rev. B}\ }\textbf {\bibinfo {volume} {103}},\ \bibinfo
  {pages} {039901(E)} (\bibinfo {year} {2021})}\BibitemShut {NoStop}%
\bibitem [{\citenamefont {Dong}\ \emph
  {et~al.}(2023{\natexlab{a}})\citenamefont {Dong}, \citenamefont {Davydova},
  \citenamefont {Ogunnaike},\ and\ \citenamefont {Levitov}}]{Dong2023}%
  \BibitemOpen
  \bibfield  {author} {\bibinfo {author} {\bibfnamefont {Z.}~\bibnamefont
  {Dong}}, \bibinfo {author} {\bibfnamefont {M.}~\bibnamefont {Davydova}},
  \bibinfo {author} {\bibfnamefont {O.}~\bibnamefont {Ogunnaike}},\ and\
  \bibinfo {author} {\bibfnamefont {L.}~\bibnamefont {Levitov}},\ }\bibfield
  {title} {\bibinfo {title} {Isospin- and momentum-polarized orders in bilayer
  graphene},\ }\href {https://doi.org/10.1103/PhysRevB.107.075108} {\bibfield
  {journal} {\bibinfo  {journal} {Phys. Rev. B}\ }\textbf {\bibinfo {volume}
  {107}},\ \bibinfo {pages} {075108} (\bibinfo {year}
  {2023}{\natexlab{a}})}\BibitemShut {NoStop}%
\bibitem [{\citenamefont {Dong}\ \emph
  {et~al.}(2023{\natexlab{b}})\citenamefont {Dong}, \citenamefont {Chubukov},\
  and\ \citenamefont {Levitov}}]{Dong2023a}%
  \BibitemOpen
  \bibfield  {author} {\bibinfo {author} {\bibfnamefont {Z.}~\bibnamefont
  {Dong}}, \bibinfo {author} {\bibfnamefont {A.~V.}\ \bibnamefont {Chubukov}},\
  and\ \bibinfo {author} {\bibfnamefont {L.}~\bibnamefont {Levitov}},\
  }\bibfield  {title} {\bibinfo {title} {Transformer spin-triplet
  superconductivity at the onset of isospin order in bilayer graphene},\ }\href
  {https://doi.org/10.1103/PhysRevB.107.174512} {\bibfield  {journal} {\bibinfo
   {journal} {Phys. Rev. B}\ }\textbf {\bibinfo {volume} {107}},\ \bibinfo
  {pages} {174512} (\bibinfo {year} {2023}{\natexlab{b}})}\BibitemShut
  {NoStop}%
\bibitem [{\citenamefont {Dong}\ \emph
  {et~al.}(2023{\natexlab{c}})\citenamefont {Dong}, \citenamefont {Levitov},\
  and\ \citenamefont {Chubukov}}]{Dong2023aa}%
  \BibitemOpen
  \bibfield  {author} {\bibinfo {author} {\bibfnamefont {Z.}~\bibnamefont
  {Dong}}, \bibinfo {author} {\bibfnamefont {L.}~\bibnamefont {Levitov}},\ and\
  \bibinfo {author} {\bibfnamefont {A.~V.}\ \bibnamefont {Chubukov}},\
  }\bibfield  {title} {\bibinfo {title} {Superconductivity near spin and valley
  orders in graphene multilayers},\ }\href
  {https://doi.org/10.1103/PhysRevB.108.134503} {\bibfield  {journal} {\bibinfo
   {journal} {Phys. Rev. B}\ }\textbf {\bibinfo {volume} {108}},\ \bibinfo
  {pages} {134503} (\bibinfo {year} {2023}{\natexlab{c}})}\BibitemShut
  {NoStop}%
\bibitem [{\citenamefont {Seiler}\ \emph {et~al.}(2022)\citenamefont {Seiler},
  \citenamefont {Geisenhof}, \citenamefont {Winterer}, \citenamefont
  {Watanabe}, \citenamefont {Taniguchi}, \citenamefont {Xu}, \citenamefont
  {Zhang},\ and\ \citenamefont {Weitz}}]{Seiler2022}%
  \BibitemOpen
  \bibfield  {author} {\bibinfo {author} {\bibfnamefont {A.~M.}\ \bibnamefont
  {Seiler}}, \bibinfo {author} {\bibfnamefont {F.~R.}\ \bibnamefont
  {Geisenhof}}, \bibinfo {author} {\bibfnamefont {F.}~\bibnamefont {Winterer}},
  \bibinfo {author} {\bibfnamefont {K.}~\bibnamefont {Watanabe}}, \bibinfo
  {author} {\bibfnamefont {T.}~\bibnamefont {Taniguchi}}, \bibinfo {author}
  {\bibfnamefont {T.}~\bibnamefont {Xu}}, \bibinfo {author} {\bibfnamefont
  {F.}~\bibnamefont {Zhang}},\ and\ \bibinfo {author} {\bibfnamefont {R.~T.}\
  \bibnamefont {Weitz}},\ }\bibfield  {title} {\bibinfo {title} {Quantum
  cascade of correlated phases in trigonally warped bilayer graphene},\ }\href
  {https://doi.org/10.1038/s41586-022-04937-1} {\bibfield  {journal} {\bibinfo
  {journal} {Nature}\ }\textbf {\bibinfo {volume} {608}},\ \bibinfo {pages}
  {298} (\bibinfo {year} {2022})}\BibitemShut {NoStop}%
\bibitem [{\citenamefont {Zhou}\ \emph {et~al.}(2022)\citenamefont {Zhou},
  \citenamefont {Holleis}, \citenamefont {Saito}, \citenamefont {Cohen},
  \citenamefont {Huynh}, \citenamefont {Patterson}, \citenamefont {Yang},
  \citenamefont {Taniguchi}, \citenamefont {Watanabe},\ and\ \citenamefont
  {Young}}]{Zhou2022a}%
  \BibitemOpen
  \bibfield  {author} {\bibinfo {author} {\bibfnamefont {H.}~\bibnamefont
  {Zhou}}, \bibinfo {author} {\bibfnamefont {L.}~\bibnamefont {Holleis}},
  \bibinfo {author} {\bibfnamefont {Y.}~\bibnamefont {Saito}}, \bibinfo
  {author} {\bibfnamefont {L.}~\bibnamefont {Cohen}}, \bibinfo {author}
  {\bibfnamefont {W.}~\bibnamefont {Huynh}}, \bibinfo {author} {\bibfnamefont
  {C.~L.}\ \bibnamefont {Patterson}}, \bibinfo {author} {\bibfnamefont
  {F.}~\bibnamefont {Yang}}, \bibinfo {author} {\bibfnamefont {T.}~\bibnamefont
  {Taniguchi}}, \bibinfo {author} {\bibfnamefont {K.}~\bibnamefont
  {Watanabe}},\ and\ \bibinfo {author} {\bibfnamefont {A.~F.}\ \bibnamefont
  {Young}},\ }\bibfield  {title} {\bibinfo {title} {Isospin magnetism and
  spin-polarized superconductivity in {{Bernal}} bilayer graphene},\ }\href
  {https://doi.org/10.1126/science.abm8386} {\bibfield  {journal} {\bibinfo
  {journal} {Science}\ }\textbf {\bibinfo {volume} {375}},\ \bibinfo {pages}
  {774} (\bibinfo {year} {2022})}\BibitemShut {NoStop}%
\bibitem [{\citenamefont {Zhou}\ \emph {et~al.}(2021)\citenamefont {Zhou},
  \citenamefont {Xie}, \citenamefont {Ghazaryan}, \citenamefont {Holder},
  \citenamefont {Ehrets}, \citenamefont {Spanton}, \citenamefont {Taniguchi},
  \citenamefont {Watanabe}, \citenamefont {Berg}, \citenamefont {Serbyn} \emph
  {et~al.}}]{Zhou2021}%
  \BibitemOpen
  \bibfield  {author} {\bibinfo {author} {\bibfnamefont {H.}~\bibnamefont
  {Zhou}}, \bibinfo {author} {\bibfnamefont {T.}~\bibnamefont {Xie}}, \bibinfo
  {author} {\bibfnamefont {A.}~\bibnamefont {Ghazaryan}}, \bibinfo {author}
  {\bibfnamefont {T.}~\bibnamefont {Holder}}, \bibinfo {author} {\bibfnamefont
  {J.~R.}\ \bibnamefont {Ehrets}}, \bibinfo {author} {\bibfnamefont {E.~M.}\
  \bibnamefont {Spanton}}, \bibinfo {author} {\bibfnamefont {T.}~\bibnamefont
  {Taniguchi}}, \bibinfo {author} {\bibfnamefont {K.}~\bibnamefont {Watanabe}},
  \bibinfo {author} {\bibfnamefont {E.}~\bibnamefont {Berg}}, \bibinfo {author}
  {\bibfnamefont {M.}~\bibnamefont {Serbyn}}, \emph {et~al.},\ }\bibfield
  {title} {\bibinfo {title} {Half-and quarter-metals in rhombohedral trilayer
  graphene},\ }\href@noop {} {\bibfield  {journal} {\bibinfo  {journal}
  {Nature}\ }\textbf {\bibinfo {volume} {598}},\ \bibinfo {pages} {429}
  (\bibinfo {year} {2021})}\BibitemShut {NoStop}%
\bibitem [{\citenamefont {{de la Barrera}}\ \emph {et~al.}(2022)\citenamefont
  {{de la Barrera}}, \citenamefont {Aronson}, \citenamefont {Zheng},
  \citenamefont {Watanabe}, \citenamefont {Taniguchi}, \citenamefont {Ma},
  \citenamefont {{Jarillo-Herrero}},\ and\ \citenamefont
  {Ashoori}}]{DeLaBarrera2022}%
  \BibitemOpen
  \bibfield  {author} {\bibinfo {author} {\bibfnamefont {S.~C.}\ \bibnamefont
  {{de la Barrera}}}, \bibinfo {author} {\bibfnamefont {S.}~\bibnamefont
  {Aronson}}, \bibinfo {author} {\bibfnamefont {Z.}~\bibnamefont {Zheng}},
  \bibinfo {author} {\bibfnamefont {K.}~\bibnamefont {Watanabe}}, \bibinfo
  {author} {\bibfnamefont {T.}~\bibnamefont {Taniguchi}}, \bibinfo {author}
  {\bibfnamefont {Q.}~\bibnamefont {Ma}}, \bibinfo {author} {\bibfnamefont
  {P.}~\bibnamefont {{Jarillo-Herrero}}},\ and\ \bibinfo {author}
  {\bibfnamefont {R.}~\bibnamefont {Ashoori}},\ }\bibfield  {title} {\bibinfo
  {title} {Cascade of isospin phase transitions in {{Bernal}} bilayer graphene
  at zero magnetic field},\ }\href {https://doi.org/10.1038/s41567-022-01616-w}
  {\bibfield  {journal} {\bibinfo  {journal} {Nat. Phys.}\ }\textbf {\bibinfo
  {volume} {18}},\ \bibinfo {pages} {771} (\bibinfo {year} {2022})},\ \Eprint
  {https://arxiv.org/abs/2110.13907} {arxiv:2110.13907 [cond-mat]} \BibitemShut
  {NoStop}%
\bibitem [{\citenamefont {Seiler}\ \emph {et~al.}(2023)\citenamefont {Seiler},
  \citenamefont {Statz}, \citenamefont {Weimer}, \citenamefont {Jacobsen},
  \citenamefont {Watanabe}, \citenamefont {Taniguchi}, \citenamefont {Dong},
  \citenamefont {Levitov},\ and\ \citenamefont {Weitz}}]{Seiler2023}%
  \BibitemOpen
  \bibfield  {author} {\bibinfo {author} {\bibfnamefont {A.~M.}\ \bibnamefont
  {Seiler}}, \bibinfo {author} {\bibfnamefont {M.}~\bibnamefont {Statz}},
  \bibinfo {author} {\bibfnamefont {I.}~\bibnamefont {Weimer}}, \bibinfo
  {author} {\bibfnamefont {N.}~\bibnamefont {Jacobsen}}, \bibinfo {author}
  {\bibfnamefont {K.}~\bibnamefont {Watanabe}}, \bibinfo {author}
  {\bibfnamefont {T.}~\bibnamefont {Taniguchi}}, \bibinfo {author}
  {\bibfnamefont {Z.}~\bibnamefont {Dong}}, \bibinfo {author} {\bibfnamefont
  {L.~S.}\ \bibnamefont {Levitov}},\ and\ \bibinfo {author} {\bibfnamefont
  {R.~T.}\ \bibnamefont {Weitz}},\ }\href
  {https://doi.org/10.48550/arXiv.2308.00827} {\bibinfo {title}
  {Interaction-driven (quasi-) insulating ground states of gapped
  electron-doped bilayer graphene}} (\bibinfo {year} {2023}),\ \Eprint
  {https://arxiv.org/abs/2308.00827} {arxiv:2308.00827 [cond-mat.str-el]}
  \BibitemShut {NoStop}%
\bibitem [{\citenamefont {Arp}\ \emph {et~al.}(2023)\citenamefont {Arp},
  \citenamefont {Sheekey}, \citenamefont {Zhou}, \citenamefont {Tschirhart},
  \citenamefont {Patterson}, \citenamefont {Yoo}, \citenamefont {Holleis},
  \citenamefont {Redekop}, \citenamefont {Babikyan}, \citenamefont {Xie},
  \citenamefont {Xiao}, \citenamefont {Vituri}, \citenamefont {Holder},
  \citenamefont {Taniguchi}, \citenamefont {Watanabe}, \citenamefont {Huber},
  \citenamefont {Berg},\ and\ \citenamefont {Young}}]{Arp2023}%
  \BibitemOpen
  \bibfield  {author} {\bibinfo {author} {\bibfnamefont {T.}~\bibnamefont
  {Arp}}, \bibinfo {author} {\bibfnamefont {O.}~\bibnamefont {Sheekey}},
  \bibinfo {author} {\bibfnamefont {H.}~\bibnamefont {Zhou}}, \bibinfo {author}
  {\bibfnamefont {C.~L.}\ \bibnamefont {Tschirhart}}, \bibinfo {author}
  {\bibfnamefont {C.~L.}\ \bibnamefont {Patterson}}, \bibinfo {author}
  {\bibfnamefont {H.~M.}\ \bibnamefont {Yoo}}, \bibinfo {author} {\bibfnamefont
  {L.}~\bibnamefont {Holleis}}, \bibinfo {author} {\bibfnamefont
  {E.}~\bibnamefont {Redekop}}, \bibinfo {author} {\bibfnamefont
  {G.}~\bibnamefont {Babikyan}}, \bibinfo {author} {\bibfnamefont
  {T.}~\bibnamefont {Xie}}, \bibinfo {author} {\bibfnamefont {J.}~\bibnamefont
  {Xiao}}, \bibinfo {author} {\bibfnamefont {Y.}~\bibnamefont {Vituri}},
  \bibinfo {author} {\bibfnamefont {T.}~\bibnamefont {Holder}}, \bibinfo
  {author} {\bibfnamefont {T.}~\bibnamefont {Taniguchi}}, \bibinfo {author}
  {\bibfnamefont {K.}~\bibnamefont {Watanabe}}, \bibinfo {author}
  {\bibfnamefont {M.~E.}\ \bibnamefont {Huber}}, \bibinfo {author}
  {\bibfnamefont {E.}~\bibnamefont {Berg}},\ and\ \bibinfo {author}
  {\bibfnamefont {A.~F.}\ \bibnamefont {Young}},\ }\href
  {https://doi.org/10.48550/arXiv.2310.03781} {\bibinfo {title} {Intervalley
  coherence and intrinsic spin-orbit coupling in rhombohedral trilayer
  graphene}} (\bibinfo {year} {2023}),\ \Eprint
  {https://arxiv.org/abs/2310.03781} {arxiv:2310.03781 [cond-mat.mes-hall]}
  \BibitemShut {NoStop}%
\bibitem [{\citenamefont {Holleis}\ \emph {et~al.}(2023)\citenamefont
  {Holleis}, \citenamefont {Patterson}, \citenamefont {Zhang}, \citenamefont
  {Yoo}, \citenamefont {Zhou}, \citenamefont {Taniguchi}, \citenamefont
  {Watanabe}, \citenamefont {{Nadj-Perge}},\ and\ \citenamefont
  {Young}}]{Holleis2023}%
  \BibitemOpen
  \bibfield  {author} {\bibinfo {author} {\bibfnamefont {L.}~\bibnamefont
  {Holleis}}, \bibinfo {author} {\bibfnamefont {C.~L.}\ \bibnamefont
  {Patterson}}, \bibinfo {author} {\bibfnamefont {Y.}~\bibnamefont {Zhang}},
  \bibinfo {author} {\bibfnamefont {H.~M.}\ \bibnamefont {Yoo}}, \bibinfo
  {author} {\bibfnamefont {H.}~\bibnamefont {Zhou}}, \bibinfo {author}
  {\bibfnamefont {T.}~\bibnamefont {Taniguchi}}, \bibinfo {author}
  {\bibfnamefont {K.}~\bibnamefont {Watanabe}}, \bibinfo {author}
  {\bibfnamefont {S.}~\bibnamefont {{Nadj-Perge}}},\ and\ \bibinfo {author}
  {\bibfnamefont {A.~F.}\ \bibnamefont {Young}},\ }\href
  {https://doi.org/10.48550/arXiv.2303.00742} {\bibinfo {title} {Ising
  {{Superconductivity}} and {{Nematicity}} in {{Bernal Bilayer Graphene}} with
  {{Strong Spin Orbit Coupling}}}} (\bibinfo {year} {2023}),\ \Eprint
  {https://arxiv.org/abs/2303.00742} {arxiv:2303.00742 [cond-mat.supr-con]}
  \BibitemShut {NoStop}%
\bibitem [{\citenamefont {Zhang}\ \emph {et~al.}(2023)\citenamefont {Zhang},
  \citenamefont {Polski}, \citenamefont {Thomson}, \citenamefont
  {{Lantagne-Hurtubise}}, \citenamefont {Lewandowski}, \citenamefont {Zhou},
  \citenamefont {Watanabe}, \citenamefont {Taniguchi}, \citenamefont {Alicea},\
  and\ \citenamefont {{Nadj-Perge}}}]{Zhang2023a}%
  \BibitemOpen
  \bibfield  {author} {\bibinfo {author} {\bibfnamefont {Y.}~\bibnamefont
  {Zhang}}, \bibinfo {author} {\bibfnamefont {R.}~\bibnamefont {Polski}},
  \bibinfo {author} {\bibfnamefont {A.}~\bibnamefont {Thomson}}, \bibinfo
  {author} {\bibfnamefont {{\'E}.}~\bibnamefont {{Lantagne-Hurtubise}}},
  \bibinfo {author} {\bibfnamefont {C.}~\bibnamefont {Lewandowski}}, \bibinfo
  {author} {\bibfnamefont {H.}~\bibnamefont {Zhou}}, \bibinfo {author}
  {\bibfnamefont {K.}~\bibnamefont {Watanabe}}, \bibinfo {author}
  {\bibfnamefont {T.}~\bibnamefont {Taniguchi}}, \bibinfo {author}
  {\bibfnamefont {J.}~\bibnamefont {Alicea}},\ and\ \bibinfo {author}
  {\bibfnamefont {S.}~\bibnamefont {{Nadj-Perge}}},\ }\bibfield  {title}
  {\bibinfo {title} {Enhanced superconductivity in spin--orbit proximitized
  bilayer graphene},\ }\href {https://doi.org/10.1038/s41586-022-05446-x}
  {\bibfield  {journal} {\bibinfo  {journal} {Nature}\ }\textbf {\bibinfo
  {volume} {613}},\ \bibinfo {pages} {268} (\bibinfo {year}
  {2023})}\BibitemShut {NoStop}%
\bibitem [{\citenamefont {Huang}\ \emph {et~al.}(2023)\citenamefont {Huang},
  \citenamefont {Wolf}, \citenamefont {Qin}, \citenamefont {Wei}, \citenamefont
  {Blinov},\ and\ \citenamefont {MacDonald}}]{Blinov2023}%
  \BibitemOpen
  \bibfield  {author} {\bibinfo {author} {\bibfnamefont {C.}~\bibnamefont
  {Huang}}, \bibinfo {author} {\bibfnamefont {T.~M.~R.}\ \bibnamefont {Wolf}},
  \bibinfo {author} {\bibfnamefont {W.}~\bibnamefont {Qin}}, \bibinfo {author}
  {\bibfnamefont {N.}~\bibnamefont {Wei}}, \bibinfo {author} {\bibfnamefont
  {I.~V.}\ \bibnamefont {Blinov}},\ and\ \bibinfo {author} {\bibfnamefont
  {A.~H.}\ \bibnamefont {MacDonald}},\ }\bibfield  {title} {\bibinfo {title}
  {Spin and orbital metallic magnetism in rhombohedral trilayer graphene},\
  }\href {https://doi.org/10.1103/PhysRevB.107.L121405} {\bibfield  {journal}
  {\bibinfo  {journal} {Phys. Rev. B}\ }\textbf {\bibinfo {volume} {107}},\
  \bibinfo {pages} {L121405} (\bibinfo {year} {2023})}\BibitemShut {NoStop}%
\bibitem [{\citenamefont {Blinov}\ \emph {et~al.}(2023)\citenamefont {Blinov},
  \citenamefont {Huang}, \citenamefont {Wei}, \citenamefont {Wei},
  \citenamefont {Wolf},\ and\ \citenamefont {MacDonald}}]{Blinov2023a}%
  \BibitemOpen
  \bibfield  {author} {\bibinfo {author} {\bibfnamefont {I.~V.}\ \bibnamefont
  {Blinov}}, \bibinfo {author} {\bibfnamefont {C.}~\bibnamefont {Huang}},
  \bibinfo {author} {\bibfnamefont {N.}~\bibnamefont {Wei}}, \bibinfo {author}
  {\bibfnamefont {Q.}~\bibnamefont {Wei}}, \bibinfo {author} {\bibfnamefont
  {T.}~\bibnamefont {Wolf}},\ and\ \bibinfo {author} {\bibfnamefont {A.~H.}\
  \bibnamefont {MacDonald}},\ }\href
  {https://doi.org/10.48550/arXiv.2303.17350} {\bibinfo {title} {Partial
  condensation of mobile excitons in graphene multilayers}} (\bibinfo {year}
  {2023}),\ \Eprint {https://arxiv.org/abs/2303.17350} {arxiv:2303.17350
  [cond-mat.mes-hall]} \BibitemShut {NoStop}%
\bibitem [{\citenamefont {Ghazaryan}\ \emph {et~al.}(2021)\citenamefont
  {Ghazaryan}, \citenamefont {Holder}, \citenamefont {Serbyn},\ and\
  \citenamefont {Berg}}]{ghazaryan2021unconventional}%
  \BibitemOpen
  \bibfield  {author} {\bibinfo {author} {\bibfnamefont {A.}~\bibnamefont
  {Ghazaryan}}, \bibinfo {author} {\bibfnamefont {T.}~\bibnamefont {Holder}},
  \bibinfo {author} {\bibfnamefont {M.}~\bibnamefont {Serbyn}},\ and\ \bibinfo
  {author} {\bibfnamefont {E.}~\bibnamefont {Berg}},\ }\bibfield  {title}
  {\bibinfo {title} {Unconventional superconductivity in systems with annular
  fermi surfaces: {{Application}} to rhombohedral trilayer graphene},\ }\href
  {https://doi.org/10.1103/PhysRevLett.127.247001} {\bibfield  {journal}
  {\bibinfo  {journal} {Phys. Rev. Lett.}\ }\textbf {\bibinfo {volume} {127}},\
  \bibinfo {pages} {247001} (\bibinfo {year} {2021})}\BibitemShut {NoStop}%
\bibitem [{\citenamefont {Chatterjee}\ \emph {et~al.}(2022)\citenamefont
  {Chatterjee}, \citenamefont {Wang}, \citenamefont {Berg},\ and\ \citenamefont
  {Zaletel}}]{chatterjee2022inter}%
  \BibitemOpen
  \bibfield  {author} {\bibinfo {author} {\bibfnamefont {S.}~\bibnamefont
  {Chatterjee}}, \bibinfo {author} {\bibfnamefont {T.}~\bibnamefont {Wang}},
  \bibinfo {author} {\bibfnamefont {E.}~\bibnamefont {Berg}},\ and\ \bibinfo
  {author} {\bibfnamefont {M.~P.}\ \bibnamefont {Zaletel}},\ }\bibfield
  {title} {\bibinfo {title} {Inter-valley coherent order and isospin
  fluctuation mediated superconductivity in rhombohedral trilayer graphene},\
  }\href {https://doi.org/10.1038/s41467-022-33561-w} {\bibfield  {journal}
  {\bibinfo  {journal} {Nat. Commun.}\ }\textbf {\bibinfo {volume} {13}},\
  \bibinfo {pages} {6013} (\bibinfo {year} {2022})}\BibitemShut {NoStop}%
\bibitem [{\citenamefont {Chichinadze}\ \emph
  {et~al.}(2022{\natexlab{a}})\citenamefont {Chichinadze}, \citenamefont
  {Classen}, \citenamefont {Wang},\ and\ \citenamefont
  {Chubukov}}]{Chichinadze2022}%
  \BibitemOpen
  \bibfield  {author} {\bibinfo {author} {\bibfnamefont {D.~V.}\ \bibnamefont
  {Chichinadze}}, \bibinfo {author} {\bibfnamefont {L.}~\bibnamefont
  {Classen}}, \bibinfo {author} {\bibfnamefont {Y.}~\bibnamefont {Wang}},\ and\
  \bibinfo {author} {\bibfnamefont {A.~V.}\ \bibnamefont {Chubukov}},\
  }\bibfield  {title} {\bibinfo {title} {Cascade of transitions in twisted and
  non-twisted graphene layers within the van {{Hove}} scenario},\ }\href
  {https://doi.org/10.1038/s41535-022-00520-z} {\bibfield  {journal} {\bibinfo
  {journal} {npj Quantum Mater.}\ }\textbf {\bibinfo {volume} {7}},\ \bibinfo
  {pages} {114} (\bibinfo {year} {2022}{\natexlab{a}})}\BibitemShut {NoStop}%
\bibitem [{\citenamefont {Chichinadze}\ \emph
  {et~al.}(2022{\natexlab{b}})\citenamefont {Chichinadze}, \citenamefont
  {Classen}, \citenamefont {Wang},\ and\ \citenamefont
  {Chubukov}}]{Chichinadze2022a}%
  \BibitemOpen
  \bibfield  {author} {\bibinfo {author} {\bibfnamefont {D.~V.}\ \bibnamefont
  {Chichinadze}}, \bibinfo {author} {\bibfnamefont {L.}~\bibnamefont
  {Classen}}, \bibinfo {author} {\bibfnamefont {Y.}~\bibnamefont {Wang}},\ and\
  \bibinfo {author} {\bibfnamefont {A.~V.}\ \bibnamefont {Chubukov}},\
  }\bibfield  {title} {\bibinfo {title} {{{SU}}(4) {{Symmetry}} in {{Twisted
  Bilayer Graphene}}: {{An Itinerant Perspective}}},\ }\href
  {https://doi.org/10.1103/PhysRevLett.128.227601} {\bibfield  {journal}
  {\bibinfo  {journal} {Phys. Rev. Lett.}\ }\textbf {\bibinfo {volume} {128}},\
  \bibinfo {pages} {227601} (\bibinfo {year} {2022}{\natexlab{b}})}\BibitemShut
  {NoStop}%
\bibitem [{\citenamefont {You}\ and\ \citenamefont
  {Vishwanath}(2022)}]{You2022}%
  \BibitemOpen
  \bibfield  {author} {\bibinfo {author} {\bibfnamefont {Y.-Z.}\ \bibnamefont
  {You}}\ and\ \bibinfo {author} {\bibfnamefont {A.}~\bibnamefont
  {Vishwanath}},\ }\bibfield  {title} {\bibinfo {title} {Kohn-{{Luttinger}}
  superconductivity and intervalley coherence in rhombohedral trilayer
  graphene},\ }\href {https://doi.org/10.1103/PhysRevB.105.134524} {\bibfield
  {journal} {\bibinfo  {journal} {Phys. Rev. B}\ }\textbf {\bibinfo {volume}
  {105}},\ \bibinfo {pages} {134524} (\bibinfo {year} {2022})}\BibitemShut
  {NoStop}%
\bibitem [{\citenamefont {Xie}\ and\ \citenamefont
  {Das~Sarma}(2023)}]{Xie2023}%
  \BibitemOpen
  \bibfield  {author} {\bibinfo {author} {\bibfnamefont {M.}~\bibnamefont
  {Xie}}\ and\ \bibinfo {author} {\bibfnamefont {S.}~\bibnamefont
  {Das~Sarma}},\ }\bibfield  {title} {\bibinfo {title} {Flavor symmetry
  breaking in spin-orbit coupled bilayer graphene},\ }\href
  {https://doi.org/10.1103/PhysRevB.107.L201119} {\bibfield  {journal}
  {\bibinfo  {journal} {Phys. Rev. B}\ }\textbf {\bibinfo {volume} {107}},\
  \bibinfo {pages} {L201119} (\bibinfo {year} {2023})}\BibitemShut {NoStop}%
\bibitem [{\citenamefont {Dong}\ \emph
  {et~al.}(2023{\natexlab{d}})\citenamefont {Dong}, \citenamefont {Ogunnaike},\
  and\ \citenamefont {Levitov}}]{Dong2023b}%
  \BibitemOpen
  \bibfield  {author} {\bibinfo {author} {\bibfnamefont {Z.}~\bibnamefont
  {Dong}}, \bibinfo {author} {\bibfnamefont {O.}~\bibnamefont {Ogunnaike}},\
  and\ \bibinfo {author} {\bibfnamefont {L.}~\bibnamefont {Levitov}},\
  }\bibfield  {title} {\bibinfo {title} {Collective {{Excitations}} in {{Chiral
  Stoner Magnets}}},\ }\href {https://doi.org/10.1103/PhysRevLett.130.206701}
  {\bibfield  {journal} {\bibinfo  {journal} {Phys. Rev. Lett.}\ }\textbf
  {\bibinfo {volume} {130}},\ \bibinfo {pages} {206701} (\bibinfo {year}
  {2023}{\natexlab{d}})}\BibitemShut {NoStop}%
\bibitem [{\citenamefont {Hossain}\ \emph {et~al.}(2021)\citenamefont
  {Hossain}, \citenamefont {Ma}, \citenamefont {{Villegas-Rosales}},
  \citenamefont {Chung}, \citenamefont {Pfeiffer}, \citenamefont {West},
  \citenamefont {Baldwin},\ and\ \citenamefont {Shayegan}}]{Hossain2021}%
  \BibitemOpen
  \bibfield  {author} {\bibinfo {author} {\bibfnamefont {{\relax Md}.~S.}\
  \bibnamefont {Hossain}}, \bibinfo {author} {\bibfnamefont {M.~K.}\
  \bibnamefont {Ma}}, \bibinfo {author} {\bibfnamefont {K.~A.}\ \bibnamefont
  {{Villegas-Rosales}}}, \bibinfo {author} {\bibfnamefont {Y.~J.}\ \bibnamefont
  {Chung}}, \bibinfo {author} {\bibfnamefont {L.~N.}\ \bibnamefont {Pfeiffer}},
  \bibinfo {author} {\bibfnamefont {K.~W.}\ \bibnamefont {West}}, \bibinfo
  {author} {\bibfnamefont {K.~W.}\ \bibnamefont {Baldwin}},\ and\ \bibinfo
  {author} {\bibfnamefont {M.}~\bibnamefont {Shayegan}},\ }\bibfield  {title}
  {\bibinfo {title} {Spontaneous valley polarization of itinerant electrons},\
  }\href {https://doi.org/10.1103/PhysRevLett.127.116601} {\bibfield  {journal}
  {\bibinfo  {journal} {Phys. Rev. Lett.}\ }\textbf {\bibinfo {volume} {127}},\
  \bibinfo {pages} {116601} (\bibinfo {year} {2021})}\BibitemShut {NoStop}%
\bibitem [{\citenamefont {Hossain}\ \emph {et~al.}(2022)\citenamefont
  {Hossain}, \citenamefont {Ma}, \citenamefont {{Villegas-Rosales}},
  \citenamefont {Chung}, \citenamefont {Pfeiffer}, \citenamefont {West},
  \citenamefont {Baldwin},\ and\ \citenamefont {Shayegan}}]{Hossain2022}%
  \BibitemOpen
  \bibfield  {author} {\bibinfo {author} {\bibfnamefont {{\relax Md}.~S.}\
  \bibnamefont {Hossain}}, \bibinfo {author} {\bibfnamefont {M.~K.}\
  \bibnamefont {Ma}}, \bibinfo {author} {\bibfnamefont {K.~A.}\ \bibnamefont
  {{Villegas-Rosales}}}, \bibinfo {author} {\bibfnamefont {Y.~J.}\ \bibnamefont
  {Chung}}, \bibinfo {author} {\bibfnamefont {L.~N.}\ \bibnamefont {Pfeiffer}},
  \bibinfo {author} {\bibfnamefont {K.~W.}\ \bibnamefont {West}}, \bibinfo
  {author} {\bibfnamefont {K.~W.}\ \bibnamefont {Baldwin}},\ and\ \bibinfo
  {author} {\bibfnamefont {M.}~\bibnamefont {Shayegan}},\ }\bibfield  {title}
  {\bibinfo {title} {Anisotropic two-dimensional disordered wigner solid},\
  }\href {https://doi.org/10.1103/PhysRevLett.129.036601} {\bibfield  {journal}
  {\bibinfo  {journal} {Phys. Rev. Lett.}\ }\textbf {\bibinfo {volume} {129}},\
  \bibinfo {pages} {036601} (\bibinfo {year} {2022})}\BibitemShut {NoStop}%
\bibitem [{\citenamefont {Hossain}\ \emph {et~al.}(2020)\citenamefont
  {Hossain}, \citenamefont {Ma}, \citenamefont {Rosales}, \citenamefont
  {Chung}, \citenamefont {Pfeiffer}, \citenamefont {West}, \citenamefont
  {Baldwin},\ and\ \citenamefont {{M. Shayegan}}}]{Hossain2020}%
  \BibitemOpen
  \bibfield  {author} {\bibinfo {author} {\bibfnamefont {M.~S.}\ \bibnamefont
  {Hossain}}, \bibinfo {author} {\bibfnamefont {M.~K.}\ \bibnamefont {Ma}},
  \bibinfo {author} {\bibfnamefont {K.~A.~V.}\ \bibnamefont {Rosales}},
  \bibinfo {author} {\bibfnamefont {Y.~J.}\ \bibnamefont {Chung}}, \bibinfo
  {author} {\bibfnamefont {L.~N.}\ \bibnamefont {Pfeiffer}}, \bibinfo {author}
  {\bibfnamefont {K.~W.}\ \bibnamefont {West}}, \bibinfo {author}
  {\bibfnamefont {K.~W.}\ \bibnamefont {Baldwin}},\ and\ \bibinfo {author}
  {\bibnamefont {{M. Shayegan}}},\ }\bibfield  {title} {\bibinfo {title}
  {Observation of spontaneous ferromagnetism in a two-dimensional electron
  system},\ }\href {https://doi.org/10.1073/pnas.2018248117} {\bibfield
  {journal} {\bibinfo  {journal} {Proc. Natl. Acad. Sci.}\ }\textbf {\bibinfo
  {volume} {117}},\ \bibinfo {pages} {32244} (\bibinfo {year}
  {2020})}\BibitemShut {NoStop}%
\bibitem [{\citenamefont {Valenti}\ \emph {et~al.}(2023)\citenamefont
  {Valenti}, \citenamefont {Calvera}, \citenamefont {Kivelson}, \citenamefont
  {Berg},\ and\ \citenamefont {Huber}}]{Valenti2023}%
  \BibitemOpen
  \bibfield  {author} {\bibinfo {author} {\bibfnamefont {A.}~\bibnamefont
  {Valenti}}, \bibinfo {author} {\bibfnamefont {V.}~\bibnamefont {Calvera}},
  \bibinfo {author} {\bibfnamefont {S.~A.}\ \bibnamefont {Kivelson}}, \bibinfo
  {author} {\bibfnamefont {E.}~\bibnamefont {Berg}},\ and\ \bibinfo {author}
  {\bibfnamefont {S.~D.}\ \bibnamefont {Huber}},\ }\href
  {https://doi.org/10.48550/arXiv.2307.15119} {\bibinfo {title} {Nematic metal
  in a multi-valley electron gas: {{Variational Monte Carlo}} analysis and
  application to {{AlAs}}}} (\bibinfo {year} {2023}),\ \Eprint
  {https://arxiv.org/abs/2307.15119} {arxiv:2307.15119 [cond-mat.str-el]}
  \BibitemShut {NoStop}%
\bibitem [{\citenamefont {Shimizu}(1964)}]{Shimizu1964}%
  \BibitemOpen
  \bibfield  {author} {\bibinfo {author} {\bibfnamefont {M.}~\bibnamefont
  {Shimizu}},\ }\bibfield  {title} {\bibinfo {title} {On the conditions of
  ferromagnetism by the band model},\ }\href
  {https://doi.org/10.1088/0370-1328/84/3/309} {\bibfield  {journal} {\bibinfo
  {journal} {Proc. Phys. Soc.}\ }\textbf {\bibinfo {volume} {84}},\ \bibinfo
  {pages} {397} (\bibinfo {year} {1964})}\BibitemShut {NoStop}%
\bibitem [{\citenamefont {Hertz}(1976)}]{Hertz1976}%
  \BibitemOpen
  \bibfield  {author} {\bibinfo {author} {\bibfnamefont {J.~A.}\ \bibnamefont
  {Hertz}},\ }\bibfield  {title} {\bibinfo {title} {Quantum critical
  phenomena},\ }\href@noop {} {\bibfield  {journal} {\bibinfo  {journal} {Phys.
  Rev. B}\ }\textbf {\bibinfo {volume} {14}},\ \bibinfo {pages} {1165}
  (\bibinfo {year} {1976})}\BibitemShut {NoStop}%
\bibitem [{\citenamefont {Millis}(1993)}]{Millis1993}%
  \BibitemOpen
  \bibfield  {author} {\bibinfo {author} {\bibfnamefont {A.~J.}\ \bibnamefont
  {Millis}},\ }\bibfield  {title} {\bibinfo {title} {Effect of a nonzero
  temperature on quantum critical points in itinerant fermion systems},\ }\href
  {https://doi.org/10.1103/PhysRevB.48.7183} {\bibfield  {journal} {\bibinfo
  {journal} {Phys. Rev. B}\ }\textbf {\bibinfo {volume} {48}},\ \bibinfo
  {pages} {7183} (\bibinfo {year} {1993})}\BibitemShut {NoStop}%
\bibitem [{\citenamefont {Moriya}(2012)}]{Moriya2012}%
  \BibitemOpen
  \bibfield  {author} {\bibinfo {author} {\bibfnamefont {T.}~\bibnamefont
  {Moriya}},\ }\href@noop {} {\emph {\bibinfo {title} {Spin Fluctuations in
  Itinerant Electron Magnetism}}},\ Vol.~\bibinfo {volume} {56}\ (\bibinfo
  {publisher} {Springer Science \& Business Media},\ \bibinfo {year}
  {2012})\BibitemShut {NoStop}%
\bibitem [{\citenamefont {Vojta}\ \emph {et~al.}(1999)\citenamefont {Vojta},
  \citenamefont {Belitz}, \citenamefont {Kirkpatrick},\ and\ \citenamefont
  {Narayanan}}]{Vojta1999}%
  \BibitemOpen
  \bibfield  {author} {\bibinfo {author} {\bibfnamefont {T.}~\bibnamefont
  {Vojta}}, \bibinfo {author} {\bibfnamefont {D.}~\bibnamefont {Belitz}},
  \bibinfo {author} {\bibfnamefont {T.}~\bibnamefont {Kirkpatrick}},\ and\
  \bibinfo {author} {\bibfnamefont {R.}~\bibnamefont {Narayanan}},\ }\bibfield
  {title} {\bibinfo {title} {Quantum critical behavior of itinerant
  ferromagnets},\ }\href {https://doi.org/10.1002/andp.199951107-908}
  {\bibfield  {journal} {\bibinfo  {journal} {Ann. Phys.}\ }\textbf {\bibinfo
  {volume} {511}},\ \bibinfo {pages} {593} (\bibinfo {year}
  {1999})}\BibitemShut {NoStop}%
\bibitem [{\citenamefont {Chubukov}\ \emph {et~al.}(2004)\citenamefont
  {Chubukov}, \citenamefont {P{\'e}pin},\ and\ \citenamefont
  {Rech}}]{Chubukov2004}%
  \BibitemOpen
  \bibfield  {author} {\bibinfo {author} {\bibfnamefont {A.~V.}\ \bibnamefont
  {Chubukov}}, \bibinfo {author} {\bibfnamefont {C.}~\bibnamefont
  {P{\'e}pin}},\ and\ \bibinfo {author} {\bibfnamefont {J.}~\bibnamefont
  {Rech}},\ }\bibfield  {title} {\bibinfo {title} {Instability of the
  {{Quantum-Critical Point}} of {{Itinerant Ferromagnets}}},\ }\href
  {https://doi.org/10.1103/physrevlett.92.147003} {\bibfield  {journal}
  {\bibinfo  {journal} {Phys. Rev. Lett.}\ }\textbf {\bibinfo {volume} {92}},\
  \bibinfo {pages} {147003} (\bibinfo {year} {2004})}\BibitemShut {NoStop}%
\bibitem [{\citenamefont {Efremov}\ \emph {et~al.}(2008)\citenamefont
  {Efremov}, \citenamefont {Betouras},\ and\ \citenamefont
  {Chubukov}}]{Efremov2008}%
  \BibitemOpen
  \bibfield  {author} {\bibinfo {author} {\bibfnamefont {D.~V.}\ \bibnamefont
  {Efremov}}, \bibinfo {author} {\bibfnamefont {J.~J.}\ \bibnamefont
  {Betouras}},\ and\ \bibinfo {author} {\bibfnamefont {A.}~\bibnamefont
  {Chubukov}},\ }\bibfield  {title} {\bibinfo {title} {Nonanalytic behavior of
  two-dimensional itinerant ferromagnets},\ }\href
  {https://doi.org/10.1103/PhysRevB.77.220401} {\bibfield  {journal} {\bibinfo
  {journal} {Phys. Rev. B}\ }\textbf {\bibinfo {volume} {77}},\ \bibinfo
  {pages} {220401(R)} (\bibinfo {year} {2008})}\BibitemShut {NoStop}%
\bibitem [{\citenamefont {Chubukov}\ and\ \citenamefont
  {Maslov}(2009)}]{Chubukov2009}%
  \BibitemOpen
  \bibfield  {author} {\bibinfo {author} {\bibfnamefont {A.~V.}\ \bibnamefont
  {Chubukov}}\ and\ \bibinfo {author} {\bibfnamefont {D.~L.}\ \bibnamefont
  {Maslov}},\ }\bibfield  {title} {\bibinfo {title} {Spin {{Conservation}} and
  {{Fermi Liquid}} near a {{Ferromagnetic Quantum Critical Point}}},\ }\href
  {https://doi.org/10.1103/PhysRevLett.103.216401} {\bibfield  {journal}
  {\bibinfo  {journal} {Phys. Rev. Lett.}\ }\textbf {\bibinfo {volume} {103}},\
  \bibinfo {pages} {216401} (\bibinfo {year} {2009})}\BibitemShut {NoStop}%
\bibitem [{\citenamefont {Aleiner}\ \emph {et~al.}(2007)\citenamefont
  {Aleiner}, \citenamefont {Kharzeev},\ and\ \citenamefont
  {Tsvelik}}]{Aleiner2007}%
  \BibitemOpen
  \bibfield  {author} {\bibinfo {author} {\bibfnamefont {I.~L.}\ \bibnamefont
  {Aleiner}}, \bibinfo {author} {\bibfnamefont {D.~E.}\ \bibnamefont
  {Kharzeev}},\ and\ \bibinfo {author} {\bibfnamefont {A.~M.}\ \bibnamefont
  {Tsvelik}},\ }\bibfield  {title} {\bibinfo {title} {Spontaneous symmetry
  breaking in graphene subjected to an in-plane magnetic field},\ }\href
  {https://doi.org/10.1103/physrevb.76.195415} {\bibfield  {journal} {\bibinfo
  {journal} {Phys. Rev. B}\ }\textbf {\bibinfo {volume} {76}},\ \bibinfo
  {pages} {195415} (\bibinfo {year} {2007})}\BibitemShut {NoStop}%
\bibitem [{\citenamefont {Kharitonov}(2012)}]{Kharitonov2012}%
  \BibitemOpen
  \bibfield  {author} {\bibinfo {author} {\bibfnamefont {M.}~\bibnamefont
  {Kharitonov}},\ }\bibfield  {title} {\bibinfo {title} {Phase diagram for the
  {$N$}=0 quantum {{Hall}} state in monolayer graphene},\ }\href
  {https://doi.org/10.1103/physrevb.85.155439} {\bibfield  {journal} {\bibinfo
  {journal} {Phys. Rev. B}\ }\textbf {\bibinfo {volume} {85}},\ \bibinfo
  {pages} {155439} (\bibinfo {year} {2012})}\BibitemShut {NoStop}%
\bibitem [{\citenamefont {Raines}\ \emph {et~al.}(2021)\citenamefont {Raines},
  \citenamefont {Fal'ko},\ and\ \citenamefont {Glazman}}]{Raines2021}%
  \BibitemOpen
  \bibfield  {author} {\bibinfo {author} {\bibfnamefont {Z.~M.}\ \bibnamefont
  {Raines}}, \bibinfo {author} {\bibfnamefont {V.~I.}\ \bibnamefont {Fal'ko}},\
  and\ \bibinfo {author} {\bibfnamefont {L.~I.}\ \bibnamefont {Glazman}},\
  }\bibfield  {title} {\bibinfo {title} {Spin-valley collective modes of the
  electron liquid in graphene},\ }\href
  {https://doi.org/10.1103/PhysRevB.103.075422} {\bibfield  {journal} {\bibinfo
   {journal} {Phys. Rev. B}\ }\textbf {\bibinfo {volume} {103}},\ \bibinfo
  {pages} {075422} (\bibinfo {year} {2021})}\BibitemShut {NoStop}%
\bibitem [{\citenamefont {Tremblay}(2021)}]{Tremblay2021}%
  \BibitemOpen
  \bibfield  {author} {\bibinfo {author} {\bibfnamefont {A.-M.}\ \bibnamefont
  {Tremblay}},\ }\bibfield  {title} {\bibinfo {title} {{{PHY-892 Quantum
  Material}}'s {{Theory}}, from perturbation theory to dynamical-mean field
  theory}} (\bibinfo {year} {2021})\BibitemShut {NoStop}%
\bibitem [{\citenamefont {Koshino}\ and\ \citenamefont
  {McCann}(2009)}]{Koshino2009}%
  \BibitemOpen
  \bibfield  {author} {\bibinfo {author} {\bibfnamefont {M.}~\bibnamefont
  {Koshino}}\ and\ \bibinfo {author} {\bibfnamefont {E.}~\bibnamefont
  {McCann}},\ }\bibfield  {title} {\bibinfo {title} {Trigonal warping and
  {{Berry}}'s phase {{N}} {$\pi$} in {{ABC-stacked}} multilayer graphene},\
  }\href {https://doi.org/10.1103/PhysRevB.80.165409} {\bibfield  {journal}
  {\bibinfo  {journal} {Phys. Rev. B}\ }\textbf {\bibinfo {volume} {80}},\
  \bibinfo {pages} {165409} (\bibinfo {year} {2009})}\BibitemShut {NoStop}%
\bibitem [{\citenamefont {Klein}\ \emph {et~al.}(2020)\citenamefont {Klein},
  \citenamefont {Maslov},\ and\ \citenamefont {Chubukov}}]{Klein2020}%
  \BibitemOpen
  \bibfield  {author} {\bibinfo {author} {\bibfnamefont {A.}~\bibnamefont
  {Klein}}, \bibinfo {author} {\bibfnamefont {D.~L.}\ \bibnamefont {Maslov}},\
  and\ \bibinfo {author} {\bibfnamefont {A.~V.}\ \bibnamefont {Chubukov}},\
  }\bibfield  {title} {\bibinfo {title} {Hidden and mirage collective modes in
  two dimensional {{Fermi}} liquids},\ }\href
  {https://doi.org/10.1038/s41535-020-0250-4} {\bibfield  {journal} {\bibinfo
  {journal} {npj Quantum Materials}\ }\textbf {\bibinfo {volume} {5}},\
  \bibinfo {pages} {55} (\bibinfo {year} {2020})}\BibitemShut {NoStop}%
\bibitem [{\citenamefont {Ma}\ and\ \citenamefont {Lee}(2024)}]{Ma2024}%
  \BibitemOpen
  \bibfield  {author} {\bibinfo {author} {\bibfnamefont {H.}~\bibnamefont
  {Ma}}\ and\ \bibinfo {author} {\bibfnamefont {S.-S.}\ \bibnamefont {Lee}},\
  }\bibfield  {title} {\bibinfo {title} {Fermi liquids beyond the
  forward-scattering limit: {{The}} role of nonforward scattering for scale
  invariance and instabilities},\ }\href
  {https://doi.org/10.1103/PhysRevB.109.045143} {\bibfield  {journal} {\bibinfo
   {journal} {Phys. Rev. B}\ }\textbf {\bibinfo {volume} {109}},\ \bibinfo
  {pages} {045143} (\bibinfo {year} {2024})}\BibitemShut {NoStop}%
\bibitem [{\citenamefont {Silin}(1958)}]{Silin1958}%
  \BibitemOpen
  \bibfield  {author} {\bibinfo {author} {\bibfnamefont {V.~P.}\ \bibnamefont
  {Silin}},\ }\bibfield  {title} {\bibinfo {title} {Oscillations of a
  {{Fermi-liquid}} in a magnetic field},\ }\href@noop {} {\bibfield  {journal}
  {\bibinfo  {journal} {Sov. Phys. JETP}\ }\textbf {\bibinfo {volume} {6}},\
  \bibinfo {pages} {945} (\bibinfo {year} {1958})}\BibitemShut {NoStop}%
\bibitem [{\citenamefont {Raines}\ \emph {et~al.}(2022)\citenamefont {Raines},
  \citenamefont {Maslov},\ and\ \citenamefont {Glazman}}]{Raines2022}%
  \BibitemOpen
  \bibfield  {author} {\bibinfo {author} {\bibfnamefont {Z.~M.}\ \bibnamefont
  {Raines}}, \bibinfo {author} {\bibfnamefont {D.~L.}\ \bibnamefont {Maslov}},\
  and\ \bibinfo {author} {\bibfnamefont {L.~I.}\ \bibnamefont {Glazman}},\
  }\bibfield  {title} {\bibinfo {title} {Spin-valley {{Silin}} modes in
  graphene with substrate-induced spin-orbit coupling},\ }\href
  {https://doi.org/10.1103/PhysRevB.105.L201201} {\bibfield  {journal}
  {\bibinfo  {journal} {Phys. Rev. B}\ }\textbf {\bibinfo {volume} {105}},\
  \bibinfo {pages} {L201201} (\bibinfo {year} {2022})},\ \Eprint
  {https://arxiv.org/abs/2107.02819} {arxiv:2107.02819 [cond-mat]} \BibitemShut
  {NoStop}%
\bibitem [{\citenamefont {Watanabe}\ and\ \citenamefont
  {Brauner}(2011)}]{Watanabe2011}%
  \BibitemOpen
  \bibfield  {author} {\bibinfo {author} {\bibfnamefont {H.}~\bibnamefont
  {Watanabe}}\ and\ \bibinfo {author} {\bibfnamefont {T.}~\bibnamefont
  {Brauner}},\ }\bibfield  {title} {\bibinfo {title} {Number of
  {{Nambu-Goldstone}} bosons and its relation to charge densities},\ }\href
  {https://doi.org/10.1103/PhysRevD.84.125013} {\bibfield  {journal} {\bibinfo
  {journal} {Phys. Rev. D}\ }\textbf {\bibinfo {volume} {84}},\ \bibinfo
  {pages} {125013} (\bibinfo {year} {2011})}\BibitemShut {NoStop}%
\bibitem [{\citenamefont {Lif{\v s}ic}\ \emph {et~al.}(2006)\citenamefont
  {Lif{\v s}ic}, \citenamefont {Pitaevskij}, \citenamefont {Landau},\ and\
  \citenamefont {Lifshitz}}]{Lifsic2006}%
  \BibitemOpen
  \bibfield  {author} {\bibinfo {author} {\bibfnamefont {E.~M.}\ \bibnamefont
  {Lif{\v s}ic}}, \bibinfo {author} {\bibfnamefont {L.~P.}\ \bibnamefont
  {Pitaevskij}}, \bibinfo {author} {\bibfnamefont {L.~D.}\ \bibnamefont
  {Landau}},\ and\ \bibinfo {author} {\bibfnamefont {E.~M.}\ \bibnamefont
  {Lifshitz}},\ }\href@noop {} {\emph {\bibinfo {title} {Statistical Physics.
  {{Part}} 2. {{Theory}} of the Condensed State}}},\ \bibinfo {edition}
  {reprinted}\ ed.,\ \bibinfo {series} {Course of {{Theoretical Physics}}},
  Vol.~\bibinfo {volume} {9}\ (\bibinfo  {publisher} {Elsevier},\ \bibinfo
  {address} {Oxford},\ \bibinfo {year} {2006})\BibitemShut {NoStop}%
\bibitem [{\citenamefont {Nozieres}\ and\ \citenamefont
  {Pines}(1999)}]{Nozieres1999}%
  \BibitemOpen
  \bibfield  {author} {\bibinfo {author} {\bibfnamefont {P.}~\bibnamefont
  {Nozieres}}\ and\ \bibinfo {author} {\bibfnamefont {D.}~\bibnamefont
  {Pines}},\ }\href@noop {} {\emph {\bibinfo {title} {Theory {{Of Quantum
  Liquids}}}}},\ Advanced {{Books Classics}}\ (\bibinfo  {publisher} {Avalon
  Publishing},\ \bibinfo {year} {1999})\BibitemShut {NoStop}%
\bibitem [{\citenamefont {McCann}\ and\ \citenamefont
  {Koshino}(2013)}]{McCann2013}%
  \BibitemOpen
  \bibfield  {author} {\bibinfo {author} {\bibfnamefont {E.}~\bibnamefont
  {McCann}}\ and\ \bibinfo {author} {\bibfnamefont {M.}~\bibnamefont
  {Koshino}},\ }\bibfield  {title} {\bibinfo {title} {The electronic properties
  of bilayer graphene},\ }\href {https://doi.org/10.1088/0034-4885/76/5/056503}
  {\bibfield  {journal} {\bibinfo  {journal} {Rep. Prog. Phys.}\ }\textbf
  {\bibinfo {volume} {76}},\ \bibinfo {pages} {056503} (\bibinfo {year}
  {2013})}\BibitemShut {NoStop}%
\bibitem [{\citenamefont {Rakhmanov}\ \emph {et~al.}(2023)\citenamefont
  {Rakhmanov}, \citenamefont {Rozhkov}, \citenamefont {Sboychakov},\ and\
  \citenamefont {Nori}}]{Rakhmanov2023}%
  \BibitemOpen
  \bibfield  {author} {\bibinfo {author} {\bibfnamefont {A.~L.}\ \bibnamefont
  {Rakhmanov}}, \bibinfo {author} {\bibfnamefont {A.~V.}\ \bibnamefont
  {Rozhkov}}, \bibinfo {author} {\bibfnamefont {A.~O.}\ \bibnamefont
  {Sboychakov}},\ and\ \bibinfo {author} {\bibfnamefont {F.}~\bibnamefont
  {Nori}},\ }\bibfield  {title} {\bibinfo {title} {Half-metal and other
  fractional metal phases in doped {{A B}} bilayer graphene},\ }\href
  {https://doi.org/10.1103/PhysRevB.107.155112} {\bibfield  {journal} {\bibinfo
   {journal} {Phys. Rev. B}\ }\textbf {\bibinfo {volume} {107}},\ \bibinfo
  {pages} {155112} (\bibinfo {year} {2023})}\BibitemShut {NoStop}%
\bibitem [{\citenamefont {Sboychakov}\ \emph {et~al.}(2023)\citenamefont
  {Sboychakov}, \citenamefont {Rozhkov},\ and\ \citenamefont
  {Rakhmanov}}]{Rakhmanov2023a}%
  \BibitemOpen
  \bibfield  {author} {\bibinfo {author} {\bibfnamefont {A.~O.}\ \bibnamefont
  {Sboychakov}}, \bibinfo {author} {\bibfnamefont {A.~V.}\ \bibnamefont
  {Rozhkov}},\ and\ \bibinfo {author} {\bibfnamefont {A.~L.}\ \bibnamefont
  {Rakhmanov}},\ }\bibfield  {title} {\bibinfo {title} {Triplet
  superconductivity and spin density wave in biased {{AB}} bilayer graphene},\
  }\href {https://doi.org/10.1103/PhysRevB.108.184503} {\bibfield  {journal}
  {\bibinfo  {journal} {Phys. Rev. B}\ }\textbf {\bibinfo {volume} {108}},\
  \bibinfo {pages} {184503} (\bibinfo {year} {2023})}\BibitemShut {NoStop}%
\bibitem [{\citenamefont {Rozhkov}\ \emph {et~al.}(2023)\citenamefont
  {Rozhkov}, \citenamefont {Sboychakov},\ and\ \citenamefont
  {Rakhmanov}}]{Rakhmanov2023b}%
  \BibitemOpen
  \bibfield  {author} {\bibinfo {author} {\bibfnamefont {A.~V.}\ \bibnamefont
  {Rozhkov}}, \bibinfo {author} {\bibfnamefont {A.~O.}\ \bibnamefont
  {Sboychakov}},\ and\ \bibinfo {author} {\bibfnamefont {A.~L.}\ \bibnamefont
  {Rakhmanov}},\ }\bibfield  {title} {\bibinfo {title} {Ordering in the
  {{SU}}(4)-symmetric model of {{AA}} bilayer graphene},\ }\href
  {https://doi.org/10.1103/PhysRevB.108.205153} {\bibfield  {journal} {\bibinfo
   {journal} {Phys. Rev. B}\ }\textbf {\bibinfo {volume} {108}},\ \bibinfo
  {pages} {205153} (\bibinfo {year} {2023})}\BibitemShut {NoStop}%
\end{thebibliography}%

\appendix
\section{Hubbard-Stratonovich derivation of mean field energies}
\label{sec:hs-deriv}

The expression for the mean-field energies can be derived via Hubbard-Stratonovich transformation of the original action.
We illustrate this here for the case of the $k^2$ dispersion and the
minimal $\mathrm{SU(4)}$ invariant model where we have set $g_{\gamma}=g$, $g_{d}=0$ in \cref{eq:Smodel}
\begin{multline}
    S =  -\sum_{k\sigma\tau}\bar{\psi}_{k\sigma\tau}(i\epsilon_{n} - \epsilon(\abs{\mathbf{k}}) + \mu)\psi_{k\sigma\tau}\\
    - \frac{1}{8}g T^{2}\sum_{kk'q} \sum_{\gamma} \bar{\psi}_{k+q/2}\hat{\Gamma}_{\gamma}\psi_{k-q/2}\bar{\psi}_{k'-q/2}\hat{\Gamma}_{\gamma}\psi_{k'+q/2}.
    \label{eq:Smodelsu4}
\end{multline}
We introduction Hubbard-Stratonovich fields $\Delta_{i}$ with action
\begin{equation}
    S_{HS}  = \frac{2}{g} \sum_{q} |\Delta_{q,i}|^{2}
\end{equation}
and perform the shift
\begin{equation}
    \Delta_{q,\gamma}  \to \Delta_{q,\gamma} - \frac{g}{4} T \sum_{k} \bar{\psi}_{k'-q/2}\hat{\Gamma}_{\gamma}\psi_{k+q/2}.
\end{equation}
to obtain
\begin{multline}
    S[\psi, \Delta]  =
    -\sum_{k\sigma\tau}\bar{\psi}_{k\sigma\tau}(i\epsilon_{n} - \epsilon(\abs{\mathbf{k}}) +
    \bar{\mu}
    )\psi_{k\sigma\tau}\\
    - T\sum_{k,q}\bar{\psi}_{k+q/2}\hat{\Gamma}_{\gamma} \Delta_{q,\gamma}\psi_{k-q/2}
    + \frac{2}{g}\sum_{q}|\Delta_{q,\gamma}|^{2}.
\end{multline}
The matrix $\sum_{\gamma}\Delta_{\gamma}\hat{\Gamma}_{\gamma}$ is Hermitian and traceless and can therefore be diagonalized
\begin{equation}
    \hat{R}^{\dagger}\sum_{\gamma}\Delta_{\gamma}\hat{\Gamma}_{\gamma}\hat{R} = \sum_{\mu}\phi_{\mu}\hat{T}_{\mu}
\end{equation}
where we have defined three new linear combinations of the order parameters $\phi_{\mu=1,2,3}$ and associated diagonal traceless matrices $\hat{T}_{\mu}$
\begin{equation}
    \begin{gathered}
        \hat{T}_{1}  = \diag(1, -1, 1, -1),\\
        \hat{T}_{2}  = \diag(1, 1, -1, -1),\\
        \hat{T}_{3}  = \diag(1, -1, -1, 1).
    \end{gathered}
\end{equation}
Performing the change of basis $\psi=\hat{R}c$ and $\Delta_{\gamma}\to\phi_{\mu}$ the action becomes
\begin{multline}
    S[\psi, \Delta]  =
    -\sum_{k\lambda\rho}\bar{c}_{k\lambda\rho}(i\epsilon_{n} - \epsilon(\abs{\mathbf{k}}) +
    \mu
    )c_{k\lambda\rho}\\
    - T\sum_{k,q,\mu}\bar{c}_{k+q/2}\hat{T}_{\mu} \phi_{\mu,q}c_{k-q/2}
    + \frac{2}{g}\sum_{q,\mu}|\phi_{q,\mu}|^{2}
\end{multline}
where here, as in the main text, $\lambda,\rho$ are quantum numbers taking the values $\pm$, which index the bands in the diagonal basis.
Defining
\begin{equation}
    \hat{G}^{-1}_{kk'} = (i\epsilon_{n} - \epsilon(\abs{\mathbf{k}}) +
    \bar{\mu}
    )\beta\delta_{k-k'} + \sum_{\mu}\phi_{k-k',\mu}\hat{T}_{\mu}
    \label{eq:hs-gf}
\end{equation}
we can integrate out the fermions to obtain
\begin{equation}
    S_{\phi} =  \frac{2}{g}\sum_{q,\mu}|\phi_{q,\mu}|^{2} - \ln\det\hat{G}^{-1}.
\end{equation}
We now will make a saddle point approximation, assuming spatially constant order,
to obtain the mean-field grand potential
\begin{equation}
    \frac{\Omega}{V}=   \frac{2}{g}\sum_{\mu}|\phi_{\mu}|^{2}
    - T\sum_{k\lambda\rho}\ln \beta G^{-1}_{k\lambda\rho}
    \label{eq:grand-can}
\end{equation}
There are two obvious routes to calculating this quantity: we may either just directly perform the frequency and momentum integrals, or we may expand in $\phi$ to obtain a Landau like free energy.
We will illustrate both here, as the first reproduces the approach of the main text, while second illuminates why quartic and higher terms are absent for a parabolic band.

Inserting the explicit form the Green's function \cref{eq:hs-gf}
\begin{equation}
    \frac{\Omega}{V}=   \frac{2}{g}\sum_{\mu}|\phi_{\mu}|^{2}
    - T\sum_{k\lambda\rho}\ln\beta\left(i\epsilon_{n}-\epsilon(\abs{\mathbf{k}}) +\mu_{\lambda\rho}\right)
\end{equation}
where we have defined the chemical potential relative to the bottom of each band $\mu_{\lambda\rho} =
    \bar{\mu}
    + \lambda \phi_{1}  + \rho \phi_{2} + \lambda \rho \phi_{3}$.
The Matsubara sum is evaluated via standard techniques and, converting from integration over momentum to integration over energy, we have
\begin{equation}
    \frac{\Omega}{V}=   \frac{2}{g}\sum_{\mu}|\phi_{\mu}|^{2}
    - \nu_{F,1} \sum_{\lambda\rho}T\int_{0}^{\infty} d\epsilon \ln\left(1 + \exp\frac{\mu_{\lambda\rho}-\epsilon}{T}\right).
\end{equation}
where, we remind, the sub-index $1$ in $\nu_{F,1}$  stands for the exponent
$\alpha =1$  in the dispersion $k^{2\alpha}$.
Taking the $T\to0$ limit
\begin{multline}
    \lim_{T\to0}\frac{\Omega}{V}=   \frac{2}{g}\sum_{\mu}|\phi_{\mu}|^{2}\\
    +\lim_{T\to0}\nu_{F,1}\sum_{\lambda\rho} T\int_{0}^{\infty} d\epsilon \frac{\epsilon - \mu_{\lambda\rho}}{T}\Theta(\mu_{\lambda\rho}-\epsilon),
\end{multline}
and performing the integration
\begin{equation}
    \frac{\Omega}{V}=   \frac{2}{g}\sum_{\mu}|\phi_{\mu}|^{2}
    -\sum_{\lambda\rho}\frac{1}{2}\mu^{2}_{\lambda\rho}
\end{equation}
we recover our quadratic energy.

Now let us, alternatively, approach the problem as a sum over terms with $l$ electron Green's functions.
The trace log in \cref{eq:grand-can} can be written
\begin{equation}
    - T\sum_{k\lambda\rho}\ln\beta G^{-1}_{k\lambda\rho}
    =
    - T\sum_{k\lambda\rho}\ln\beta G^{-1}_{k0}
    + T\sum^{\infty}_{l=1}\sum_{k\lambda\rho}\frac{G^{l}_{k0}}{l}(-\phi_{\lambda\rho})^{l}
    \label{eq:powers-of-phi}
\end{equation}
where $G_{k0}$ is the Green's function of the unpolarized state, $G^{-1}_{k0} = i\epsilon_{n}-\epsilon(\abs{\mathbf{k}})+\bar{\mu}$.
In this way, we can write the free energy as a power series in the fields $\phi_{\mu}$.
Let us then consider the integrals appearing in the second term of \cref{eq:powers-of-phi}
\begin{equation}
    I_{l}(T) =  \nu_{F,1} \int d\epsilon T \sum_{n}\frac{1}{(i\epsilon_{n}-\epsilon +
        \bar{\mu})^{l}}.
\end{equation}
Here we have already taken $\omega_{m}=0,\mathbf{q}\to0$, corresponding the equilibrium limit.
There are no worries about order of integration for $l>2$ since the integral and sum are absolutely convergent.
Performing the Matsubara sum first
\begin{multline}
    I_{l}(T)=\nu_{F,1}\int_{0}^{\infty} d\epsilon  \frac{1}{l-1}\partial^{l-1}_{\epsilon}n_{F}(\epsilon-
    \bar{\mu}
    )\\
    = \frac{\nu_{F,1}}{l-1}\partial^{l-2}_{\epsilon}n_{F}\bigg|^{\infty}_{-
    \bar{\mu}
    }
    = -\frac{\nu_{F,1}}{l-1}\partial^{l-2}_{\epsilon}n_{F}\bigg|_{\epsilon=-
    \bar{\mu}
    }
\end{multline}
Thus $\lim_{T\to0}I_{l>2}(T)=0$.
Plugging this back into \cref{eq:grand-can} we see that all terms beyond quadratic order in $\phi$ vanish in the $T\to0$ limit.
Note that this is a direct consequence of the constant density of states
for a parabolic dispersion in 2D.
For a general $\nu(\epsilon)$ we would instead have $\lim_{T\to0}I_{l>2}(T)\propto \partial^{l-2}\nu(\epsilon) \neq0$,
in which case $\Omega$ in \cref{eq:grand-can} contains higher order terms in $\phi$.

\section{Diagnosing the transition from the collective modes}\label{sec:analytic-collective}
In the Heisenberg case, the discontinuous nature of the transition can be seen in the analytic structure of the collective modes across the transition line.
This is most clearly seen by looking at the behavior of the longitudinal mode in the vicinity of the Stoner critical point $g_{c}$.
As discussed above, the unpolarized state mode equation \cref{eq:normal-state-modes} has a linear-in-$q$ solution $\omega \approx i\delta v_{F} q$ in the vicinity of the critical point, where $\delta \equiv \nu_{F}g_{z}-1$ parameterizes the distance from the Stoner criterion.
As $\delta$ crosses $0$, this overdamped mode naively moves into the upper half plane, indicating the instability of the unpolarized state.

Let us first see how this is resolved in the case of the usual 3D Stoner transition.
We introduce the order parameter $\Delta$ which is related to the densities and Fermi energies of the bands in the same way as \cref{sec:mf}, i.e., $\Delta = gn_{0}\zeta$, $E_{F\lambda} = \bar{\mu} \pm \Delta$.
The d-dimensional analog of \cref{eq:long-susc} for a generic polarization in the ordered state is
\begin{multline}
    \chi^{R} _{L\lambda}(\omega, \mathbf{q})
    =
    \frac{1}{2}\int \frac{d\mathbf{k}}{(2\pi)^{d}}\\
    \times
    \frac{\Theta(E_{F\lambda}-\epsilon_{\mathbf{k}+\mathbf{q}/2}) - \Theta(E_{F\lambda}-\epsilon_{\mathbf{k}-\mathbf{q}/2})}{\omega + i 0- \mathbf{v} \cdot \mathbf{q}}.
    \label{eq:long-d-dim}
\end{multline}
In terms of the susceptibility $\chi$, the longitudinal mode equation is
\begin{multline}
    \begin{vmatrix}
        \chi_{+} + \chi_{-} & \chi_{+} - \chi_{-}               \\
        \chi_{+} - \chi_{-} & -\frac{1}{g}+ \chi_{+} + \chi_{-}
    \end{vmatrix}= 0 \\
    \implies
    -g\sum_{\lambda} \chi_{\lambda}
    + (g\sum_{\lambda}\chi_{\lambda})^{2}
    - (g\sum_{\lambda}\lambda \chi_{\lambda})^{2}
    =0.
    \label{eq:mode-det-2nd}
\end{multline}
This corresponds to the two eigenvalue equations
\begin{equation}
    1 - 2g \sum_{\lambda} \chi_{\lambda} \pm \sqrt{1 + (2g \sum_{\lambda}\lambda \chi_{\lambda})^{2}} =0.
    \label{eq:mode-ev-eqn}
\end{equation}
The positive sign is connected to the longitudinal mode solution in the unpolarized state while the negative sign
is associated with the particle-hole excitations of the unpolarized state.
We thus focus on the $+$ sign solution.

As before expanding in $q/k_{F}$ and taking the $T\to0$ limit we have
\begin{multline}
    \chi_{L\lambda}(s)\approx
    -
    \frac{1}{2}
    \nu_{F\lambda}\oint_{FS}
    \frac{\cos\theta}{\frac{v_{F}}{v_{F\lambda}}s+ i 0- \cos\theta}\\
    =
    \frac{1}{2}\nu_{F\lambda}\left(1
    -
    s\frac{v_{F}}{v_{F\lambda}}\oint_{FS}
    \frac{1}{\frac{v_{F}}{v_{F\lambda}}s+ i 0- \cos\theta}
    \right).
    \label{eq:chi-L-lambda}
\end{multline}

\subsection{3D}\label{sec:partial-coll-3d}

We first solve for the longitudinal mode in three dimensions.
We now evaluate
\begin{equation}
    f(z) \equiv z\oint_{FS}
    \frac{1}{z+ i 0- \cos\theta}
    \label{eq:fs}
\end{equation}
so that we can express \cref{eq:chi-L-lambda}
\begin{equation}
    \chi_{L\lambda} =
    \frac{1}{2}\nu_{F\lambda}(1-f\left(\frac{v_{F}}{v_{F\lambda}}s+i0\right)).
\end{equation}
We first note that there is a pole infinitesimally close to the integration contour for $-1<s<1$ indicating a branch cut just below the real axis on $[-1,1]$.
For a three-dimensional system we have
\begin{equation}
    f^{3D}(z) \equiv \frac{z}{2}
    \int_{-\frac{\pi}{2}}^{\frac{\pi}{2}}\frac{\sin\theta}{z- \cos\theta}
    = \frac{z}{2}\ln\frac{z + 1}{z-1}.
    \label{eq:f3D}
\end{equation}
Before returning to \cref{eq:mode-det-2nd,eq:chi-L-lambda} we also compute the functions
\begin{equation}
    \tilde{\nu}_{\pm}  = \nu_{F\pm}/\nu_{F}, \quad \tilde{v}_{\pm} = v_{F\pm}/v_{F}.
\end{equation}
In terms of the Fermi energies and momenta
\begin{equation}
    \tilde{\nu}^{3D}_{\pm}  = \sqrt{\frac{E_{F\pm}}{E_{F0}}}.
\end{equation}
$E_{F\pm}$ are related to $E_{F0}$ through the densities and polarizations and therefore so are $k_{F\pm}$ and $k_{F0}$,
\begin{equation}
    \frac{n^{3D}_{\pm}}{n_{0}} = \frac{k^{3}_{F\pm}}{k^{3}_{F0}} = \frac{E^{3/2}_{F\pm}}{E^{3/2}_{F0}}.
    \label{eq:nvsn03d}
\end{equation}
Using, as in the main text, $n_{\pm} = n_{0}(1\pm\zeta)$ we have
\begin{equation}
    \tilde{\nu}^{3D}_{\pm}  = (1\pm\zeta)^{1/3},\quad \tilde{v}^{3D}_{\pm} = (1\pm\zeta)^{1/3}.
\end{equation}
And plugging this into \cref{eq:f3D} we have
\begin{equation}
    \tilde{\nu}^{3D}_{\lambda}
    f^{3D}\left(\frac{s}{\tilde{v}^{3D}_{\lambda}}\right)
    = s\coth^{-1}\frac{s}{(1 + \lambda\zeta)^{1/3}},
\end{equation}
where we recall that $s$ is being analytically continued from upper half plane.
We can thus write
\begin{equation}
    g\chi^{3D} _{\lambda} =
    \frac{1}{2}\nu_{F}g\left((1+\lambda\zeta)^{1/3} - s \coth^{-1}\frac{s}{(1+\lambda\zeta)^{1/3}} \right).
\end{equation}

We expand $g \chi$ to second order in $\zeta$
\begin{equation}
    g\chi^{3D}  _{\lambda} \approx
    \frac{1}{2}\nu_{F}g\left(
    1 - s\coth^{-1}s + \frac{\lambda\zeta}{3(1-s^{2})} + \frac{\zeta^{2}}{9(1-s^{2})^{2}}
    \right)
\end{equation}
We therefore have
\begin{equation}
    \begin{gathered}
        \sum_{\lambda} g\chi^{3D} _{\lambda}
        =
        \nu^{3D} g \left(
        1 - s\coth^{-1}s - \frac{\zeta^{2}}{9(1-s^{2})^{2}}
        \right)\\
        \sum_{\lambda} \lambda g\chi^{3D} _{\lambda}
        =
        \nu^{3D} g \frac{\zeta}{3(1-s^{2})}
    \end{gathered}
    \label{eq:chi3D}
\end{equation}
Since $\sum_{\lambda}g\chi_{\lambda}$ is order $\zeta$ we can expand the square root
in \cref{eq:mode-ev-eqn}
\begin{equation}
    2 \left( 1 -
    \sum_{\lambda} g\chi_{\lambda}
    + \left[\sum_{\lambda} \lambda g\chi_{\lambda}\right]^{2}
    \right) = 0.
    \label{eq:sqrt-expansion}
\end{equation}
Plugging in \cref{eq:chi3D}
\begin{multline}
    1 -
    \nu^{3D} g \left(
    1 - s\coth^{-1}s - \frac{\zeta^{2}}{9(1-s^{2})^{2}}
    \right)\\
    +
    \left(
    \nu^{3D} g \frac{\zeta}{3(1-s^{2})}\right)^{2} = 0.
\end{multline}
Combining terms
\begin{equation}
    s\coth^{-1}s =
    -\frac{1 -
        \nu^{3D} g}{4\nu^{3D} g}
    - (1+
    \nu^{3D} g)\frac{\zeta^{2}}{9(1-s^{2})^{2}}.
\end{equation}
For $0<\delta = \nu_{F}g-1 \ll 1$, we should have $|s|\ll1$.
We thus expand to lowest non-trivial order in $\delta$ and $s$ to find
\begin{equation}
    -\frac{i\pi}{2}s\approx
    \frac{-\delta}{-1}
    - 2\frac{\zeta^{2}}{9}
    \implies
    s \approx -i\frac{2}{\pi}
    \left(
    \frac{2}{9}
    \zeta^{2}
    -\delta
    \right).
    \label{eq:3dlongmode}
\end{equation}
So the longitudinal mode will be stable for $\zeta^{2}> (9/2)\delta$.
We can confirm that this is the case by solving the mean-field equations.

Recalling that
\begin{equation}
    n_{\lambda}  = \frac{k_{F\lambda}^{3}}{6\pi^{2}},\quad E_{F\lambda} = \frac{k^{2}_{F\lambda}}{2m},\quad
    \nu_{F} = \frac{2m (6\pi^{2}n_{0})^{1/3}}{4\pi^{2}}.
\end{equation}
we have
\begin{multline}
    u_{K}(n_{\lambda})
    = \frac{1}{2m} \int^{n_{\lambda}}_{0} dn' \left(6\pi^{2}n'\right)^{2/3}\\
    = \frac{1}{\nu_{F}} \frac{3n_{0}^{1/3}}{2}\int^{n_{\lambda}}_{0} dn' \left(n'\right)^{2/3}
    = \frac{1}{\nu_{F}} \frac{9n_{0}^{1/3}n^{5/3}_{\lambda}}{10}.
\end{multline}

So the Landau energy is
\begin{multline}
    \delta U_{MF} = 2 \sum_{\pm}(u_{K}(n_{0}(1\pm\zeta)) - u_{K}(n_{0}))
    - 2g_{z}n^{2}_{0} \zeta^{2}\\
    =
    \frac{2n^{2}_{0}}{\nu_{F}}
    \left(
    \frac{9[\sum_{\pm}(1\pm\zeta)^{5/3}-2]}{10}
    - g_{z}\nu_{F}\zeta^{2}
    \right)\\
    \approx
    \frac{2n^{2}_{0}}{\nu_{F}}
    \left(
    -\delta \zeta^{2} + \frac{\zeta^{4}}{3^{3}}
    \right).
\end{multline}
Minimizing with respect to $\zeta$ we find
\begin{equation}
    \zeta^{2}  \approx \frac{3^{3}\delta}{2} > \frac{9}{2}\delta,
\end{equation}
and the longitudinal mode is stable, since
$s$ in \cref{eq:3dlongmode} remains in the lower half plane.

This stable mode persists for finite polarization but eventually vanishes
when the order parameter reaches its maximum value, and one of the bands becomes fully depleted.
Taking the derivative of the Landau functional we have that the system reaches full polarization when
\begin{equation}
    \left.\frac{d U_{MF}}{d\zeta}\right|_{\zeta=1} = 0
    \implies
    \delta =\delta_{\text{full}} = g\nu - 1 =  \frac{3 - 2^{4/3}}{2^{4/3}} \approx 0.2.
\end{equation}
For $\delta$ slightly below this point we have that the occupied Fermi surface is much larger than the unoccupied.
From our parameterization
\begin{equation}
    \frac{k^{3}_{F+}}{k^{3}_{F}} = 1 + \zeta,\quad\frac{k^{3}_{F-}}{k^{3}_{F}} = 1 -\zeta.
\end{equation}
The longitudinal sound mode exists for the range of momenta $q \ll k_{F\pm}$ where were able to linearize the difference of Fermi functions in \cref{eq:long-d-dim}.
At
$\delta  \to \delta_{\text{full}}$, where the minority Fermi surface disappears, $\zeta \to \pm 1$, and the region, over which the longitudinal mode is defined, shrinks and vanishes at
$\delta = \delta_{\text{full}}$.
At larger $\delta$, longitudinal mode does not exist.
Note that the ``velocity'' of the mode and the residue of the longitudinal pole in the susceptibility remain finite at $\delta \to  \delta_{\text{full}}$.

\subsection{2D}\label{sec:partial-coll-2d}

We now repeat the calculation in two dimension
In the same way
as in 3D,
we
write
\begin{equation}
    \chi^{2D}_{L\lambda} =
    \frac{1}{2}\nu_{F\lambda}(1-f^{2D}\left(\frac{v_{F}}{v_{F\lambda}}s+i0\right)).
\end{equation}
in terms of
\begin{equation}
    f^{2D} (z) \equiv
    z\int^{2\pi}_{0}\frac{d\theta}{2\pi}
    \frac{1}{z- \cos\theta},
\end{equation}
which can be rewritten by breaking up the integration region
\begin{equation}
    f^{2D} (z) \equiv
    4z^{2}\int^{\pi/2}_{0}\frac{d\theta}{2\pi}
    \frac{1}{z^{2}- \cos^{2}\theta}
    = \sqrt{\frac{z^{2}}{z^{2}-1}}.
    \label{eq:f2D}
\end{equation}
In terms of the Fermi energies and momenta
\begin{equation}
    \tilde{v}_{\pm} =\frac{k_{F\pm}}{k_{F0}},
\end{equation}
and
\begin{equation}
    \frac{n^{2D}_{\pm}}{n_{0}} = \frac{k^{2}_{F\pm}}{k^{2}_{F0}} = \frac{E_{F\pm}}{E_{0}}.
\end{equation}
So in terms of $\zeta$
\begin{equation}
    \tilde{\nu}^{2D}_{\pm} = 1,\quad \tilde{v}^{2D}_{\pm} = (1\pm\zeta)^{1/2}.
\end{equation}
And plugging this into \cref{eq:f2D} we have
\begin{equation}
    \tilde{\nu}^{2D}_{\lambda}
    f^{2D}\left(\frac{s}{\tilde{v}^{2D}_{\lambda}}\right)
    =
    \sqrt{\frac{s^{2}}{s^{2} - 1 - \lambda\zeta}}.
\end{equation}
We can thus write
\begin{equation}
    g\chi^{2D} _{\lambda} =
    \frac{1}{2}\nu_{F}g\left(1 - \sqrt{\frac{s^{2}}{s^{2} - 1 - \lambda\zeta}}\right)
\end{equation}

From this we can evaluate
\begin{equation}
    \begin{gathered}
        \sum_{\lambda} g\chi^{2D} _{\lambda}
        \approx
        \nu^{2D}_F g \left(1 -\sqrt{\frac{s^{2}}{s^{2}-1}}\left[1 + \frac{3}{8}\frac{\zeta^{2}}{(1-s^{2})^{2}}\right]\right)\\
        \sum_{\lambda} \lambda g\chi^{2D} _{\lambda}
        =
        \nu^{2D}_F g\sqrt{\frac{s^{2}}{s^{2}-1}}\frac{\zeta}{2(1-s^{2})}.
    \end{gathered}
\end{equation}
We can again expand the square root in the mode equation and plug into \cref{eq:sqrt-expansion}
\begin{multline}
    1 -
    \nu^{2D}_F g +
    \nu^{2D}_F g\left(\sqrt{\frac{s^{2}}{s^{2}-1}}\left[1 + \frac{3}{8}\frac{\zeta^{2}}{(1-s^{2})^{2}}\right]\right)\\
    + \left(
    \nu^{2D}_F g\right)^{2}\left(\sqrt{\frac{s^{2}}{s^{2}-1}}\frac{\zeta}{2(1-s^{2})}\right)^{2} = 0.
\end{multline}

If we again expand in $s$ and $\delta$ assuming a 2nd order transition,
noting that the term  $ \sum_{\lambda} \lambda g\chi^{2D} _{\lambda} $ is already order $s^{2}$, we find
\begin{equation}
    -i s\left[1 + \frac{3}{8}\zeta^{2}\right]
    = \delta
    \implies
    s
    = \frac{i\delta}{1 + \frac{3}{8}\zeta^{2}}
\end{equation}
Note that we cannot choose $\zeta^{2}\propto \delta$ such that the pole is in the lower half-plane.
We thus must conclude that the state we have assumed is unstable.
If we were to prepare a state with initially small polarization as we would expect from a second order transition, the longitudinal mode will proliferate exponentially and we will flow toward a state of maximal polarization as expected from the discussion in the main text.
As we approach full polarization, however,
\begin{equation}
    k^{2}_{F+} = 2k_{F}^{2} -k_{F-}^{2},\quad k_{F-} \ll k_{F}.
\end{equation}
As $k_{F-}\to0$ the region $q \ll k_{F-}$ where the unstable mode exists shrinks to $0$, and eventually the at $k_{F-}=0$ the system becomes stable.

\section{Collective modes for non-equivalent couplings}\label{app:coll_m}

\subsection{Longitudinal mode}

The behavior
in the longitudinal sector is
the same for the easy-axis, easy-plane, and Heisenberg cases, only depending on the value of the coupling constant $g_{z}$.
To obtain
the longitudinal modes we solve
\begin{equation}
    \det\hat{D}^{-1}_{L}= -\frac{1}{g_{d}g_{z}} + \left(\frac{1}{g_{d}} - \frac{1}{g_{z}}\right)\chi^{R}_{L}= 0.
    \label{eq:detDL}
\end{equation}
Comparing the expressions for the retarded susceptibilities in this case and in the unpolarized state, \cref{eq:long-susc,eq:chi-normal},  we see that the only difference is the replacement $E_{F}\to E_{F+}$.
Using the unpolarized state result for $\chi^R$ and plugging
the susceptibility into \cref{eq:detDL} we obtain
the equation for the dispersion of the longitudinal mode
\begin{equation}
    1
    -
    \frac{i s_{+}}{\sqrt{1 - (s_{+}+i0)^{2}}} = \frac{2}{\nu_{F+}(g_{z} - g_{d})}
\end{equation}
where $s_{+}\equiv \omega/v_{F+}q$  is defined in terms of the Fermi velocity of the occupied band.
This is the equation for the zero sound mode for a single partly occupied band of fermions with an effective
interaction $g_\text{eff} = (g_{z}-g_{d})/2$.
Because $g_d >0$, $g_\text{eff} < g_z/2$.
Near the transition, at $\nu g_z \approx 1$, we then have $\nu g_\text{eff} <1/2$.
The zero mode in the density channel is then either a
`mirage' mode ($0<\nu_{F+}g_{\text{eff}}<1/2$) or a propagating zero sound mode ($g_{\text{eff}}<0$)~\cite{Klein2020,Raines2021}.
We emphasize that this is the only collective mode of the longitudinal sector per spin projection,   because one of the bands is fully depleted.
We therefore `lose' a mode when going from the unpolarized state to the full polarization state.

For larger $g_z$ we may have $\nu g_z >1/2$, in which case the zero-sound mode becomes an overdamped one.
For even stronger $g_z$,
it is possible to reach $\nu_{F+}g_\text{eff}>1$, at which point there a secondary instability towards phase separation within the ordered state.

\subsection{Transverse mode}

We next analyze transverse modes.
In the unpolarized state, these are modes in non-critical channels (in the IVC channel near a VP instability,  and in the VP channel near an IVC instability).
In the ordered state (VP or IVC), these modes are modified by the order.

To analyze these modes, we set $T=0$ and expand
$\chi^{R}_{-+}$ to lowest order in $q/k_{F+}$ as
\begin{multline}
    \chi^{R}_{-+}(\omega, \mathbf{q}) \approx\\
    -\frac{1}{2}
    \int_{0}^{E_{F+}} d\epsilon\oint\frac{d\theta}{2\pi}
    \frac{\nu(\epsilon)}{\omega + i 0- v q \cos\theta- 2\Delta}\\
    -
    \frac{1}{4}
    \nu_{F+}\oint \frac{d\theta}{2\pi}
    \frac{v_{F+}q \cos\theta}{\omega + i 0- v_{F+}q\cos\theta- 2\Delta}
    \label{eq:chi-pm}
\end{multline}
where $v_{F+}$ and $\nu_{F+}$ are the Fermi velocity and density of states per spin and valley of the occupied band.

It is useful to consider each of the three cases discussed in \cref{sec:mf} (easy-axis, easy-plane, Heisenberg) separately, as the mode dispersion in the transverse sector depends on the relations between the interaction constants.

\subsubsection{Easy-axis}
We begin with the \emph{easy-axis} case in which $g_{x} = g_{y} = g_{\text{IVC}}$ and $g_{z}=g_{\text{VP}}$.
Here the ordering happens along the easy-axis with coupling $g_{\text{VP}}$ while there is a weaker interaction $g_{\text{IVC}}$ in the easy-plane,
$g_{\text{VP}} > g_{\text{IVC}} > 0$
The mode condition,
\cref{eq:rpa-trans},
may be simplified by noting that
in the easy-axis case,
the circular polarization vectors $\mathbf{e}_{\pm} \equiv (1, \pm i)^{T}$ are eigenvectors of $\hat{D}^{-1}_{T}$ with eigenvalues $g^{-1}_{\text{IVC}} - 2\chi^{\pm\mp}$.
The mode condition then is simply
\begin{equation}
    1 -  2 g_{\text{IVC}}\chi^{R}_{\mp\pm}(\pm\omega_\text{IVC}(\mathbf{q}), \mathbf{q}) = 0
    \label{eq:ising-mode-eqn}
\end{equation}
where we have defined the transverse mode frequency $\omega_\text{IVC}(\mathbf{q})$ and used $\chi^{R}_{+-}(\omega, \mathbf{q})=\chi^{A}_{-+}(-\omega,-\mathbf{q})$.
In the easy-axis state, there is no Goldstone mode, as we are only breaking a discrete symmetry ($\tau_{z}\to-\tau_{z}$).
As we shall see, the transverse modes of the easy-axis state are indeed gapped.

We find the energy gap for the transverse mode by solving \cref{eq:ising-mode-eqn} at $vq\ll\omega$.
In this limit \cref{eq:chi-pm} becomes
\begin{equation}
    \chi^{R}_{-+}(\omega, 0) \approx
    -\frac{1}{2}
    \int_{0}^{E_{F+}} d\epsilon\oint\frac{d\theta}{2\pi}
    \frac{\nu(\epsilon)}{\omega + i 0- 2\Delta}.
\end{equation}
The angular integration is trivial while the energy integration simply gives the total density.
As a result,
\begin{equation}
    \chi^{R}_{-+}(\omega, 0) \approx
    -\frac{1}{4}
    \frac{n}{\omega + i 0- 2\Delta}.
    \label{eq:chi-pm-0}
\end{equation}
Plugging \cref{eq:chi-pm-0} back into \cref{eq:ising-mode-eqn} and solving for $\omega_\text{IVC}(0)$ we find
\begin{equation}
    \omega_\text{IVC}(0) = 2\Delta
    \left(
    1  - \frac{g_{\text{IVC}}}{g_{\text{VP}}}
    \right) =
    \frac{n}{2}
    (g_{\text{VP}}- g_{\text{IVC}})
    \label{eq:ising-rest-mass}
\end{equation}
where in the last equality we have used the self-consistency condition \cref{eq:self-consistency} in the form $\Delta= (1/4) g_{z}n $.
Because $g_{\text{VP}} > g_{\text{IVC}}$, $\omega_\text{IVC}(0)$ is positive.

Expanding next \cref{eq:ising-mode-eqn} to order $q^2$, we obtain after simple algebra the mode dispersion in the form
\begin{equation}
    \omega_\text{IVC}(q) =
    \frac{n}{2}
    (g_{\text{VP}}- g_{\text{IVC}}) + \frac{q^{2}}{2M_\text{IVC}},
    \label{eq:ising-dispersion}
\end{equation}
with the kinetic mass term
\begin{equation}
    M^{-1}_{\text{IVC}} \equiv
    \frac{1}{n}
    \left[
        \frac{\partial C}{\partial E_{F+}}
        -
        \frac{4C}{n g_{\text{IVC}}}
        \right]
    \label{eq:Midef}
\end{equation}
where
\begin{equation}
    C\equiv
    \int_{0}^{E_{F+}} d \epsilon \nu(\epsilon) v^{2}(\epsilon)
    = \int_{0}^{2n_{0}} dn' v^{2}_{F}(n').
    \label{eq:C}
\end{equation}
For our power-law model \cref{eq:eps-power} we can straightforwardly calculate,
\begin{equation}
    \begin{gathered}
        C = \frac{c^{2 }\alpha^{2} (2n^{2}_{0})^{2\alpha}}{2 \pi}\\
        \frac{\partial C}{\partial E_{F+}} = \nu_{F+}v_{F+}^{2} = \frac{c\alpha}{\pi} (2n_{0})^{\alpha}
    \end{gathered}
\end{equation}
and thus obtain the explicit form of the kinetic mass
\begin{equation}
    M^{-1}_{\text{IVC}} \equiv
    \frac{1}{2\pi \nu_{F+}}
    \left[
    1
    -
    \frac{1}{
    \alpha \nu_{F+}g_{\text{IVC}}}
    \right].
    \label{eq:Mipower}
\end{equation}

The transverse modes of the easy-axis state are qualitatively similar to Silin modes in a Fermi-liquid with an applied Zeeman field~\cite{Silin1958}.
Silin modes are excitations of a Fermi liquid in the presence of a Zeeman field and can be thought of as precession of the collective spin about the external field.
This idea has previously been generalized to materials with spin and valley degrees of freedom in Ref.~\onlinecite{Raines2022}.
The transverse modes are seen to be precession modes as they are the circularly polarized eigenvectors of the valley isospin, and like the usual Silin mode, they are quadratically dispersing.
Here the central difference is that there is no external field, but rather the isospin precesses around its spontaneously generated expectation value.
Unlike in the case of the Silin mode, the precession frequency is renormalized down from the Larmor frequency.

\subsubsection{Easy-plane}

In the easy plane case, we have $g_{x}=g_{z}=g_{\text{IVC}}$ and $g_{y}=g_{\text{VP}}$, so that the ordering vector lies in the easy-plane.
We again can write $\hat{D}^{-1}_{T}$ in the basis of circularly polarized vectors, but in this case the matrix is no longer diagonal.
In terms of
$g^{-1}_{\pm}=(1/2)(g^{-1}_{\text{VP}}\pm g^{-1}_{\text{IVC}})$, \cref{eq:rpa-trans} becomes
\begin{multline}
    0 =
    \begin{vmatrix}
        \frac{1}{g_{+}} - 2 \chi^{R}_{-+} & \frac{1}{g_{-}}                  \\
        \frac{1}{g_{-}}                   & \frac{1}{g_{+}} - 2\chi^{R}_{+-}
    \end{vmatrix}\\
    = (\frac{1}{g_{+}} - 2 \chi^{R}_{+-})(\frac{1}{g_{+}} - 2 \chi^{R}_{-+}) -\frac{1}{g_{-}^{2}}.
    \label{eq:easy-circular}
\end{multline}

We first note that
\cref{eq:easy-circular} is satisfied at $q=0, \omega\to0$ because
\begin{equation}
    \chi^{R}_{+-} (\omega\to0,\mathbf{q})
    = \chi^{R}_{-+} (\omega\to0,\mathbf{q})
    = \frac{n}{8 \Delta} = \frac{1}{2 g_{\text{IVC}}}
\end{equation}
where we used $\Delta = g_{\text{IVC}} n_0 =  g_{\text{IVC}}  n/4$.
This solution corresponds to the eigenvector $\mathbf{e}_{x}$.
As we shall see this, this is the Goldstone mode describing rotation of the mean field state in the easy-plane.
To see this, we express the susceptibility as
$\chi^{R}_{-+}(\omega,q) = \chi^{A}_{+-}(-\omega,q) = \tilde{\chi}^{R}_{T}(\omega,q) + \frac{1}{2g_{\text{IVC}}}$  and
express the mode condition as
\begin{equation}
    0 = \left(\frac{1}{g_{-}} - 2 \tilde{\chi}^{R}_{T}(\omega,q)\right)\left(\frac{1}{g_{-}} - 2 \tilde{\chi}^{A}_{T}(-\omega,q)\right) - \frac{1}{g_{-}^{2}}.
    \label{eq:gsdet}
\end{equation}
Using the expansion \cref{eq:chi-pm} and $\Delta = g_{\text{IVC}}  n/4$, we can express $\tilde{\chi}^{R}_{T}(\omega,q)$ as
\begin{multline}
    \tilde{\chi}^{R}_{T}(\omega, q) =
    -\frac{1}{2g_{\text{IVC}}}\frac{\omega}{\omega + i0 -2\Delta}\\
    -
    \frac{1}{2}\smashoperator{\int_{0}^{E_{F+}}} d\epsilon\oint\frac{d\theta}{2\pi}
    \nu(\epsilon)
    \frac{v q \cos\theta}{(\omega + i 0- v q \cos\theta- 2\Delta) (\omega + i 0- 2\Delta)}\\
    -
    \frac{1}{4} \nu_{F+}\oint \frac{d\theta}{2\pi}
    \frac{v_{F+}q \cos\theta}{\omega + i 0- v_{F+}q\cos\theta- 2\Delta}
    \label{eq:chi-tilde}
\end{multline}
Expanding to second order in $\omega/(2\Delta)$ and $v_{F+}q/\Delta$, we obtain
\begin{multline}
    \tilde{\chi}^{R}_{T}(\omega, q)\approx
    \frac{1}{8}
    \frac{\omega}{2g_{\text{IVC}}\Delta}
    \left[
        1 + \frac{\omega}{2\Delta}
        \right]\\
    -\frac{1}{2}
    \frac{q^{2}}{4\Delta^{2}}
    \left[
    \nu_{F+}v^{2}_{F+}
    -\frac{
        4C}{ng_{\text{IVC}}}
    \right]
    \label{eq:chiT-expansion}
\end{multline}
where $C$ is the same integral as in \cref{eq:C}.
Working to this same order, \cref{eq:gsdet} can be rewritten as
\begin{equation}
    \frac{\omega^{2}}{4\Delta^{2}}
    \frac{1}{2g_{\text{IVC}}g_{\text{VP}}}
    =\frac{1}{4g_{-}}
    \frac{q^{2}}{4\Delta^{2}}
    \left[
    \nu_{F+}v^{2}_{F+}
    -\frac{4C}{ng_{\text{IVC}}}
    \right].
\end{equation}
We thus find a linearly dispersing Goldstone mode
\begin{equation}
    \omega^{2}_{GS} = v^{2}_{GS}q^{2}
    \label{eq:gsdis}
\end{equation}
with velocity
\begin{equation}
    v^{2}_{GS}  = \frac{1}{2} \frac{g_{\text{IVC}}g_{\text{VP}}}{g_{-}}\left[\nu_{F+}v^{2}_{F+} - \frac{4C}{g_{\text{IVC}}n}\right].
\end{equation}
Note the similarity to the expression for the kinetic mass of the mode in the easy-axis case \cref{eq:Midef}.
We may in fact use
\cref{eq:ising-rest-mass} and
rewrite the velocity as
\begin{equation}
    v^{2}_{GS} = \frac{-\omega_\text{IVC}(0)}{2M_\text{IVC}}.
\end{equation}
In the easy-plane state, $\omega_\text{IVC}(0)<0$, so the velocity of the Goldstone mode is real.
This reflects the fact that the precession mode in the easy-axis case becomes soft as $g_{\text{IVC}}\to g_{\text{VP}}$ and becomes the Goldstone mode in the easy-plane state.

Note that while the Goldstone mode intuitively corresponds to the broken generator $\hat{\tau}_{x}$, there is no other transverse mode corresponding to $\hat{\tau}_{y}$.
This can be understood by considering the action of collective mode operators
\begin{equation}
    \hat{\Phi}_{\mathbf{q};x,y} \equiv \sum_{\mathbf{k}}c^{\dagger}_{\mathbf{k}-\mathbf{q}\sigma}\hat{\tau}_{x,y}c_{\mathbf{k}\sigma}
\end{equation}
on the ground state
\begin{equation}
    \ket{\Omega}  \equiv \sum_{\mathbf{k}}c^{\dagger}_{\mathbf{k}\sigma}\frac{1}{2}\left(\hat{\tau}_{0}+\hat{\tau}_{z}\right)c_{\mathbf{k}\sigma}\ket{0}.
\end{equation}
Because only the $+$ band is occupied, we have
\begin{multline}
    \hat{\Phi}_{\mathbf{q},x}\ket{\Omega} =
    \sum_{\mathbf{k}}c^{\dagger}_{\mathbf{k}-\mathbf{q}\sigma,\tau=-}c_{\mathbf{k}\sigma,\tau=+}\ket{\Omega}\\
    =
    -i \sum_{\mathbf{k}}c^{\dagger}_{\mathbf{k}-\mathbf{q}\sigma,\tau=-}ic_{\mathbf{k}\sigma,\tau=+}\ket{\Omega}
    = -i \hat{\Phi}_{\mathbf{q},y}\ket{\Omega}.
\end{multline}
So the $\hat{\Phi}_{x}$ and $\hat{\Phi}_{y}$ modes correspond to the same state and there is thus only the single mode.

\subsubsection{Heisenberg}

In the Heisenberg case all couplings are equal, $g_{x}=g_{y}=g_{z} =g$.
The Heisenberg limit
may be reached from the easy-axis state solution~\cref{eq:ising-dispersion}, taking $g_{\text{IVC}} \to g_{\text{VP}}$, or from the easy-plane solution,~\cref{eq:gsdet}, taking $g_{\text{VP}}\to g_{\text{IVC}}$.
Either way, we obtain
\begin{equation}
    \omega_{M}(q) = \frac{q^{2}}{2M_\text{IVC}},
\end{equation}
This is a quadratic magnon dispersion, expected in the Heisenberg model, and is the Goldstone mode of the Heisenberg state.

\section{Comparison of the ground state energy in \cref{eq:energy} with the one in Ref.~\onlinecite{Chichinadze2022}}
\label{app:comparison}

It is instructive to compare the explicit result for the ground state energy, expanded to quartic order,
\cref{eq:energy}, with the phenomenological expression, obtained in Ref.~\onlinecite{Chichinadze2022} using symmetry considerations.
For the convenience of the reader we reproduce here \cref{eq:energy}:
\begin{multline}
    \delta u =
    \frac{1}{2}
    (1-\nu_{F}g)(\zeta^{2}_{1}+\zeta^{2}_{2}+\zeta^{2}_{3})
    + (\alpha - 1) \zeta_{1}\zeta_{2}\zeta_{3}\\
    + \frac{(\alpha - 1)(\alpha-2)}{8}
    \left[(\zeta_{1}^{2}+\zeta^{2}_{2}+\zeta^{2}_{3})^{2}
        -
        \frac{2}{3}
        \left(
        \zeta_{1} ^{4}+ \zeta_{2}^{4}+ \zeta_{3}^{4}
        \right)
        \right]\\
    + \cdots.
    \label{energy_1}
\end{multline}
where $\zeta_i$,
each subject to $-1 \leq \zeta_i \leq 1$,
are the coefficients in the expression for the density
in each band
\begin{equation}
    n_{\lambda\rho} = n_{0}(1 + \lambda \zeta_1 + \rho \zeta_2 + \lambda \rho \zeta_3)
\end{equation}
and where $n_{0}$ is the normal state density per spin and valley and $\lambda, \rho = \pm 1$.

In [\onlinecite{Chichinadze2022}], the energy is expressed in terms of three component of the traceless $\mathrm{SU(4)}$ matrix $\hat{\Phi}$ in the diagonal basis: $\hat{\Phi} = \diag(\lambda_1, \lambda_2,\lambda_{3}, -(\lambda_1+\lambda_2+\lambda_{3}))$.
The energy in terms of $\lambda_i$ is
\begin{multline}
    \mathcal{F} = - \frac{
        a}{2} \left( \sum_j^{3} \lambda_j^2 + (\sum_j^{3} \lambda_j)^2 \right) + \frac{\gamma}{3}
    \left( \sum_j^{3} \lambda_j^3 - (\sum_j^{3} \lambda_j)^3 \right)  \\
    + \frac{\beta}{4} \left( \sum_j^{3} \lambda_j^4 + (\sum_j^{3} \lambda_j)^4 \right)
    + \frac{\beta'}{4}\left( \sum_j^{3} \lambda_j^2 + (\sum_j^{3} \lambda_j)^2 \right)^2 + \ldots
    \label{Lagr2}
\end{multline}
Here $a \propto 1-\nu_{F\alpha} g$, and $\gamma$, $\beta$, and $\beta'$ are phenomenological parameters.

To compare the two expressions we realize that the quadratic form in \cref{Lagr2} can be brought
to sum of the three squares, as in \cref{energy_1}, if we choose
\begin{equation}
    \zeta_1 = \lambda_2+\lambda_3, \quad\zeta_2 = \lambda_1+\lambda_3, \quad\zeta_3 = \lambda_1+\lambda_2.
\end{equation}
Using these relations we find
\begin{equation}
    \begin{gathered}
        \sum_j^{3} \lambda_j^2 + (\sum_j^{3} \lambda_j)^2 = \zeta^2_1 + \zeta^2_2 + \zeta^2_3                                                                                        \\
        \sum_j^{3} \lambda_j^3 - (\sum_j^{3} \lambda_j)^3 = -3 \zeta_1 \zeta_2 \zeta_3 \\
        \sum_j^{3} \lambda_j^4 + (\sum_j^{3} \lambda_j)^4 = \frac{3}{2} \left(\zeta^2_1 + \zeta^2_2 + \zeta^2_3\right) - \frac{1}{2} \left(\zeta^4_1 + \zeta^4_2 + \zeta^4_3\right)
    \end{gathered}
\end{equation}
Substituting into \cref{Lagr2} and comparing with \cref{energy_1} we immediately obtain
\begin{equation}
    \begin{gathered}
        a= - (1 -\nu_{F,\alpha} g),\quad \gamma = 1-\alpha, \\
        \beta = \frac{2(\alpha -1)(\alpha -2)}{3},\quad \beta' = - \frac{9 \beta}{4}.
    \end{gathered}
\end{equation}
Observe that $\beta <0$ when $\alpha$ is between $1$ and $2$.

\end{document}